\documentclass[12pt,a4j,epsf]{report}

\usepackage{amsmath, amssymb}
\usepackage[dvips]{graphicx}
\usepackage{calrsfs}
\usepackage{fancybox}
\def\tr{\mathop{\rm tr}\nolimits}
\def\Tr{\mathop{\rm Tr}\nolimits}

\def\mylimit#1{\mathrel{\mathop{\kern0pt\longrightarrow}\limits_{#1}}}
\def\myequiv#1{\mathrel{\mathop{\kern0pt\longleftrightarrow}\limits_{#1}}}

\newcommand{\VEV}[1]{\left\langle #1 \right\rangle}
\newcommand{\del}{\partial}
\newcommand{\lag}{{\cal L}}
\newcommand{\nn}{\nonumber}
\newcommand{\diag}{{\rm diag.}}

\newcommand{\Mp}{M_P}
\newcommand{\MGUT}{\Lambda_G}
\newcommand{\MU}{\Lambda_U}
\newcommand{\MSB}{\Lambda_{SB}}
\newcommand{\order}[1]{{\cal O}(#1)}

\newcommand{\TeV}{\mbox{TeV}}
\newcommand{\GeV}{\mbox{GeV}}

\newcommand{\ie}{{\it i.e.}}
\newcommand{\eg}{{\it e.g.}}

\newcommand{\dtf}{\int{\rm d}^4\theta}
\newcommand{\dt}{\int{\rm d}^2\theta}

\newcommand{\Kahler}{K\"ahler}

\newcommand{\GSM}{SU(3)\sub C$\times$SU(2)\sub L$\times$U(1)$_Y$}
\newcommand{\Ga}{SU(3)\sub C$\times$SU(2)\sub L$\times$SU(2)\sub R%
                 $\times$U(1)$_{B-L}$}
\newcommand{\msGUT}{the minimal SU(5) SUSY-GUT\ }
\newcommand{\aGUT}{anomalous U(1) GUT}
\newcommand{\E}[1]{E$_#1$}

\newcommand{\abs}[1]{\left| #1 \right|}
\newcommand{\s}[1]{\tilde{#1}}

\newcommand{\cc}[1]{\overline{#1}}
\newcommand{\sub}[1]{$_{\rm{#1}}$}
\newcommand{\eff}{{\rm {eff}}}

\newcommand{\bequ}{\begin{equation}}
\newcommand{\eequ}{\end{equation}}
\newcommand{\beqn}{\begin{eqnarray}}
\newcommand{\eeqn}{\end{eqnarray}}
\newcommand{\bctr}{\begin{center}}
\newcommand{\ectr}{\end{center}}
\newcommand{\bit}{\begin{itemize}}
\newcommand{\eit}{\end{itemize}}
\newcommand{\Ls}{\left(}
\newcommand{\Rs}{\right)}
\newcommand{\Lm}{\left\{}
\newcommand{\Rm}{\right\}}
\newcommand{\Ll}{\left[}
\newcommand{\Rl}{\right]}
\newcommand{\LL}{\left.}
\newcommand{\RR}{\right.}
\newcommand{\hsp}[1]{\hspace {#1cm}}
\newcommand{\vsp}[1]{\vspace {#1cm}}
\newcommand{\half}{{1\over2}}

\title{\Huge{\bf{Anomalous U(1) GUT}}}
\author{\LARGE Toshifumi Yamashita}
\date{}
\begin{document}
\maketitle
%
%
%
\begin{abstract}\quad
We have proposed a very attractive scenario of Grand Unified 
 Theories (GUTs).
It employs the supersymmetry (SUSY) and an anomalous U(1) 
 symmetry whose anomaly is canceled via the Green-Schwarz 
 mechanism. 
In this scenario, the doublet-triplet splitting problem 
 is solved and the success of the gauge coupling unification 
 in the minimal SU(5) GUT is naturally explained 
 with sufficiently stable nucleon. 
Realistic fermion Yukawa matrices can also be realized 
 simultaneously.
In addition, a horizontal symmetry helps to solve the 
 SUSY-flavor problem. 
\end{abstract}


\tableofcontents

\chapter{Introduction}

The standard model (SM) of particle physics is a very successful model 
 which explains hundreds of precise measurements. 
Theoretically, however, it has many issues to explain. 
For instance, we have not understood the reason why 
 the absolute value of the electric charge of electron coincides 
 with that of proton very accurately. 
The radiative correction induces very large mass to the Higgs scalar, 
 and thus it looks unnatural that the electro-weak (EW) symmetry 
 breaking scale is so small ($\sim\order{100\GeV}$). 
In addition, the SM contains many parameters ($\sim\order{10}$), 
 some of which have hierarchical structure, \eg\ top Yukawa is 
 much larger than up Yukawa as $\frac{Y_u}{Y_t}\sim10^{-5}$. 
Also, it does not treat gravity at all. 
By such reasons, many authors do not consider it 
 as the most fundamental theory, and have proposed various scenarios 
 for the physics beyond the SM. 
Among them, the supersymmetric grand unified theory (SUSY-GUT) 
 is one of the most famous scenarios.  

SUSY-GUTs realize very beautiful unifications 
 of matter fields and forces, and 
 can give natural solutions for many problems of the SM. 
In addition, the simplest scenario realizes 
 the unification of the gauge coupling constants 
 though, unfortunately, 
 it is (almost) excluded by non-observation of nucleon decay. 
However they still have some problems to solve. 
One of the biggest problems is the so-called 
 doublet-triplet splitting (DTS) problem. 
And realizing realistic Yukawa matrices of 
 quarks and leptons is also a big issue. 

I and my collaborators showed almost all problems of 
 SUSY-GUTs are solved with the aid of an anomalous U(1) 
 gauge symmetry\ (U(1)\sub A)\cite{U(1),IR}, 
 whose anomaly is assumed to be canceled 
 via the Green-Schwarz\ (GS) mechanism\cite{GS}. 
In this scenario, we introduce all the possible interactions 
 that respect the symmetry of the theory, and their 
 coupling constants are assumed to be of order one in unit 
 of the cutoff scale of the theory. 
This means that the definition of a model is given by the 
 definition of a symmetry, and there is no need to fix each 
 coupling constant if precise analysis is not needed. 
From such a natural assumption, the DTS problem is solved%
\cite{maekawa}-\cite{sliding}
 and the success of the gauge coupling unification (GCU) 
 in the minimal SU(5) GUT is naturally explained 
 while nucleon is sufficiently stable\cite{1-loop,2-loop,NGCU}. 
Realistic fermion Yukawa matrices, including the neutrino 
 bi-large mixing angles, can also be realized simultaneously%
\cite{maekawa,E6,reduced,BM}. 
In addition, a horizontal symmetry helps to solve the SUSY-flavor 
 problem in E$_6$ models\cite{horiz,HorizHiggs}. 
This analysis may help to construct a realistic E$_8$ unification 
 model.

In this thesis, we summarize these studies of SUSY-GUTs with 
 an anomalous U(1) symmetry(\aGUT s).
In \S\ref{review}, some fundamental ideas of the SM, GUT, SUSY and 
 anomalous U(1) symmetry are briefly reviewed. 
In \S\ref{aGUT}, the starting point and some significant features 
 of the \aGUT\ scenario are explained. 
Some concrete models based on SO(10) or \E6 gauge symmetry 
 are examined in \S\ref{Models}. 
The role of horizontal symmetries in the \aGUT\ scenario 
 is discussed in \S\ref{Modelsw/Horiz}. 
\S\ref{Summary} is for summary.

\chapter{SM, GUT, SUSY and 
            Anomalous U(1) Symmetry}
\label{review}

\section{Standard Model}
\label{SM}

The SM is a renormalizable gauge theory based on 
 $G$\sub{SM}$=$\GSM\ gauge symmetry. 
The corresponding gauge bosons $G$, $W$ and $B$ mediate 
 the strong, weak and electro-magnetic\ (EM) interactions. 
It contains a matter sector of quarks and leptons and 
 a Higgs sector that breaks SU(2)\sub L$\times$U(1)$_Y$ down to 
 the electro-magnetic U(1)\sub{EM} symmetry.

The matter sector consists of three sets of the left-handed quark 
 doublet $Q=(u_{\rm L},d_{\rm L})$, 
 the right-handed up-type quark $U^c=u_{\rm R}^c$, 
 the right-handed down-type quark $D^c=d_{\rm R}^c$, 
 the left-handed lepton doublet $L=(\nu_{\rm L},e_{\rm L})$ and 
 the right-handed charged lepton $E^c=e_{\rm R}^c$. 
It is convenient to add the right-handed neutrino $N^c$ 
 in order to explain non-vanishing neutrino masses 
 reported in Refs.\cite{atmos,solar}. 
Their quantum numbers are shown in the Table\ \ref{QuantumNum} 
 in the left-handed basis. 
\begin{table}
\bctr
\begin{tabular}{|c|c|l|}
\hline
 & spin & (SU(3)\sub C,SU(2)\sub L)$_{{\rm U(1)}_Y}$ \\
\hline\hline
 $G$ &   & \hsp1(${\bf8}$,${\bf1}$)$_0$ \\
 $W$ & 1 & \hsp1(${\bf1}$,${\bf3}$)$_0$ \\
 $B$ &   & \hsp1(${\bf1}$,${\bf1}$)$_0$ \\ 
\hline
 $Q$ &   & \hsp1(${\bf3}$,${\bf2}$)$_\frac16$ \\
 $U^c$&  & \hsp1(${\bf\bar3}$,${\bf1}$)$_{-\frac23}$ \\
 $D^c$&1/2&\hsp1(${\bf\bar3}$,${\bf1}$)$_\frac13$ \\ 
 $L$ &   & \hsp1(${\bf1}$,${\bf2}$)$_{-\half}$ \\
 $E^c$&  & \hsp1(${\bf1}$,${\bf1}$)$_1$ \\
 ($N^c$)& &\hsp1(${\bf1}$,${\bf1}$)$_0$ \\ 
\hline
 $H$ & 0 & \hsp1(${\bf1}$,${\bf2}$)$_{\half}$ \\
\hline
\end{tabular}
\ectr
\caption{
The quantum numbers for the participants of the SM.
}
\label{QuantumNum}
\end{table}
Hereafter, we often use the characters in the first column of the table 
 for representing the quantum numbers in the corresponding 
 third column. 
Interestingly, this set of fermions is anomaly free. 
Namely, all the triangle anomalies including the mixed  anomalies 
 are canceled. 
Anomaly cancellations of SU(3)$^3$ and SU(2)$^3$ are trivial and 
 U(1)$^3$, [SU(3)]$^2\times$U(1), [SU(2)]$^2\times$U(1) and 
 [gravity]$^2\times$U(1) anomalies are evaluated by 
\beqn
 \tr(Q_Y^3) 
  &=& 6\times\frac16^3+3\times\Ls-\frac23\Rs^3+3\times\frac13^3 
     +2\times\Ls-\half\Rs^3+1^3 = 0, \\
 \tr(\Lm T_{\rm C}^a,T_{\rm C}^b\Rm Q_Y) 
  &=& \half\Ls 2\times\frac16+\Ls-\frac23\Rs+\frac13 \Rs = 0, \\
 \tr(\Lm T_{\rm L}^a,T_{\rm L}^b\Rm Q_Y) 
  &=& \half\Ls 3\times\frac16+\Ls-\frac12\Rs \Rs = 0, \\
 \tr(Q_Y) 
 &=& 6\times\frac16+3\times\Ls-\frac23\Rs+3\times\frac13 
    +2\times\Ls-\half\Rs+1 = 0. 
\eeqn
In this way, anomalies of the quark sector and those of 
 the lepton sector cancel out. 
This fact seems to indicate that quarks and leptons have 
 something to do with each other. 
This gives one of the motivations to consider GUTs, which 
 realize unification between quarks and leptons. 

The Higgs sector consists of only one doublet scalar, 
 $H=(H^+,H^0)$. 
The renormalizable Higgs potential is 
\bequ
 V=\mu^2\abs H^2+\frac\lambda2 \Ls\abs H^2\Rs^2.
\label{HiggsPot}
\eequ
If the parameter $\mu^2$ is negative, \ie\ $\mu^2=-m^2$, 
 the equation of motion\ (EOM) requires 
 $\abs H^2={m^2}/\lambda\equiv v^2/2$. 
($\lambda$ should be positive for the stability of the theory.) 
The gauge transformation of SU(2)\sub L$\times$U(1)$_Y$ can 
 make the vacuum expectation values\ (VEVs) of the other components 
 than the real part of $H^0=\frac1{\sqrt2}\Ls h+i\chi\Rs$
 vanishing. 
This means that these modes are the Numb-Goldstone\ (NG) modes 
 which are eaten through the Higgs mechanism. 
In fact by this VEV, three gauge bosons acquire masses 
 proportional to the VEV $v$. 
This is understood by examining the gauge interaction of the Higgs, 
 which is given as 
\bequ
 \abs{D_\mu H}^2=
 \abs{\Ls \del_\mu+ig\frac{\tau^a}2W^a_\mu+ig'Q_YB\mu\Rs H}^2, 
\eequ
 where $\tau^a$'s are the Pauli matrices and $Q_Y$ is 
 the generator of the hypercharge. 
We can see that the $W$ boson 
 $W_\mu^\pm=\frac1{\sqrt2}(W_\mu^1\pm iW_\mu^2)$ 
 and the $Z$ boson which is a linear combination of $W^3$ and $B$, 
 $Z_\mu\propto gW^3_\mu-g'B_\mu$ acquire $m_W=\frac{g}2v$ and 
 $m_Z=\frac{\sqrt{g^2+{g'}^2}}2v$, respectively. 
The remaining gauge boson $A_\mu\propto g'W_\mu^3+gB_\mu$, which 
 is the gauge boson of the EM interaction, remains 
 massless. 

Note that quarks and leptons are vector-like under the 
 remaining EM symmetry. 
Thus, they are expected to acquire masses proportional to the positive 
 power of $v$.
In fact, they acquire masses through the Yukawa interactions: 
\bequ
  {\cal L}_{\rm Yukawa}=(Y_U)_{ij}Q_iU^c_jH+(Y_D)_{ij}Q_iD^c_jH^\dagger
                     +(Y_E)_{ij}L_iE^c_jH^\dagger,
\label{Yukawa}
\eequ
 where $Y_f$'s, $f=U,D,E$, are $3\times3$ Yukawa matrices 
 related with the mass matrices by $M_f=Y_fv$. 
Their elements are complex and therefore they have eighteen real 
 parameters per a matrix. 
On the other hand, we can rename the fields that have a common quantum 
 number,\eg\ ${u_{\rm R}}_i\rightarrow 
             {V_{u_{\rm R}}}_{ij}{u_{\rm R}}_j$ by a unitary matrix 
 $V_{\rm R}$. 
By using these degrees of freedom, we can diagonalize the Yukawa 
 matrices as $Y_f\rightarrow V_{f_{\rm L}}Y_fV_{f_{\rm R}}^\dagger=Y_f^\diag$. 
In this mass basis, mixings between generations appear only in 
 the charged current interactions generated by the $W$ boson, 
 as $\bar u_{\rm L}\Gamma^\mu W_\mu^+ d_{\rm L}\rightarrow
     \bar u_{\rm L}V_{u_{\rm L}}^\dagger\Gamma^\mu W_\mu^+ 
     V_{d_{\rm L}}d_{\rm L}$. 
Thus, the quark mixing is parameterized by only one unitary matrix 
\bequ
 V_{\rm CKM}=V_{u_{\rm L}}^\dagger V_{d_{\rm L}}. 
\label{CKM}
\eequ
Note that we still have degrees of freedom to rotate the phases of quarks 
 although the common phase rotations of quarks because of the 
 accidental baryon number symmetry. 
This means that five phases of $V_{\rm CKM}$ are not physical parameters 
 but three angles and one phase are the physical parameters. 
Note that the neutral current interaction does not change the flavors 
 at the tree level. 
Furthermore, even if we consider quantum corrections, 
 flavor changing neutral current\ (FCNC) processes are suppressed 
 very much by the GIM mechanism, which is consistent with the 
 experimental results. 

As for the lepton sector, if the right-handed neutrinos are not 
 introduced, neutrino cannot have mass. 
Then, all the neutrinos are degenerate and we can rename them freely 
 and thus the lepton mixing vanishes, which is inconsistent 
 with the experiments\cite{atmos,solar}. 
Thus we introduce three right-handed neutrinos, $\nu$\sub R. 
Because these $\nu_{\rm R}$'s are neutral under $G$\sub{SM}, 
 they can acquire Majorana masses much larger than the weak scale. 
Then in the effective theory at the weak scale, 
 $\nu_{\rm R}$'s are almost absent, leading to 
 very tiny left-handed neutrino masses (Seesaw mechanism). 
This is consistent with the neutrino experiments if the Majorana 
 scale is around $\order{10^{13\mbox{-}15}}\GeV$. 

\subsubsection{Problems}

The SM explained above is consistent with almost all the 
 experimental results if we assign the appropriate values 
 to the parameters by hand. 
The number of parameters is much less than that of the 
 experimental results and thus the SM is very successful. 
Nevertheless, we are not satisfied with it. 
It looks unnatural that the Higgs mass parameter $\mu$ 
 in (\ref{HiggsPot}) is the order of the weak scale when there are 
 other scale much larger than the weak scale, such as the Majorana 
 scale and Planck scale.
This is because scalar masses are not protected by any symmetry 
 against the quantum correction and they suffer very large 
 correction. 
This means that we need fine-tuning between a tree level mass and 
 quantum corrections. 
This problem is called as the hierarchy problem. 
Furthermore, 
 the Yukawa matrices also have hierarchical structures, although 
 the degree of the hierarchy is much milder than that 
 in the Higgs mass. 
The hypercharges are assigned to be integers when they are multiplied 
 by six, in spite of the fact that charges of a U(1) symmetry can 
 take any values. 
In addition, it cannot explain why the strong interaction 
 conserves the $\cal{CP}$ symmetry. 
Cosmologically, it cannot explain the baryon asymmetry in universe, 
 it has no candidates for the dark matter, 
 and we do not understand why the dark energy is so small. 
Of course, the quantum gravity is not treated at all.

\section{GUT}
\label{GUT}

As in the SM, the gauge symmetry that is observed 
 by experiments may be different from the symmetry of the theory. 
And chiral fields of the original symmetry may acquire masses 
 if they become vector-like under the reduced symmetry. 
Such masses should be proportional to the symmetry breaking scale, 
 and the vector-like pairs decouple from the low energy effective 
 theory (if the masses are not so small).
This idea is employed in GUTs.
Namely, we can extend the gauge symmetry $G$\sub{SM} to a GUT symmetry 
 $G$ that contain $G$\sub{SM} as a subgroup. 
If $G$ is a semi-simple group, the charge quantization can 
 be explained. 
The simplest example is SU(5). 
In this case, three forces of the SM are unified into a single force. 
The additional gauge bosons $X({\bf3},{\bf2})_{5/6}$ and $\bar X$ of 
 the adjoint representation of SU(5)
\bequ
 {\bf24}\rightarrow G+W+B+X+\bar X
\label{SU(5)24}
\eequ
 acquire masses of the order of the SU(5) breaking scale.

\begin{figure}
\begin{center}
\leavevmode
\put(300,50){{\large $\bf{\log \mu (GeV)}$}}
\put(0,260){{\Large $\bf{\alpha^{-1}}$}}
\put(29,240){$\alpha_1^{-1}$}
\put(31,155){$\alpha_2^{-1}$}
\put(31,95){$\alpha_3^{-1}$}
\includegraphics[width=11cm]{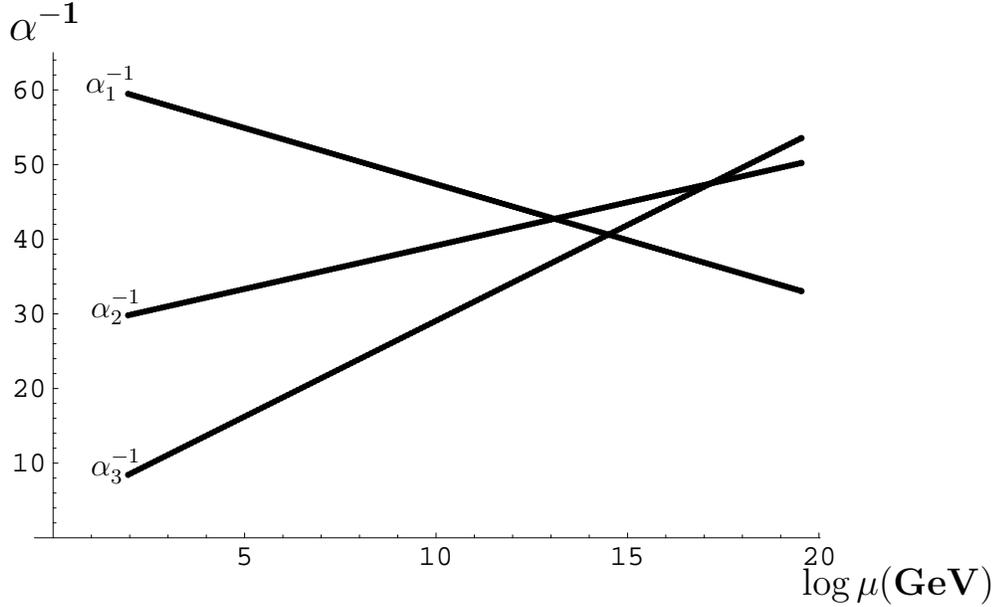}
\vspace{-2cm}
\caption{
The gauge coupling flows in the SM: 
Here we adopt 
$\alpha_1^{-1}(M_Z)=59.47$, $\alpha_2^{-1}(M_Z)=29.81$,
$\alpha_3^{-1}(M_Z)=8.40$.
} 
\label{GCUSM}
\end{center}
\end{figure}

This unification of forces requires unification of the gauge coupling 
 constants. 
Unfortunately, the gauge coupling unification\ (GCU) is not so good 
 in non-SUSY GUTs, although the gauge couplings tend to approach 
 each other, as shown in the Figure\ \ref{GCUSM}.

The unification of forces also requires unification of matter or 
 introduction of additional matter. 
Surprisingly, the matter sector of the SM can be unified without 
 introducing any additional matter fields in SU(5) GUTs: 
\beqn
 {\bf10}&\rightarrow& Q+U^c+E^c, 
\label{SU(5)10}\\
 {\bf\bar5}&\rightarrow& L+D^c, 
\label{SU(5)5bar}\\
 {\bf1}&\rightarrow& N^c. 
\label{SU(5)1}
\eeqn 
In SO(10) GUTs, they can be unified further as 
\bequ
 {\bf16}\rightarrow{\bf10}+{\bf\bar5}+{\bf1}, 
\label{SO(10)16}
\eequ
 that is, each generation, including the right-handed neutrino, 
 can be unified into a single multiplet. 
In \E6 GUT, we need additional matter fields ${\bf10}$ and ${\bf1}$ 
 of SO(10) to embed ${\bf16}$ into the fundamental multiplet 
 of \E6 ${\bf27}$ which is decomposed in terms of SO(10) as 
\bequ
 {\bf27}\to{\bf16}+{\bf10}+{\bf1}.
\label{E627}
\eequ
Note that, even in SU(5) GUTs, 
 quarks and leptons are contained a common multiplet. 
This means the baryon number symmetry is not valid, and nucleon 
 is no longer stable. 
This is the most impressive prediction of GUTs. 
In fact, the $X$ boson induces the proton decay $p^+\to e^++\pi^0$. 
Because nucleon decay has not been observed\cite{SKproton}, 
 the mass of the $X$ boson must be larger 
 than $\order{10^{15}}\GeV$.
Thus, the SU(5) breaking scale has to also be larger than that scale, 
 as indicated by the Figure\ \ref{GCUSM}. 

In contrast to the matter sector, Higgs cannot be unified 
 because there are only one doublet Higgs in the SM, and we have to 
 introduce additional Higgs fields. 
The simplest possibility is to embed the doublet Higgs into ${\bf5}$ 
 Higgs.
In this case, from the decomposition (\ref{SU(5)5bar}) we find that 
 the partner is color triplet Higgs $H$\sub C. 
This colored Higgs also induces nucleon decay, although their 
 coupling to the matter field is Yukawa interaction and thus 
 the contribution is very small. 
In SO(10) GUTs, the smallest multiplet is a real 
 ${\bf10}$ representation which is decomposed in terms of SU(5) as 
\bequ
 {\bf10}\to{\bf5}+{\bf\bar5} ,
\label{SO(10)10}
\eequ
and can be embedded into ${\bf27}$ in \E6 models. 

The Yukawa interactions are given as 
\beqn
 {\bf10}_i{\bf10}_j{\bf5} &\to& 
   \Ll Q_iU^c_jH+U^c_iE^c_jH_{\rm C}+\Lm i\leftrightarrow j\Rm\Rl
   +Q_iQ_jH_{\rm C}, 
\label{SU(5)10Yukawa}\\
 {\bf10}_i{\bf\bar 5}_j{\bf5}^\dagger &\to& 
   Q_iD^c_jH^\dagger+E^c_iL_jH^\dagger 
    +Q_iL_jH_{\rm C}^\dagger+U^c_iD^c_jH_{\rm C}^\dagger,
\label{SU(5)5barYukawa} 
\eeqn
 in the SU(5) model, which are further unified as 
\bequ
 {\bf16}_i{\bf16}_j{\bf10} \to 
   {\bf10}_i{\bf10}_j{\bf5}+\Ll{\bf10}_i{\bf\bar5}_j{\bf5}^\dagger 
    +{\bf\bar5}_i{\bf1}_j{\bf5}+\Lm i\leftrightarrow j\Rm\Rl
\eequ
in the SO(10) model, which is further unified as
\bequ
 {\bf27}_i{\bf27}_j{\bf27} \to 
  {\bf16}_i{\bf16}_j{\bf10}+{\bf10}_i{\bf10}_j{\bf1}
 +\Ll{\bf16}_i{\bf10}_j{\bf16}+{\bf10}_i{\bf1}_j{\bf10}
 +\Lm i\leftrightarrow j\Rm\Rl
\eequ
in the \E6 model. 
Thus, the Yukawa matrices of the down-type quarks and charged leptons 
 are related with each other at the SU(5) breaking scale as $Y_L=Y_D^t$ 
 in the SU(5) model. 
In the SO(10) and \E6 models, the Yukawa matrix of the up-type quarks is 
 also related as $Y_D=Y_U$ at the SO(10) breaking scale. 
These relations are bad because of the experimental relation 
\beqn
 &&\frac{m_\mu}{m_\tau}\sim3\frac{m_s}{m_b}\gg\frac{m_c}{m_t}, \\
 &&\frac{m_e}{m_\tau}\sim\frac13\frac{m_d}{m_b}\gg\frac{m_u}{m_t}, 
\eeqn
 and they are called as wrong GUT relation.
Note that from Eqs.(\ref{SU(5)10Yukawa}) and (\ref{SU(5)5barYukawa}), 
 we can find that we get baryon number violating interactions 
 $QQQL$ and $U^cE^cU^cD^c$ after the colored Higgs $H$\sub C 
 is integrated out. 

In this way, GUTs can realize very beautiful unifications of forces 
 and of matter fields and can give a natural reason why the hypercharges 
 are quantized. 
In addition, they give testable predictions, such as nucleon decay 
 and GUT relations of gauge coupling unifications and 
 of Yukawa couplings, although these are not good unfortunately. 
On the other hand, some problems of the SM are still not solved. 
Among them, the hierarchy problem becomes a more concrete problem, 
 because there appears the very large GUT scale as indicated 
 by the Figure\ \ref{GCUSM}. 
In the next section, we introduce SUSY in order to solve the hierarchy 
 problem.

\section{SUSY}
\label{SUSY}

SUSY is the symmetry that exchanges a boson and a fermion 
 (See, for example, Refs.\cite{SUSY,Martin}). 
This means that each field in the SM has a superpartner 
 which has a different spin from that of the SM particle by $\half$, 
 as shown in the Table\ \ref{SUSYMultiplet}.
\begin{table}
\bctr
\begin{tabular}{|ccccc||c|}
\hline
  scalar & $\myequiv{Q}$ & spinor & $\myequiv{Q}$ & vector
   & supermultiplet\\ \hline
  && {gaugino} && gauge boson & vector \\ 
\vsp{-0.4} &&&&&\\
  {squark} && quark &&
   & chiral\\
  {slepton} && lepton &&& chiral\\ \vsp{-0.4} &&&&&\\
  Higgs && {higgsino} &&
   & chiral \\
\hline
\end{tabular}
\ectr
\caption{
Superpartner and supermultiplet.
}
\label{SUSYMultiplet}
\end{table}
There appears two kinds of supermultiplets: the vector supermultiplet, $V$, 
 which contains a gauge boson and the corresponding Majorana 
 gaugino (and an auxiliary field $D$), and the chiral supermultiplet, 
 $\Phi$, 
 which contains a Wyle fermion and a complex scalar field 
 (and an auxiliary field $F$). 
Note that, the sign of the quantum correction to the Higgs mass 
 by a boson is opposite to that by a fermion. 
Thus, when the boson and fermion have the same quantum numbers, 
 \eg\ mass and coupling, the quantum correction vanishes 
 so that the hierarchy problem is solved.

At this stage, the Higgs field is treated equally as 
 quarks and leptons. 
We have to introduce the fermionic partner of Higgs which contribute 
 the anomalies so that the anomalies become non-zero. 
The simplest way to cancel the anomalies is to introduce 
 one more doublet Higgs that has opposite hypercharge $-\half$.
These two Higgs fields are also required for giving masses 
 to both the up-type and down-type quarks because of the holomorphy 
 of the superpotential. 
Thus, we introduce two Higgs doublets $H_u({\bf1},{\bf2})_\half$ and 
  $H_d({\bf1},{\bf2})_{-\half}$.\footnote{
Because these two Higgs are vector-like, they could have 
 a very large mass. 
Then, they would decouple from the low energy effective theory. 
Thus, the mass parameter $\mu$ has to be around the weak scale. 
But it looks unnatural that the SUSY parameter $\mu$ has 
 the same scale as the SUSY breaking scale. 
This is called as the $\mu$ problem. 
A solution for this problem is discussed in Ref.\cite{mu} 
 in the context of \aGUT.
} 
This supersymmetric extension of the SM is called as the minimal 
 supersymmetric standard model\ (MSSM). 
In the MSSM, the Yukawa interaction is given by the same expression as 
 Eq.(\ref{Yukawa}) if we exchange $\lag$ by superpotential $W$, 
 $H$ by $H_u$ and $H^\dagger$ by $H_d$ and we interpret 
 each field as a superfield. 
Because there are two Higgs doublets, there appears an additional 
 parameter that is the ratio of the VEVs of these Higgs fields, 
 $\tan\beta\equiv\frac{\VEV{H_u}}{\VEV{H_d}}$. 
When $\tan\beta$ is large, bottom Yukawa and tau Yukawa is also 
 large.

\begin{figure}
\begin{center}
\leavevmode
\put(300,-5){{\large $\bf{\log \mu (GeV)}$}}
\put(0,205){{\Large $\bf{\alpha^{-1}}$}}
\put(29,185){$\alpha_1^{-1}$}
\put(31,100){$\alpha_2^{-1}$}
\put(31,40){$\alpha_3^{-1}$}
\includegraphics[width=11cm]{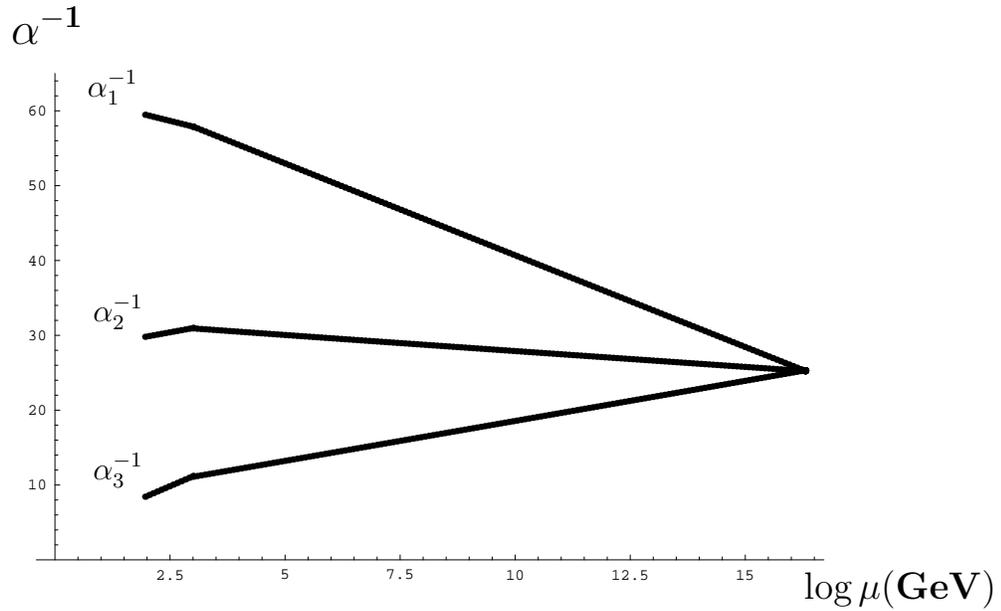}
\caption{
The gauge coupling flows in the MSSM: 
Here we adopt, 
$\alpha_1^{-1}(M_Z)=59.47$, $\alpha_2^{-1}(M_Z)=29.81$,
$\alpha_3^{-1}(M_Z)=8.40$.
} 
\label{GCUMSSM}
\end{center}
\end{figure}

Because we have not observed any superpartners yet, 
 SUSY must be broken. 
In order to keep the quantum correction smaller than $\TeV$ scale, 
 SUSY should be broken softly, that is, only the interactions 
 whose coefficients have positive mass dimensions are allowed, 
 and the mass scale is around the weak scale ($\lesssim\order1\TeV$). 
This assumption for solving the hierarchy problem leads to 
 an amazing result. 
The gauge couplings meet with each other very accurately 
 at a very high scale, the usual GUT scale 
 $\MGUT\sim2\times10^{16}\GeV$, 
 if we take the normalization of U(1)$_Y$ same as that of 
 the SU(5) model, as shown in the Figure\ \ref{GCUMSSM}. 
This fact seems to imply the existence of SUSY-GUTs 
 as a more fundamental theory.

\subsection{SUSY-Flavor Problem}

\begin{figure}
\begin{center}
\includegraphics[width=15cm]{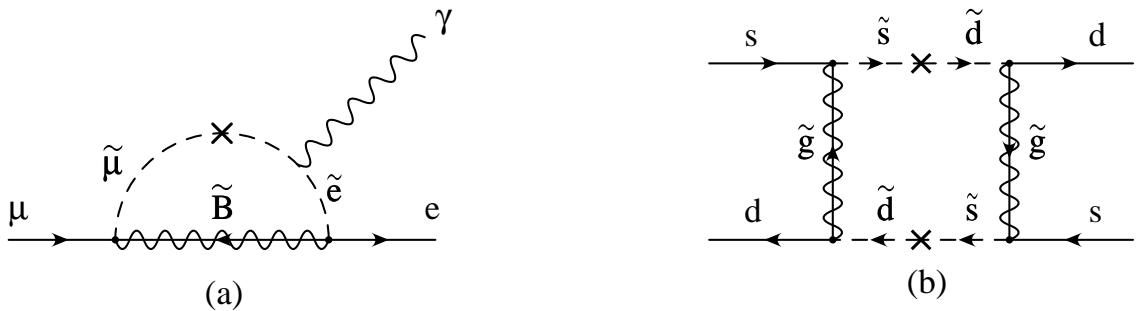}
\caption{
Diagrams that may induce FCNC processes: 
This figure is shown in the Figure 12 of Ref.\cite{Martin}.
} 
\label{FCNC}
\end{center}
\end{figure}

If we believe the success of GCU seriously, 
 the SUSY breaking scale should not be far away from the weak scale.
This means the superpartners will be discovered by $\TeV$ scale 
 experiments. 
In addition, they can give considerable contributions to the low energy 
 precise measurements through loop effects, especially 
 to the processes that the SM has small contributions. 
Note that the FCNC processes are very suppressed in the SM as mentioned 
 in \S\ref{SM}. 
In fact, the FCNC processes have already given severe constraints 
 to some soft parameters. 
For example, the off-diagonal elements of sfermion mass-squared 
 matrices can make large contributions to FCNC processes 
 through the diagrams shown in the Figure\ \ref{FCNC} 
The most severe constrains come from the $K^0$-$\bar K^0$ mixing 
 and the $\mu\to e\gamma$ process. 
Also, the $D^0$-$\bar D^0$ mixing, $B^0$-$\bar B^0$ mixing, 
 $b\to s\gamma$ and $\tau\to\mu\gamma$ processes give constraints. 
However they are much weaker than the former constraints, and thus 
 we consider only the former constrains in this thesis 
 for simplicity. 
Thus, the constrained parameters are $(1,2)$ elements of sfermion 
 mass-squared matrices in the mass basis where 
 the fermion Yukawa matrices are diagonalized. 
They are defined by using the fermion mixing matrices $V_{f_\chi}$ 
 as 
\bequ
 {m^2_{f_\chi}}^\diag=V_{f_\chi}^\dagger m^2_{f_\chi} V_{f_\chi} 
\eequ
 for each flavor $f=U,D,E$ and chirality $\chi={\rm L},{\rm R}$.

Roughly speaking, there are three ways to suppress the off diagonal 
 elements of ${m^2_{f_\chi}}^\diag$:
\bit
 \item The degenerate solution: \\
       If $m^2_{f_\chi}\propto{\bf1}_3$, where ${\bf1}_3$ is the 
       $3\times3$ identity matrix, then the off diagonal elements 
       are not induced even after $V_{f_\chi}$ are operated.
 \item The alignment solution: \\
       If $m^2_{f_{\rm L}}\propto Y_fY_f^\dagger$ and 
       $m^2_{f_{\rm R}}\propto Y_f^\dagger Y_f$, then $m^2_{f_\chi}$'s are 
       diagonalized by $V_{f_\chi}$'s.
 \item The decoupling solution: \\
       The first and second generations which couple to Higgs 
       very weakly may be heavy as $\order{10}\TeV$. 
       Then, the contributions are suppressed by the heavy masses.
\eit
The second solution is not easy to realize, and the third one 
 is not sufficient by itself. 
Thus, we consider the first solution. 

The degeneracy can be realized if a flavor-blind mediation mechanism 
 of the SUSY breaking, such as the gauge mediation, gaugino mediation 
 and anomaly mediation, is realized.
In this case, the SUSY-flavor problem is a problem of 
 the SUSY breaking and/or mediation mechanism, which we do not treat 
 so much in this thesis. 
There are another way to realize the degeneracy. 
It is to introduce non-abelian flavor symmetry\ (horizontal symmetry) 
 and to embed the first and second generations into a single multiplet 
 of the horizontal symmetry. 
Then, the horizontal symmetry ensures that they have a common mass 
 in the symmetric limit because they belong to a common multiplet. 
In fact, we have not observed such a symmetry, and therefore the 
 horizontal symmetry is broken, lifting the degeneracy. 
It is discussed whether sufficient degeneracy can be obtained or not 
 in \aGUT\ scenario in \S\ref{HorizFCNC}.

\subsection{Nucleon Decay}
\label{SUSYProton}

In SUSY theories, there appear two kinds of nucleon decaies: 
 the nucleon decay via dimension 4 operators and 
 that via dimension 5 operators. 
Of course, nucleon decay would occur via effective 4-Fermi operators 
 whose mass dimension is 6. 
In non-SUSY theories, there are only fermionic matter fields 
 and thus the effective operators are suppressed by the second 
 power of a very large scale, such as the GUT scale and Planck scale. 
In contrast, there are also scalar partners and thus the large scale 
 may be replaced by the SUSY breaking scale.

\subsubsection{via Dim. 4 operators}

Because both Higgs and matter fields are chiral superfields and 
 $H_d$ has the same quantum number as $L$, $H_d$ and $L$ cannot 
 be distinguished. 
This means the following baryon number violating Yukawa interaction 
 is allowed by the symmetry: 
\beqn
  &&  W_{\Delta {L} =1} =
      {1\over 2} \lambda^{ijk} L_i L_j E^c_k
      + \lambda^{\prime ijk} L_i Q_j D^c_k
      + \mu^{\prime i} L_i H_u. 
\label{DelL}\\
  &&  W_{\Delta {B}= 1} = {1\over 2} \lambda^{\prime\prime ijk}
      U^c_i D^c_j D^c_k. 
\label{DelB}
\eeqn
Integrating out $D^c$ whose mass is around the SUSY breaking scale, 
 we get effective 4-Fermi interactions which are suppressed by 
 only the second power of the SUSY breaking scale, leading to 
 destructively rapid nucleon decay. 
These dangerous interactions can be forbidden by introducing a 
 $Z_2$ symmetry, called as $R$-parity.\footnote{
This $Z_2$ symmetry is not an $R$-symmetry in this sence. 
But we can assign the parities so that each element of a 
 superfield does not have a common parity, 
 by redefining $R$-parity as $Z_2(-1)^{2s}$ where $s$ is the spin. 
This additional factor has no physical meanings as far as 
 we consider only Lorentz invariant interactions.
}
The parity assignment is that all the matter superfields 
 have odd parities and the other superfields have even parities.
Then, all the terms in the superpotential (\ref{DelL}) and 
 (\ref{DelB}) are odd under $R$-parity and therefore are forbidden.

Note that $R$-parity can remain exact. 
Then the lightest superpartner\ (LSP) which is the lightest particle 
 possessing odd $R$-parity defined in above footnote cannot decay 
 and thus is stable. 
The LSP is a candidate for the cold dark matter, 
 favored by the observations.

\subsubsection{via Dim. 5 operators}

The effective operators $QQQL$ and $U^cE^cU^cD^c$, induced by the 
 colored Higgs exchange in \S\ref{GUT}, become dimension 5 operators 
 when they appear in superpotential, namely two of the fields are 
 scalars and the other two are fermions. 
Then the operators are suppressed by only first power of a large 
 scale $M$, leading to rapid nucleon decay.
Because the scalar matter fields have to be transformed into 
 the fermionic matter fields through a superpartner exchange, 
 the rate of nucleon decay depends strongly on 
 the SUSY breaking parameters%
 \cite{ProtonCancel}. 
However, if we do not allow fine-tuning, the large scale 
 ($\times$ coefficient $y$) should satisfy the relation 
 $y/M<1/\order{10^{26}}\GeV$\cite{KakizakiProton}. 
In the case of the colored Higgs exchange, the coefficient $y$ 
 is small due to the small Yukawa coupling, but we need typically 
 $M>\Mp\sim\order{10^{19}}\GeV$. 
Note that, even when the colored Higgs is absent, 
 the physics of the Planck scale may induce the effective operators. 
And if the coefficient $y$ is $\order1$, nucleon decay would occur 
 too rapidly. 
Thus, it looks natural that each field has a suppression factor 
 as in the case where the Froggatt-Nielsen\ (FN) mechanism acts 
 (See \S\ref{U(1)A}).

\section{SUSY-GUT}
\label{SUSY-GUT}

Employing SUSY, the hierarchy problem can be solved also in GUTs, 
 keeping the beautiful structures of GUTs, \eg\ the unifications 
 of forces and of matter fields. 
In addition, GCU is realized accurately as shown 
 in Figure\ \ref{GCUMSSM}, supposing the colored Higgs has a mass 
 around the GUT scale $\MGUT\sim2\times10^{16}\GeV$.

Unfortunately, such a colored Higgs mass is too light to 
 suppress nucleon decay sufficiently. 
On the other hand, if the colored Higgs mass is sufficiently large 
 $\sim\Mp$ so that the nucleon decay via the colored Higgs exchange 
 is suppressed, GCU is spoiled. 
Of course it is possible to restore it by introducing other parameters
 and adjusting them. 
For example, generally GUTs based on a symmetry that has 
 a rank larger than 4, \eg\ SO(10) or \E6, have several 
 symmetry breaking scales which can be used for restoring GCU. 
Alternatively, we can introduce additional multiplets whose mass spectrum 
 does not respect the SU(5) symmetry. 
In such cases, however, GCU is not a prediction but a constraint 
 of models, and one of the motivations to consider SUSY-GUTs is 
 lost. 
This is one of the problems of SUSY-GUTs. 
This issue is discussed in \S\ref{GCU}.

\subsection{Doublet-Triplet Splitting Problem} 
\label{DTS}

Another problem of SUSY-GUTs is the so-called DTS problem. 
As mentioned above, the doublet Higgs fields have to be light 
 ($\sim\order{100}\GeV$) while the colored partners must be 
 superheavy ($>\MGUT$). 
It is difficult to realize such a mass splitting within 
 multiplets. 
Let us illustrate this by using \msGUT\ as an example.
Here, we introduce a pair of ${\bf5}$ and ${\bf\bar5}$ Higgs 
 which contain the MSSM doublet Higgs, $H_u$ and $H_d$, respectively: 
\beqn
\bar H({{\bf\bar5}}) &=& ( \bar H_{\rm C}, H_d ), \\
H({{\bf5}}) &=& ( H_{\rm C}, H_u ).
\eeqn
SU(5) is broken down to $G$\sub{SM} by the following VEV of an 
 adjoint Higgs $A({\bf24})$: 
\bequ
  \VEV {A}
   = \left(
    \begin{array}{cc}
      2v{\bf1}_3 & 0 \\
       0 &-3v{\bf1}_2\\
    \end{array}
  \right)  
\,\,,\quad
v\sim\order{10^{16}}\GeV.
\label{SU(5)VEVA}
\eequ
Then, the mass term of $H$ and $\bar H$ is given as 
\bequ
W_{\rm DT}=\bar H[m+A]H, 
\eequ
 where $m$ is a mass parameter. 
From this mass term, we find that the colored Higgs mass $m$\sub C and 
 the doublet Higgs mass $\mu$ are given as 
\beqn
 m_{\rm C} &=& m+2v \,>\,10^{{16}}\GeV, \\
 \mu &=& m-3v \,\sim\,10^{2}\GeV.  
\eeqn
This required fine-tuning of at least $\order{10^{-14}}$ 
 between the parameter $m$ and the VEV $v$. 

In the following, for simplicity, 
 we aim to realize $\mu=0$ as the first approximation, 
 and assume that $\mu$ becomes $\order{\MSB}$ when we take the 
 SUSY breaking into account as shown in Ref.\cite{mu}.
This is one of the biggest problems of SUSY-GUTs. 
And many authors have proposed solutions for this problem. 
Here, we show some of them, 
 although there are other possible solutions\cite{GIFT,orbifold}.

\subsubsection{Dimopoulos-Wilczek mechanism}

The solution that we mainly employ in \aGUT s is 
 the Dimopoulos-Wilczek\ (DW) mechanism\cite{DW}. 
This mechanism may be realized GUTs based on a symmetry that 
 contains the U(1)$_{B-L}$ symmetry, such as SO(10) or \E6.
Because the doublet Higgs fields have vanishing U(1)$_{B-L}$ charges 
 while the colored Higgs fields have non-vanishing charges, the generator 
 of U(1)$_{B-L}$ operates only on the colored Higgs. 
This means that if an adjoint Higgs $A$ acquires a non-vanishing VEV only 
 in the direction, that is, the VEV is proportional to the generator, 
 then the VEV contributes only to the colored Higgs mass. 
This is easily understood if we write an explicit form of the DW-VEV: 
\bequ
 \VEV A_{B-L}=\tau_2\times\diag(v,v,v,0,0), 
\label{DWVEV}
\eequ
 in SO(10) models.

Note that the vector multiplets of SO(10) couple to the adjoint 
 multiplet anti-symmetrically:
\bequ
 {\bf10}\times{\bf10}={\bf1}_{\rm s}+{\bf45}_{\rm a} 
                     +{\bf54}_{\rm s},
\eequ
where the index ``s'' denotes that the coupling is symmetric and 
 ``a'' denotes that the coupling is anti-symmetric. 
This means we need one more vector Higgs $H'({\bf10})$ in addition 
 to $H({\bf10})$ that contains $H({\bf5})$ and $\bar H({\bf\bar5})$ 
 of \msGUT.
The mass term $H'H'$ is required to give mass to the additional 
 doublet Higgs of $H'$, while the mass terms $HH$ and $HH'$ 
 must be forbidden because they contribute to the mass of the 
 MSSM doublet Higgs: 
\beqn
W_{\rm DT}&=&\bar HAH'+mH'H'(+H'AH') \\
 &=&\Ls H({\bf5}),\,\,H'({\bf5})\Rs 
    \Ls\begin{array}{cc}
         0 & \VEV A \\
         \VEV A & m(+\VEV A) 
       \end{array}\Rs
    \Ls\begin{array}{c}
         \bar H({\bf\bar5})\\
         \bar H'({\bf\bar5})
       \end{array}\Rs.
\eeqn
Because $\VEV A$ does not contribute to the doublet masses, 
 we find that one pair of the doublets are indeed massless. 

In this way, the DTS problem can be solved by the DW mechanism. 
However, it is difficult to realize the DW-VEV (\ref{DWVEV}), 
 and usually it needs fine-tuning.

\subsubsection{Sliding Singlet mechanism}
\label{Sliding}

The sliding singlet mechanism\cite{SS} is 
 the smartest solution that dynamically achieves DTS.

This mechanism was originally proposed in the context 
of SU(5)\cite{SS}, in which an singlet field $Z({\bf1})$ is introduced 
 and the following terms are allowed in the superpotential:
\begin{equation}
W_{\rm{SS}} = \bar{H} ( A + Z ) H.
\label{su5ss}
\end{equation}
Here, the adjoint Higgs $A(\bf{24})$ is assumed 
to have the VEV (\ref{SU(5)VEVA}).
Since the doublet Higgs fields have non-vanishing VEVs $\VEV{H_u}$ and 
 $\VEV{H_d}$ to break SU(2)\sub L$\times$U(1)$_Y$ into U(1)\sub{EM}, 
the minimization of the potential,
\bequ
V_{{\mbox{\scriptsize{SUSY}}}}=\abs{F_H}^2+\abs{F_{\bar H}}^2=
\Ls\abs{\VEV{\bar H}}^2+\abs{\VEV H}^2\Rs\times\abs{-3v+\VEV Z}^2,
\label{potential}
\eequ
leads to the vanishing doublet Higgs mass  
$\mu=(\VEV A+\VEV Z)_{\bf2}=-3v+\VEV Z=0$
by {\it sliding} the VEV of $Z$.\footnote{
Here, the contributions to the potential from 
$F_A$ and $F_Z$ are neglected because they are of order $(\VEV{\bar HH})^2$.
In this sense, the doublet Higgs mass $\mu$ does not vanish exactly but may 
become
of order SUSY breaking scale $\MSB$.
}
For these VEVs, $\VEV A+\VEV Z=\diag(5,5,5,0,0)v$, the color 
triplet partners of doublet Higgs have a large mass 
$5v\sim\order{10^{16}}\GeV$.

Unfortunately, this DTS is known to fail if the SUSY breaking is 
taken into account. For example, 
the soft SUSY breaking mass term $\s m^2\abs Z^2$ 
$(\s m\sim\MSB)$ 
shifts the VEV $\VEV Z$ by an amount of 
$\delta\VEV Z\sim\frac{\s m^2v}{{\VEV H}^2+\s m^2}\sim\MGUT$ to minimize
the potential.
Thus the DTS is spoiled by the SUSY breaking effect
in this mechanism.%
\footnote{
The soft term $\s mZF_Z$ also destabilizes the sliding singlet
mechanism because this term alters the contribution of 
$F_Z$ to the scalar potential as $\abs{\bar HH+\s mZ}^2$, which
is the order of $\MSB^2\MGUT^2(\gg\MSB^4)$ if $\VEV H$ is 
order of the weak scale.
Such a term is induced by loop effects through the coupling between
$Z$ and the color triplet Higgs. Therefore, even if the terms
$\s mZF_Z$ and $\s m^2|Z|^2$ are absent at the tree level, this problem
cannot be avoided.
}

This is caused by the fact that 
the terms $|F_H|^2+|F_{\bar H}|^2$ give only a mass of 
order ${\VEV H}$ to $Z$,
which is the same order as (or smaller than) 
the SUSY breaking contribution.
Because this mass parameterizes the stability of $\VEV Z$ 
against other contributions to the potential, \eg\ 
SUSY breaking effects
$\s m^2 |Z|^2$, 
soft terms of order $\MSB$ easily shift the VEV from that 
in the SUSY limit by a large amount.
This can be avoided if 
large VEVs $\VEV H$ and $\VEV {\bar H}$ (larger than
$\sqrt{\MSB\MGUT}$) which give a larger mass to
$Z$ are used, and the VEV of $Z$ is stabilized against SUSY breaking
effects. 
Of course, such large VEVs of the MSSM doublet Higgs are 
 not consistent with the experiments.
But large VEVs are acceptable for other Higgs fields that break
 a larger gauge group into $G$\sub{SM}, and 
 SU(6) models are examined in Ref.\cite{Sen,su6}.

We have extracted the essence of this mechanism to make this mechanism 
 applicable in more generic cases 
 by a group theoretical analysis\cite{sliding}.
The essential idea of this mechanism is as follows:
\bit
 \item A mass term of a certain component that has a non-vanishing VEV 
       may vanish because of the EOM as in the previous SU(5) example. 
 \item If a mass parameter of a certain component is guaranteed to be
       the same as that of the component with non-vanishing VEV, 
       then the mass parameter also vanishes. 
 \item If the non-vanishing VEVs are sufficiently large, 
       the mass hierarchy is stable against possible 
       SUSY breaking effects.
\eit
For instance in \E6 models, the U(1)$_{B-L}$ charges of the doublets 
 contained in the SO(10) ${\bf10}$ component of ${\bf27}$ are 
 the same as that of 
 the SO(10) ${\bf1}$ component, namely they are zero. 
Thus, if the mass term of $\Phi({\bf27})$ and $\bar\Phi'({\bf\cc{27}})$ 
 is a function of an adjoint Higgs $A({\bf78})$ and a singlet Higgs 
 $Z({\bf1})$ as 
\bequ
 W_{\rm SS}=\bar\Phi' f(A,Z) \Phi, 
\eequ
 $A$ acquires non-vanishing VEV in the direction of 
 U(1)$_{B-L}$ generator\ (DW-VEV) and ${\bf1}_\Phi$ acquires 
 non-vanishing VEV, then the sliding singlet mechanism can work. 
Namely, EOM of ${\bf1}_{\bar\Phi'}$ makes $\VEV Z$ slided 
 so that $f(0,Z)=0$, leading to the vanishing mass term of 
 the doublets. 
In Ref.\cite{sliding}, a more detailed analysis has been made.

\subsubsection{Missing Partner mechanism}

The missing partner mechanism\cite{MP} is also proposed in 
 the context of SU(5). 
The idea is that, if we introduce additional Higgs possessing 
 representations that contain a triplet but not doublets, 
 it is possible to give mass only to the triplet Higgs. 
For instance, ${\bf50}$ contains a triplet but does not 
 contain doublets, and ${\bf75}$ Higgs can connect the ${\bf50}$ Higgs 
 to the ${\bar\bf5}$ Higgs through the interaction 
 ${\bf50}\cdot{\bf75}\cdot{\bar\bf5}$. 
We can show the situation schematically as  
\begin{equation}
\left( \begin{array}{c} 
         {\bf\bar{3}} \\
         {\bf\bar{2}}
       \end{array} \right)_{\bf{\bar{5}}} 
\begin{array}{c} \longleftrightarrow \\ \vsp0 \end{array}
\left( \begin{array}{c} 
         {\bf3} \\ 
         {\rm others}
       \end{array} \right)_{\bf50} 
\begin{array}{c} \longleftrightarrow\\ \longleftrightarrow \end{array}
\left( \begin{array}{c} 
         {\bf\bar{3}} \\ 
         {\rm others}
       \end{array} \right)_{\bf{\cc{50}}} 
\begin{array}{c} \longleftrightarrow  \\ \vsp0 \end{array}
\left( \begin{array}{c} 
         {\bf3} \\ 
         {\bf2}
       \end{array} \right)_{\bf5}, 
\end{equation}
 where the arrows represent that the pointed components 
 have non-vanishing mass terms. 
Of course, we have to forbid the direct mass term 
 ${\bf5}\cdot{\bf\bar5}$ which gives a large mass to 
 the doublet Higgs.
In this way, relatively large representations, such as 
 ${\bf50}$ and ${\bf75}$, are required to realize this mechanism 
 in SU(5) models. 
These fields may make the unified gauge coupling divergent 
 below the Planck scale. 

On the other hand, this mechanism can be realized in a simpler way 
 in flipped SU(5) models\cite{FlippedSU(5)}, where the gauge symmetry 
 is not a simple group: SU(5)\sub F$\times$U(1)\sub F. 
In this model, the SM fields are embedded in a way the right-handed fields, 
 \ie\ SU(2)\sub R doublets, are ``flipped'' as 
\begin{eqnarray}
  {\bf 10}_1&=&(Q,D^c,N^c),  \\
{\bf \bar 5}_{-3}&=&(U^c,L), \\
{\bf 1}_5&=&E^c. 
\end{eqnarray}
SU(5)\sub F$\times$U(1)\sub F is broken down into $G$\sub{SM} 
 by the VEV of $C({\bf10}_1)$ (and $C({\bf\cc{10}}_{-1})$ 
 for the $D$-flatness). 
It is interesting that in the flipped SU(5) models, adjoint Higgs 
 fields are not required.
As for the MSSM Higgs, $H_u$ and $H_d$ are ``flipped'', while 
 the colored partners, $H$\sub C and $\bar H$\sub C 
 are not ``flipped'' as 
\begin{eqnarray}
  \bar H({\bf \bar 5}_{-2})&=&(\bar H_{\rm C}, H_u), \\
  H({\bf 5}_2)&=&(H_{\rm C} ,H_d),  
\end{eqnarray}
where $(\bar H_{\rm C}, H_u, H_{\rm C} ,H_d)$ have the same 
 quantum number of the SM gauge group as $(D^c,\bar L,\bar D^c,L)$, 
 respectively.
Note that ${\bf10}_1$ contains $D^c$ but does not contain $L$.
In fact, the superpotential 
\bequ
 W_{\rm MP}=CCH+\bar C\bar C\bar H,
\eequ
 gives masses proportional to $\VEV C$ to the triplet Higgs while 
 the doublet Higgs fields remain massless. 
We have generalized the flipped model to 
 an SO(10)\sub F$\times$U(1)$_{{\rm F}'}$ model\cite{FlippedSO(10)}.

\section{Anomalous U(1) Symmetry} 
\label{U(1)A}

Finally, we make comments on U(1)\sub A. 
This is often introduced as a simple way to realize the FN mechanism%
 \cite{FN}, which can explain the hierarchy in the Yukawa matrices. 

It is known U(1)\sub A sometimes appears in the low energy effective 
 theory of a string theory\cite{IR}. 
The anomaly of U(1)\sub A is cancelled via the GS mechanism\cite{GS}, 
 for example a non-linear transformation of 
 the dilaton multiplet $S$ (axion) 
\bequ
 S \to S+\frac{i}2\delta_{\rm GS}\Lambda{}, 
\label{Dilaton}
\eequ
 where $\Lambda$ is a gauge transformation parameter,  
 cancels the anomaly. 
In this sence, the anomalous U(1) symmetry is not anomalous in total. 

To be more precise, the gauge kinetic terms in SUSY theories are 
 written as 
\begin{equation}
  {\cal L}_{\mbox{\scriptsize gauge}}
   = \frac{1}{4} \dt
       \left[\,k_{\rm A}S\,W_{\rm A}{}^\alpha W_{\rm A}{}_\alpha 
       + k_a S\, W_a{}^\alpha W_a{}_\alpha \,\right] + \mbox{h.c.}, 
\end{equation}
 by using the SUSY field strengths of U(1)\sub A and the gauge 
 symmetry $G_a$, $W$\sub A and $W_a$. 
Here, $k$\sub A and $k_a$'s are the Kac-Moody levels of U(1)\sub A and 
 $G_a$'s, respectively.
The gauge coupling constants $g_a$ are given by the VEV of the Dilaton 
 field $S$ as $k_a\VEV S=1/g_a^2$. 

The ``anomalous'' U(1) transformation induces a shift
$
 \Delta{\cal L}\propto \dt\Ll iC_a\Lambda W_a^\alpha W_{a\alpha}\Rl
$
for each $G_a$, where $C_a=\Tr_{G_a}T({\bf R})^2Q$\sub A 
 is the mixed-anomaly. 
Because the shift induced by (\ref{Dilaton}) is common for 
 each $G_a$ except for $k_a$, the ratio $C_a/k_a$ must be common 
 for each $G_a$ in order that the anomalies are canceled. 
Taking the [U(1)\sub A]$^3$ and gravitational anomalies, 
 we get the following GS relations\cite{maekawa}: 
\begin{equation}
  2\pi^2\delta_{\rm {GS}}\,= \,\frac{C_a}{k_a}
                        \,= \,\frac{1}{3k_{\rm A}}
                                       \tr{Q_{\rm A}}^3 
                        \,= \,\frac{1}{24}\tr Q_{\rm A}, 
\label{GSRelation}
\end{equation}
where $Q$\sub A is U(1)\sub A charge.

It is known that the Fayet-Iliopoulos\ (FI) $D$ term proportional to 
 the anomaly $\delta$\sub{GS} is induced radiatively. 
 Because the \Kahler\ potential $K$ for $S$ must be a function of 
 $S+S^\dagger-\delta_{\rm GS}V$\sub A where $V$\sub A is the 
 vector superfield of U(1)\sub A, the FI $D$-term is given as 
\begin{equation}
\dtf \,K(S+S^\dagger-\delta_{\rm {GS}}V_{\rm A})
=  \left(-\frac{\delta_{\rm {GS}}K'}{2}\right)D_{\rm A} + \cdots
  \equiv \xi^2 D_{\rm A} +\cdots,
\label{FIterm}
\end{equation}
 where we fix the sign of $Q$\sub A so that $\xi^2>0$.
If $K(x)=-\ln(x)$ as calculated in Ref.\cite{U(1)}, 
 $\xi^2$ is approximated as 
 \begin{equation}
  \xi^2=\frac{g_s^2\tr Q_{\rm A}}{192\pi^2}, 
\label{FItree}
\end{equation}
 where $g_s^2=1/\VEV S$.

The relation (\ref{GSRelation}) is an important relation 
 in models that employs U(1)\sub A.  
But they can be adjusted by introducing some singlet fields 
 with appropriate charges. 
In particular in GUT models, $a$ runs only one index, and it looks 
 easier to satisfy this relation. 
Thus, in the following, we do not take care of this relation, 
 for simplicity.
As for FI $D$-term, the charge assignments of the models 
 discussed in this thesis give a large $\xi^2$ if we calculate 
 it by the relation (\ref{FItree}). 
Thus, we have to assume $K'$ is smaller than the simple case.%
\footnote{
It may be possible to assume a tree level FI $D$-term that 
 cancel with the loop contribution.
}
In any case, we simply assume $\xi^2$ is a desirable value
 in the following.

\subsubsection{Froggatt-Nielsen mechanism}

U(1)\sub A is often used for realizing the FN mechanism 
 which can give hierarchical factor in effective coefficients 
 even if the original theory does not have such hierarchy.

Let us illustrate this mechanism by using the up-type Yukawa 
 interactions $Q_iU^c_jH$ as an example. 
If an additional U(1) symmetry is imposed and charges 
 $(q_i,u^c_j,h)$ are assigned for $(Q_i,U^c_j,H)$, 
 the Yukawa interactions are generally forbidden 
 by the U(1) invariance.
Their charges may be compensated by a singlet field $\Theta$,  
 called as the FN field, whose U(1) charge is $-1$ as 
\beqn
 {\cal L}&\ni& y_{ij}\Ls\frac{X}{\Lambda}\Rs
                     ^{\abs{q_i+u^c_j+h}}Q_iU^c_jH,  
\label{FNYukawa}\\
  X &=& \left\{
\begin{array}{cr}
\Theta     & \hsp1  q_i+u^c_j+h > 0  \\
\Theta^\dagger & \hsp1 q_i+u^c_j+h < 0  \\
\end{array}
\right.,
\eeqn
 where the interactions become non-renormalizable and thus 
 suppressed by some power of the cutoff scale $\Lambda$ 
 and $y_{ij}$'s are the original parameters of $\order1$.
Suppose that the FN field acquires a non-vanishing VEV much smaller 
 than $\Lambda$ as 
\bequ
  {\VEV\Theta\over\Lambda} \equiv \lambda \ll 1.
\eequ
Then, in the effective theory where $\Theta$ is integrated out, 
 the effective Yukawa coupling can be written as 
\bequ
  {\cal L}\,\ni\, y_{ij}\lambda^{\abs{q_i+u^c_j+h}}Q_iU^c_jH.
\eequ
In this way, we get suppression factors $\lambda^{\abs{q_i+u^c_j+h}}$ 
 from the theory that has no hierarchy originally.
If we assign, for instance, 
 $h=-2n,(q_1,q_2,q_3)=(u^c_1,u^c_2,u^c_3)=(n+3,n+2,n)$, 
 the up-type Yukawa matrix is given as 
\bequ
Y_U=\left(
\begin{array}{ccc}
\lambda^6 & \lambda^5 & \lambda^3 \\
\lambda^5 & \lambda^4 & \lambda^2 \\
\lambda^3 & \lambda^2 & 1    
\end{array}
\right),
\eequ
where we omit the $\order1$ coefficients $y_{ij}$. 
This gives tolerable mass spectrum and mixings when 
 $\lambda\sim\sin\theta_C\sim0.2$, although 
 $m_u/m_c$ is rather large. 

Note that the suppression factors are determined by U(1) charges, 
 and thus we can consider that each field carries its own suppression 
 factor. 
This leads to the factorizable form of $Y_U$ if we forget 
 the $\order1$ coefficients $y_{ij}$:
\bequ
Y_U\sim\Ls\lambda^3,\lambda^2,1\Rs\times
\left(
\begin{array}{c}
\lambda^3 \\
\lambda^2 \\
1
\end{array}
\right).
\eequ
And this helps to suppress the dangerous nucleon decay 
 via dimension 5 operators induced through the physics around $\Mp$ 
 in SUSY theories, as discussed in \S\ref{SUSYProton}.

\subsubsection{SUSY-zero mechanism}

In SUSY theories, $\Theta^\dagger$ cannot appear in the 
 superpotential $W$, due to the holomorphy of $W$. 
This means that negative charges in $W$ cannot compensated 
 by $\Theta^\dagger$. 
Thus, negatively charged operators are forbidden 
 by the U(1) invariance 
 while positively charged operators can appear in the effective 
 theory below $\VEV\Theta$.
This is the SUSY-zero mechanism. 
This mechanism constrain the form of the superpotential strongly. 
As shown below, \aGUT\ scenario makes full use of this mechanism.

\subsubsection{Problems} 

As shown above, U(1)\sub A can be used as a powerful tool 
 for analyzing models, especially in SUSY theories. 
But it also has some issues. 
In order that interactions are allowed by U(1)\sub A, the 
 U(1)\sub A charges must be quantized, but we have no reason 
 the charges are quantized.\footnote{
This may give a hint for a more fundamental theory 
 as the quantization of the hypercharges does. 
}

Another problem is related with the SUSY-flavor problem.
Usually we assign different charges to different generations 
 to reproduce generation dependent hierarchies. 
Then, sfermions have generation dependent masses if U(1)\sub A 
 $D$ term is not zero, and thus the $D$ term must be very small 
 compared to a universal contribution.
Thus we have to consider a SUSY breaking mechanism and a mediation 
 mechanism that does not induce a large U(1)\sub A $D$ term.

\chapter{Anomalous U(1) GUT}
\label{aGUT}

In this chapter, we show some amazing features of the 
 \aGUT\ scenario. 

\section{Starting Point}
One of the most basic assumptions of this scenario is that 
 we introduce ``generic interaction''. 
Here ``generic interaction'' implies that
 we introduce all possible interaction terms that 
 respect the symmetry of the model, and that their coupling 
 constants are $\order1$ in the unit of the cutoff scale $\Lambda$
 of the model.%
\footnote{
Hereafter, we often use this unit without notice.
}
This means that infinite number of interaction terms are 
 introduced. 
Thus, it is indeed difficult to give predictions on 
 many of precision measurements.
However, we can still predict the order of magnitude of 
 the parameters related 
 to the low energy physics at the level of order magnitude.
Another consequence of the assumption is that the definition 
 of a model is given, except for a few parameters, by the 
 definition of a symmetry: 
 a symmetry group, matter content and representations 
 of the matter fields under the symmetry.
This means that, as far as the order-of-magnitude arguments 
 are concerned, the parameters of the models are essentially 
 the anomalous U(1) charges, whose number is the same as 
 that of the superfields.%
\footnote{
If we wish to make a precise analysis, we should fix a large 
 number of parameters (more than those in the SM). 
Thus, we concentrate ourselves on discussions 
 of order of magnitude in the following. 
Note that, however, it is still non-trivial whether it is 
 possible or not to reproduce the correct values of the 
 parameters in the SM, because we assume they are all $\order1$.
}

Another important assumption is made for the vacuum structure 
 of the model: 
\bequ
 \VEV{O_i} \sim \Lm
  \begin{array}{lll}
   \lambda^{-o_i} \quad & \mbox{for} & o_i\leq0 \\
   0                    & \mbox{for} & o_i>0
  \end{array}\RR.
\label{VEV}
\eequ
Here, we denote GUT singlet operators(``$G$-singlets'') 
 as $O_i$'s and their anomalous U(1) charge as $o_i$'s,%
\footnote{
Throughout this thesis, we denote all the superfields and chiral 
operators by uppercase letters and their anomalous U(1) 
charges by the corresponding lowercase letters.
}and 
 $\lambda(\ll1)$ is the VEV of the FN field.
The validity of this assumption is discussed in \S\ref{Ansatz}. 
In such vacua, the SUSY-zero mechanism acts, and the number 
 of the relevant interaction terms is reduced, so that 
 we can make analysis of models in spite of the infinite number 
 of interactions. 
In addition, if the symmetry contains the SU(5) symmetry of 
 Georgi-Glashow and the MSSM is realized below a certain 
 energy, the gauge coupling unification(GCU) of \msGUT is 
 naturally explained as shown in \S\ref{GCU}. 
In this way, this assumption plays a crucial role 
 in this scenario.

\subsection{Vacuum Structure}
\label{Ansatz}

\subsubsection{Singlet fields}

At first, let us consider the simplest case, where the gauge 
 symmetry is only the anomalous U(1) symmetry and 
 there are no fractional charges. 
We denote the fields with positive charges as $Z_i^+$'s\ 
 ($i=1,\cdots,n^+$), and the fields with negative ones 
 as $Z_i^-$'s ($i=1,\cdots,n^-$).

Supposing $\VEV{Z_i^+}=0$ for all $i$ as in Eq.(\ref{VEV}), 
 each $F$-flatness condition with respect to $Z_i^-$ is 
 automatically satisfied, 
 because the terms in the $F$-flatness condition contain 
 at least one of $Z_j^+$'s 
 to compensate the negative charge, $z_i^-$. 
In addition, terms that contain more than two of $Z_i^+$'s  
 never contribute to the $F$-flatness conditions, so that we can 
 analyze the vacuum structure only by considering terms that 
 contain one of $Z_i^+$'s. 
It is worthwhile to note that because the number of such 
 terms is finite, we can make an analysis in spite of the 
 infinite number of terms. 
Now, we can write the superpotential that would give 
 non-trivial $F$-flatness conditions as 
\bequ
  W_1 = \sum_i W_{Z_i^+}, 
\label{WforVEV}
\eequ
 where $W_{Z_i^+}$ consists of all the terms that contain 
 one $Z_i^+$ and no other positively charged fields.
The non-trivial $F$-flatness conditions and $D$-flatness condition 
 are given by 
\bequ
  F_{Z_i^+} = \frac{\del W_{Z_i^+}}{\del Z_i^+} = 0,\qquad
  D_A = g_A \Ls \sum_i z_i^-\abs{Z_i^-}^2 + \xi^2 \Rs = 0, 
\label{FDcondition}
\eequ
 where $g_A$ is the gauge coupling constant of the anomalous 
 U(1) symmetry.
Among the $F$-flatness conditions, one of them is written by others, 
 because the anomalous U(1) invariance makes 
 $\sum_iz_i^+Z_i^+F_{Z_i^+}=0$ hold. 
Although the $D$-flatness condition is a real condition 
 while the $F$-flatness conditions are complex ones, 
 the Higgs mechanism eats one real degree of freedom,\footnote{
Here, we ignore the degree of freedom of the dilaton multiplet. 
If we take account of the dilaton multiplet, there would be a
 superlight axion.
} 
 and we can see the $D$-flatness condition also gives 
 one complex condition.
Thus, the number of the independent conditions is $n^+$. 
On the other hand, the number of degrees of freedom is $n^-$. 
Hence, we can expect that when $n^-\geq n^+$, all the conditions 
 can be satisfied. 
This means the vacua with $\VEV{Z_i^+}=0$ can be one of the 
 SUSY preserving vacua. 
When $n^->n^+$, there would appear some flat directions. 
When $n^-=n^+$, there would be no flat directions and all the 
 fields would have superheavy masses.

Of course, the other vacua with $\VEV{Z_i^+}\neq0$ also exist. 
In such vacua, however, the SUSY-zero mechanism and the FN 
 mechanism do not act, and we cannot know there is an anomalous 
 U(1) symmetry at high energy.
Thus, we assume here that the vacua with $\VEV{Z_i^+}=0$ 
 are selected so that we can examine the implication of 
 the anomalous U(1) symmetry to low energy physics.

As for the magnitudes of the VEV of $Z_i^-$'s, 
 they have to be smaller than the coefficient of 
 the Fayet-Iliopoulos term due to the $D$-flatness condition. 
If the coefficient is small, as in string theories 
 where the term is induced radiatively, $\VEV{Z_i^-}$'s are 
 also small. 
In this case, if we assume the relation in Eq.(\ref{VEV}), 
 the VEVs become larger as the (negative) charges become larger. 
Thus, $\xi^2$ in Eq.(\ref{FDcondition}) is mainly compensated 
 by the VEV of the field with the largest negative charge.
Let us call the field as the FN field, $\Theta$, and fix the 
 normalization of the anomalous U(1) charge so that $\theta=-1$. 
Then, the VEV of the FN field is given as 
\bequ
 \VEV\Theta = \lambda \sim \xi \ll1.
\label{FNVEV}
\eequ
Below this scale, the $F$-flatness conditions are written as 
\bequ
 0=F_{Z_i^+} = \lambda^{Z_i^+}
             \Ls \sum_j\lambda^{z_j^-}Z_j^- 
               + \sum_j\sum_k\lambda^{z_j^-+z_k^-}Z_j^-Z_k^-
               + \cdots \Rs, 
\eequ
 where we omit all the coefficients which are $\order1$ 
 due to the assumption of the generic interaction.\footnote{
Hereafter, we often omit $\order1$ coefficients without notice.
}
Defining $\s Z_i^-$'s by $\s Z_i^- \equiv\lambda^{z_i^-}Z_i^-$, 
 the above conditions becomes 
 \bequ
 0=F_{Z_i^+} = \lambda^{Z_i^+}
             \Ls \sum_j\s Z_j^- 
               + \sum_j\sum_k\s Z_j^-\s Z_k^-
               + \cdots \Rs, 
\eequ
 which generally leads to solutions with $\VEV{\tilde Z_i^-}\sim1$ 
 if these $F$-flatness conditions determine the VEVs. 
Thus, the $F$-flatness conditions require
\bequ
  \VEV{Z_i^-} \sim \lambda^{-z_i^-}. 
\eequ
This relation is exactly the same one in Eq.(\ref{VEV}).
Thus, the assumption of the relation (\ref{VEV}) is self-consistent, 
 and therefore such vacua may be SUSY vacua. 
And we assume one of the vacua is selected as the vacuum of 
 the model. 

The above argument is for the simplest case.
The argument should be changed slightly, 
 when the Higgs sector has a structure by which the difference
 between the number of non-trivial $F$-flatness conditions and 
 that of the degrees of freedom of non-vanishing VEVs is changed. 
Such a structure can be realized 
 by imposing a certain symmetry, such as $Z_2$ parity, or 
 by introducing rational number charges.
For example, when the number of $Z_i^+$'s with odd $Z_2$-parity 
 is different from that of $Z_i^-$'s with odd $Z_2$-parity,
 the difference is changed by taking vanishing VEVs of 
 the $Z_2$-odd fields.
In such cases, the number of the relevant fields should be considered. 
Then an essentially same argument can be applied.

\subsubsection{Non-singlet fields}

Next, let us consider more general cases where the symmetry 
 contains a GUT symmetry in addition to the anomalous U(1) 
 symmetry, and Higgs fields possessing non-trivial 
 representations are introduced.
Even in such cases, the same arguments can be applied 
 if we use a set of independent $G$-singlets instead of 
 the singlet fields $Z_i$'s. 
We can determine the VEVs of $G$-singlets $O_i$'s from 
 the same superpotential as in Eq.(\ref{WforVEV}), 
 replacing $Z_i^+$ 
 by a set of independent $O_i$'s with positive charges,  
 although this is not easy.
The calculation is simplified if all the fields 
 $\Phi_i^+$'s (including non-singlets) with positive charges 
 have vanishing VEVs.
In such cases, the VEVs are determined only by the following 
 part of the superpotential:
\begin{equation}
  W=\sum_i^{n_+}W_{\Phi_i^+},
\label{WforVEVnS}
\end{equation}
 where $W_{\Phi_i^+}$ consists of the terms that are linear 
 in $\Phi_i^+$ and does not contain the other fields with 
 positive charges.
Note that, however, some of non-singlet fields with positive 
 charge can have non-vanishing VEVs, while all the $G$-singlets 
 with positive charge should have vanishing VEVs.
For example, let us introduce a pair of fields possessing 
 (anti-)complex representation ${\bf R}$, $\Phi({\bf R})$ and 
 $\bar\Phi({\bf\bar{R}})$.\footnote{
Hereafter, we denote a field $\Phi$ possessing 
 a representation ${\bf R}$ under the symmetry of the model 
 as $\Phi({\bf R})$
} 
If we set $\phi=-3$ and $\bar \phi=2$, then the $G$-singlet 
 $\bar \Phi\Phi$ can have non-vanishing VEV, 
 which means that $\bar \Phi$ with positive charge 
 $\bar \phi=2$ has a non-vanishing VEV. 
In such cases, it is not guaranteed that 
 the $F$-flatness conditions of fields with negative charges are 
 automatically satisfied. 
We have to take account of the part of the superpotential 
 that includes positively charged fields with non-vanishing VEVs, 
 \eg\ $\bar\Phi$, in addition to those liner in fields with 
 vanishing VEVs, in order to determine the VEVs. 
In both cases, the $G$-singlets $O_i$ with negative charges 
 have non-vanishing VEVs, $\VEV{O_i}\sim\lambda^{-o_i}$, 
 if the $F$-flatness conditions determine the VEVs. 
For example, 
 the VEV of the $G$-singlet $\bar\Phi\Phi$ is given as 
 $\VEV{\bar \Phi\Phi}\sim\lambda^{-(\phi+\bar\phi)}$.

An essential difference appears in the $D$-flatness condition
 of the GUT symmetry, which requires 
\begin{equation} 
 \abs{\VEV{\Phi}}=\abs{\VEV{\bar\Phi}}
              \sim\lambda^{-(\phi+\bar\phi)/2}.
\end{equation}
Note that these VEVs are also determined by the anomalous U(1)
 charges, but they are different from the naive expectation
 $\VEV{\Phi}\sim\lambda^{-\phi}$.
We can interpret the difference to be generated by the 
 FN mechanism induced by a U(1) symmetry which is a 
 subgroup of the symmetry group. 
More detailed analysis is made in \S\ref{Qeff}.

Another important difference may appear in the $D$-flatness 
 condition of the anomalous U(1) symmetry, 
\bequ
 D_A=g_A\Ls \xi^2 + \sum_i\phi_i\abs{\Phi_i}^2 \Rs =0.
\eequ
When the $G$-singlet that has the largest negative charge 
 is a composite operator, such as $\bar\Phi\Phi$ where 
 $\bar\phi+\phi=-1$, the $D$-flatness condition is 
 approximated as 
 $\xi^2+\bar\phi\abs{\bar\Phi}^2+\phi\abs{\Phi}^2\sim0$.
The $D$-flatness condition of the GUT symmetry leads to 
 $\abs\Phi=\abs{\bar\Phi}$.
This means $\abs\Phi=\abs{\bar\Phi}=\xi$.
In this case, because $\bar \Phi\Phi$ plays the role of the 
 FN field, the unit of the hierarchy becomes 
 $\VEV{\bar\Phi\Phi}=\lambda\sim\xi^2$. 
This relation is different from the previous one 
 in Eq.(\ref{FNVEV}) and implies that even if $\xi$ is not 
 so small, the unit of the hierarchy is strong.

In summary, we have the following: 
\bit
\item
We assume $G$-singlets with positive total charge have 
 vanishing VEVs so that the FN mechanism and the SUSY-zero 
 mechanism act effectively. 
\item
The $F$-flatness conditions of $G$-singlets with positive 
 charges determine the VEVs of $G$-singlets $O_i$'s 
 with negative charges $o_i$'s as 
 $\VEV{O}\sim \lambda^{-o}$, while the $F$-flatness conditions 
 of $O_i$'s with negative charges are automatically 
 satisfied. 
\item
The part of the superpotential that determines the VEVs 
 is expressed as $W=\sum_i W_{O_i^+}$, where $W_{O_i^+}$ is 
 linear in $O_i^+$ that has positive charges, and does not 
 contain any other fields with positive charges. 
When all the fields $\Phi_i^+$'s (including non-singlets) with 
 positive charges have vanishing VEVs, the part can 
 be written as in Eq.(\ref{WforVEVnS}). 
If some of $\Phi_i^+$'s have non-vanishing VEVs, however, 
 the part of the superpotential $W_{NV}$ that includes 
 only the fields with non-vanishing VEVs 
 must be taken into account. 
\item
When the operator is a composite operator, \eg\ $\bar\Phi\Phi$, 
 a $D$-flatness condition of the GUT symmetry requires 
 $\abs{\VEV{\Phi}}=\abs{\VEV{\bar\Phi}}
  \sim\lambda^{-(\phi+\bar\phi)/2}$.
\item
The $G$-singlet with the largest negative charge plays 
 the role of the FN field, $\Theta$. 
When the $G$-singlet is just a singlet field, the VEV is given 
 as $\VEV{\Theta}\sim\xi$, which is determined from $D_A=0$. 
When the $G$-singlet is a composite operator, \eg\ 
 $\Theta\sim\bar \Phi\Phi$, the VEV is given by 
 $\VEV{\Theta}\sim\xi^2$.
\item
If the number of the independent $G$-singlets with negative 
 charges equals that of the independent $G$-singlets positive 
 charges, generically no massless fields appear.
\eit

\section{Gauge Coupling Unification}
\label{GCU}

We have shown that the success of the gauge coupling 
 unification (GCU) in \msGUT is 
 naturally explained in the \aGUT\ scenario 
 in Ref.\cite{NGCU}.
In this section, we show the argument.

As mentioned in \S\ref{SUSY}, the hierarchical three gauge 
 couplings meet with each others at the usual GUT scale $\MGUT$ 
 in the MSSM, if the suitable SUSY breaking scale $\MSB$ is assumed. 
This is a very significant result and it is sometimes regarded as 
 an evidence supporting the validity of the existence of SUSY-GUT. 
Unfortunately, however, the nucleon decay via the colored Higgs 
 exchange is predicted to occur rapidly enough to be observed 
 in present experiments, while it has not been observed yet%
 \cite{SKproton}. 
In many GUTs, suppression of this nucleon decay is incompatible 
 with the success of GCU\cite{Murayama,Goto,lattice}. 
It may be possible to realize both the suppression and GCU 
 by adjusting parameters by hand, but in such models, 
 GCU is not a prediction but a constraint on the models. 
It is, however, desirable to construct a model where 
 such adjustments emerge in a natural manner. 
A few models that realize such adjustments have been 
 proposed\cite{orbifold,babu}.
In these models, the MSSM is realized as the effective theory 
 below the unification scale $\MU$ that is defined as the scale 
 where the three gauge coupling constants meet with each other. 
In contrast to these models, in \aGUT s, 
 the MSSM is realized not around $\MU$ but below a scale much smaller 
 than $\MU$, and the mass spectrum of superheavy 
 fields does not respect the SU(5) symmetry. 
In addition, the unified gauge group $G$ may have a higher 
 rank than SU(5) and there are several gauge symmetry 
 breaking scales. 
Nevertheless, there appear no adjustable parameters that 
 affect the condition for GCU, except for one parameter. 
This parameter corresponds to the mass of the colored Higgs
 in \msGUT. 
We can show GCU occurs if this parameter take an appropriate 
 value which corresponds to the usual GUT scale mass of the 
 colored Higgs. 
In this sence, we can say that the success of GCU in \msGUT
 is completely reproduced in the \aGUT\ scenario. 

We introduce a useful concept of ``effective charge'' 
 in \S\ref{Qeff}. 
This concept makes very clear the discussion of GCU 
 in the \aGUT\ scenario and the following analyses. 
Then we examine GCU at 1-loop level. 
Numerical analyses at 2-loop level are shown in Ref.\cite{2-loop}

\subsection{Effective Charge}
\label{Qeff}

In many models where the FN mechanism acts, magnitudes of 
 coefficients of interactions are determined by 
 the simple sum of the charges of the relevant fields. 
For example, the coefficient of the Yukawa interaction 
 $HQU$ is given as $\lambda^{h+q+u}$.

This feature is common for \aGUT s as far as VEVs 
 of $G$-singlets are given by Eq.(\ref{VEV}). 
Let us consider an effective interaction $X_1X_2\cdots X_N$. 
This interaction term may have contributions from several 
 interaction terms such as $X_1X_2\cdots X_N$$Z_1Z_2\cdots Z_N$ 
 where $G$-singlets $Z_i$ acquire non-vanishing VEVs. 
An important point is that orders of magnitude of 
 such contributions are common, $\lambda^{x_1+x_2+\cdots+x_N}$. 
This is because $X_1X_2\cdots X_N$$Z_1Z_2\cdots Z_N$ has a 
 coefficient of order 
 $\lambda^{x_1+x_2+\cdots+x_N+z_1+z_2+\cdots+z_N}$, and 
 the VEVs of $Z_i$'s cancel the extra factor 
 $\lambda^{z_1+z_2+\cdots+z_N}$ thanks to the VEV relation 
 in Eq.(\ref{VEV}).
In this way, all the effective interactions have coefficients 
 determined by the simple sum of the charges of the relevant 
 fields, as far as VEVs of $G$-singlets are related. 

Note that such a feature is generally broken when VEVs of 
 non-singlet operators are related. 
As shown in \S\ref{Ansatz}, non-singlet fields may acquire 
 VEVs different from the naively expected values in the 
 same way as $G$-singlets. 
For instance, let us consider an SO(10) model. 
A pair of (anti-)spinor Higgs, $C({\bf{16}})$ and 
 $\bar C({\bf\cc{16}})$ acquire non-vanishing VEVs 
 when $c+\bar c\leq0$:\ 
 $\abs{\VEV C}=\abs{\VEV{\bar C}}\sim\lambda^{-\half(c+\bar c)}$. 
To be more precise, $\VEV C$ is directed in the ${\bf1}_{5}$ component 
 of the decomposition
\bequ
 {\bf16} \rightarrow {\bf10}_1+{\bf\bar5}_{-3}+{\bf1}_{5}
\label{16}
\eequ
 in terms of SU(5)\sub{GG}$\times$U(1)$_V$($\subset$SO(10)), 
 and $\VEV{\bar C}$ is directed in the conjugate component ${\bf1}_{-5}$.
They are generally different from the expected values 
 $\lambda^{-c}$ and $\lambda^{-\bar c}$ for $G$-singlets.
Note that they are also written by anomalous U(1) charges, 
 but they are not written only by their own charge. 
Let us examine the effect of VEVs of such non-singlet fields
 on effective interactions. 
The effective mass term between 
 the SU(5) ${\bf5}$ component of a field $T({\bf10})$ and 
 the SU(5) ${\bf{\bar5}}$ component of a field $\Psi({\bf16})$
 is given by the interaction $T\Psi C$. 
Substituting the VEV of $C$, we find that the effective mass 
 is written as 
\bequ
 \lambda^{\psi+t+c}\VEV{C}\sim \lambda^{\psi+t+\half(c-\bar c)}, 
\label{tpsi}
\eequ
 which is not written by the simple sum of the charges of 
 the relevant fields. 
The discrepancy $\Delta c\equiv\half(c-\bar c)$ appears 
 through the VEV of non-singlet field $C$, 
 especially through that in a direction with non-vanishing charge 
 of the additional U(1)$_V$. 
And the magnitude of the discrepancy is proportional to the U(1)$_V$ 
 charge.
This is led from the assumption that all the  $G$-singlets, which 
 have vanishing U(1)$_V$ charge, acquire VEVs given 
 by Eq.(\ref{VEV}). 
In fact, the magnitudes of the discrepancies in the VEVs of $C$ 
 and $\bar C$ have opposite signs: $\VEV C\sim\lambda^{-c+\Delta c}$ 
 and $\VEV{\bar C}\sim\lambda^{-\bar c-\Delta c}$. 
This observation shows that the discrepancies are generated through 
 the FN mechanism by U(1)$_V$, and thus discrepancies in 
 coefficients of effective interactions can be written 
 by the simple sum of the U(1)$_V$ charges of the relevant fields.
This means we can consider each field has characteristic discrepancy 
 proportional to its U(1)$_V$ charge, and we can define 
 ``effective charge'' as $\tilde\phi\equiv\phi+Q_\phi\Delta_V$ 
 for a field $\Phi$, where $Q_\phi$ is the U(1)$_V$ charge of 
 $\Phi$. 
Namely, a new hierarchical factor $\lambda^{\Delta_V}$ is 
 generated by the other FN mechanism for each U(1)$_V$ charge. 
Then, we can determine coefficients of effective interactions 
 by the simple sum of these ``effective charges'' of the relevant 
 fields, even if VEVs of non-singlets are related. 
The coefficient $\Delta_V$ is determined as
 $\Delta_V=-\frac{\Delta c}5$ so that the similar 
 relation to that in Eq.(\ref{VEV}) holds for non-vanishing VEVs:
 $\VEV C\sim\lambda^{-\tilde c}$ and 
 $\VEV{\bar C}\sim\lambda^{-\s{\bar c}}$. 
Then, the effective charges of ${\bf5}$ of $T$ and of 
 ${\bf{\bar5}}$ of $\Psi$ are given as $\s t=t+\frac25\Delta c$ and 
 $\s \psi=\psi+\frac35\Delta c$. 
Thus, we can see that the effective mass of this 
 ${\bf5}$-${\bf{\bar5}}$ pair, (\ref{tpsi}), is indeed given by 
 the simple sum of $\s t$ and $\s\psi$ as $\lambda^{\s t+\s\psi}$.

The extension of the concept of effective charges to more general 
 situation is straightforward.
If there are several Higgs fields that break $U(1)_V$, the new 
 hierarchical factor $\lambda^{\Delta_V}$ is determined 
 by the Higgs fields with the largest VEVs which dominate the
 $D$-flatness condition of SO(10). 
In the case where the GUT symmetry $G$ has larger rank than 4, 
 such discrepancies appear through VEVs of non-singlet fields, 
 especially through those in directions with non-vanishing 
 charges of additional U(1)$_k$'s of 
 $G\supset$SU(5)$\times\prod_k$U(1)$_k$. 
And the magnitude of the discrepancy in an effective operator 
 is given by a linear combination of the set of the charges of 
 the effective operator. 
Thus, we can define ``effective charge'' of $\Phi$ with 
 U(1)$_k$ charge $Q^k_\phi$ as\footnote{
Hereafter, we denote the effective charge of a field 
 by using the tilded lowercase letter.}
\bequ
 \s\phi\equiv \phi+\sum_k Q_{\phi}^k \Delta_k.
\label{qeff}
\eequ
Namely, a new hierarchical factor, $\lambda^{\Delta_k}$, is 
 generated by each U(1)$_k$, and its magnitude is determined 
 by the Higgs fields with the largest VEVs that break U(1)$_k$
 so that the similar relation to that in Eq.(\ref{VEV}), 
\bequ
 \VEV\Phi\sim\lambda^{-\s\phi}
\label{effVEV}
\eequ
 holds for non-vanishing VEVs. 
Then, we can determine coefficients of effective interactions 
 by the simple sum of these ``effective charges'' of the relevant 
 fields, even if VEVs of non-singlets are related. 
Note that the effective charges respect SU(5) symmetry,
 because all the U(1)$_V$'s respect this symmetry.\footnote{
We have flexibility to define the effective charge other than 
 (\ref{qeff}).  
Even if we introduce a new hierarchy for each unbroken U(1), \eg\ 
 the hypercharge U(1)$_Y$, such a hierarchy does not appear 
 in the SM invariant interactions. 
By using this flexibility, we can define the effective charge 
 in a more convenient manner.
}
These features of the effective charge play an essential role 
 in the following discussion of GCU. 

For example, the masses of superheavy fields $X_i$ are 
easily evaluated as
\begin{equation}
 m_\eff^{x_ix_j}\lambda^{\tilde x_i+\tilde x_j},
\end{equation}
unless the mass terms are forbidden by some mechanism, such as the 
SUSY-zero mechanism.
Therefore, the determinants of the mass matrices $M_I$ of superheavy 
fields, which appear in the expressions of the gauge coupling flows, 
are written as 
\bequ
  \det M_I = \lambda^{\Sigma_i \tilde x_i^I},
\eequ
where $I$ is the index denoting the SM irreducible representations.
Note that $\det M$ can be calculated using the simple sum of 
the effective charges of the massive fields.
The ratio of the determinants for each pair of the SM multiplets
$I$ and $I'$ contained in a single multiplet of SU(5),
$\frac{\det M_I}{\det M_{I^\prime}}$,
appears in the relations for GCU.
Because the effective charges respect SU(5) symmetry, 
the contributions of the SU(5) multiplet whose $I$ and 
$I^\prime$ components are both massive cancel.
Hence, only the effective charges of massive modes 
whose SU(5) partners are massless contribute.
This can be reinterpreted as meaning that only the effective charges of the 
massless modes appear in the ratios, that is,
\bequ
  \frac{\det M_I}{\det M_{I^\prime}}
 = \frac{1/\lambda^{\Sigma_i \tilde x_i^I}}
        {1/\lambda^{\Sigma_i \tilde x_i^{I^\prime}}}, 
\label{ratio}
\eequ
where $i$ runs over the massless modes.

\subsection{Gauge Coupling Unification}
\label{1-loop}

Now, we carry out an analysis based on the renormalization group
equations (RGEs) up to 1-loop level. 
Here, we consider the most general situation, in which the GUT symmetry $G$ 
is successively broken into $G$\sub{SM} as
\bequ
  G(\equiv H_0)
   \mylimit{\Lambda_1}H_1
   \mylimit{\Lambda_2}\dots
   \mylimit{\Lambda_N}G_{\rm{SM}}
                                   (\equiv H_N).
\eequ
First, the conditions for GCU are given by
\bequ
\alpha_3(\Lambda)=\alpha_2(\Lambda)=
\frac{5}{3}\alpha_Y(\Lambda)\equiv\alpha_1(\Lambda),
\eequ
and the gauge couplings at the cutoff scale $\Lambda$ are given by
\beqn
\alpha_a^{-1}(\Lambda)&=&\alpha_a^{-1}(\MSB)
  +\frac{1}{2\pi}\Ls b_a\ln\Ls\frac{\MSB}{\Lambda}\Rs\RR\nn\\
  &&\hsp{1.89}\LL+\sum_i \Delta b_{ai}\ln\Ls\frac{m_i}{\Lambda}\Rs
+\sum_n \Delta_{an}\ln\Ls\frac{\Lambda_n}{\Lambda}\Rs
                 \Rs\,,
\eeqn
where $a=1,2,3$, $\MSB$ is the SUSY breaking scale, 
$(b_1,b_2,b_3)=(33/5,1,-3)$ are the 
renormalization group coefficients of the MSSM,
$\Delta b_{ai}$'s are the corrections to the coefficients 
caused by the massive fields with masses $m_i$'s
(see Table\ \ref{bcoeff} for concrete values),
 and the last term 
is the correction due to the restoration of the gauge symmetry 
above each symmetry breaking scale $\Lambda_n$:
\bequ
  \Delta_{an}=-3T_a\Ll H_{n-1}/H_n\Rl 
              +T_a\Ll\mbox{NG}_n\Rl=-2T_a\Ll\mbox{NG}_n\Rl.
\label{GCU3rdCorrection}
\eequ
Here, NG$_n$ denotes the NG modes that are absorbed through the Higgs 
mechanism at the scale 
$\Lambda_n$, and $T_a[{\bf R}]$'s are the Dynkin indices of a 
 representation ${\bf R}$, defined as 
\bequ
  \Tr(T_AT_B)=T[{\bf R}]\delta_{AB}, 
\eequ
where $T_A$'s are the generators in ${\bf R}$. 
The $n$-th NG modes, NG$_n$, reside in $H_{n-1}/H_n$, and thus 
 the second equality in Eq.(\ref{GCU3rdCorrection}) is derived.

\begin{table}
\label{bcoeff}
\bctr
\begin{tabular}{|c|c|c|c|c|c|c|c|c|}
\hline
$I$      & $Q+\bar Q$ & $U^c+\bar U^c$ & $E^c+\bar E^c$ & $D^c+\bar D^c$ 
         & $L+\bar L$ & $G $& $W$ & $X+\bar X$ \\
\hline
$\Delta b_{1I}$ & $\frac{1}{5}$& $\frac{8}{5}$& $\frac{6}{5}$& $\frac{2}{5}$
         & $\frac{3}{5}$ & 0 &  0 & 5 \\
\hline
$\Delta b_{2I}$ & 3 & 0 & 0 & 0 & 1 & 0 & 2 & 3 \\
\hline
$\Delta b_{3I}$ & 2 & 1 & 0 & 1 & 0 & 3 & 0 & 2 \\ 
\hline
\end{tabular}
\caption{
The correction to the renormalization coefficients $\Delta b_{aI}$ 
 by each vector-like pair of chiralmultiplet.}
\ectr
\end{table}

By using the fact that in the MSSM the three gauge couplings meet at 
the scale $\Lambda_G\sim2\times10^{16}\GeV$, the relations expressing 
unification, $\alpha_a(\Lambda)=\alpha_b(\Lambda)$, become
\beqn
  (b_a-b_b)\ln(\Lambda_G)
  &+&\sum_I(\Delta b_{aI}-\Delta b_{bI})\ln\Ls\det M_I\Rs
  \nn\\
  &+&\sum_n(\Delta_{an}-\Delta_{bn})\ln\Ls\Lambda_n\Rs=0,
\label{Cond}
\eeqn
where $I$ runs over the SM irreducible representations.
Because the sum of $\Delta b_{aI}$'s over an SU(5) 
multiplet is independent of $a$, the second term in Eq. (\ref{Cond})
can be written in terms of 
the ratios of the determinants of the mass matrices in (\ref{ratio}), 
and therefore in terms of the contributions from the massless modes, 
as mentioned above.
For example for ${\bf{\bar5}}$ representation, 
 the second term is
\bequ
 (\Delta b_{aL}-\Delta b_{bL})\ln\Ls\det M_L\Rs
  +(\Delta b_{aD^c}-\Delta b_{bD^c})\ln\Ls\det M_{D^c}\Rs, 
\label{5barexample}
\eequ
 and we know
\bequ
 \Delta b_{aL}+\Delta b_{aD^c}=\Delta b_{bL}+\Delta b_{bD^c}, 
\eequ
 thus (\ref{5barexample}) becomes
\bequ
 (\Delta b_{aL}-\Delta b_{bL})\ln\Ls\frac{\det M_L}{\det M_{D^c}}\Rs, 
\eequ
 which is written by the ratio of the determinants.
In terms of the ``effective mass'' of massless modes, which is defined 
as $m_\eff\equiv\lambda^{\tilde x+ \tilde y}$ even when 
$\tilde x+ \tilde y<0$, the second term in (\ref{Cond}) can be written 
 as
\bequ
  \sum_{i=\mbox{\scriptsize{massless}}}
  (T_a\Ll i\Rl -T_b\Ll
          i
      \Rl)
  \ln\Ls 1/m_\eff^{i}\Rs.
\nn
\eequ
These massless modes consist of two types, physical massless modes, 
such as the MSSM doublet Higgs ($H_u$ and $H_d$), and unphysical NG modes.
From (\ref{GCU3rdCorrection}), we can see that the contribution of 
the latter type is cancelled by that of the last term in Eq.(\ref{Cond})
if the conditions  
\bequ
  m_\eff^{\mbox{\scriptsize{NG}}_n}\sim\Lambda_n^{-2}
\label{condi}
\eequ
hold.
These conditions are satisfied when the 
vacuum structure satisfies (\ref{VEV}),
because $m_\eff^{\mbox{\scriptsize{NG}}_n}$ is the coefficient of 
the bilinear term of the $n$-th NG modes, 
$\Phi$ and $\bar\Phi\ (\tilde {\bar\phi}=\tilde \phi)$, 
and therefore 
$m_\eff^{\mbox{\scriptsize{NG}}_n}\sim\lambda^{2\tilde \phi}$, 
and from (\ref{effVEV}), $\Lambda_n\sim\lambda^{-\tilde \phi}$.

When (\ref{condi}) holds, only the physical massless 
modes contribute to the conditions for GCU, 
and they are independent of the details of the Higgs sector, 
such as the field content and the symmetry breaking pattern.
In particular, if all the fields other than those in the MSSM become 
superheavy, only the MSSM doublet Higgs fields $H$ contribute, and
we have 
\bequ
  (b_a-b_b)\ln(\Lambda_G)
  +(\Delta b_{aH}-\Delta b_{bH})\ln\Ls {1/m_\eff^H}\Rs =0,
\eequ
for all combinations $(a,b)$. These relations lead to 
$\ln(\Lambda_G)=\ln(m_\eff^H) =0$, and thus 
\bequ
 \quad\Lambda\sim \Lambda_G\,,\,\,\tilde h_u+\tilde h_d\sim 0.
\label{unification}
\eequ
The first relation here simply defines the scale of the theory:
The cutoff scale $\Lambda$ is taken as the usual GUT scale, 
$\Lambda_G$.
This is also the case in the minimal SU(5) SUSY-GUT, where
the scale at which SU(5) is 
broken is also taken as $\Lambda_G$. 
The second relation in (\ref{unification}) corresponds to that for the 
colored Higgs mass in the minimal SU(5) GUT, because the effective 
colored Higgs mass is obtained as 
$m^{H^c}_\eff\sim \lambda^{\tilde h_u+\tilde h_d}$. 
Therefore, we have no tuning parameters for GCU other than those 
in the minimal SU(5) SUSY-GUT.
Note that (when $\tilde h_u+\tilde h_d= 0$) 
if we calculate gauge couplings at a low energy scale in
the anomalous U(1) GUT scenario with any cutoff scale\ 
(for example, the Planck scale)
and use them as the initial values, the three running gauge
couplings calculated in the MSSM meet with each others 
at the cutoff scale.
In this way, we can naturally explain 
GCU in the minimal SU(5) SUSY-GUT.

\subsection{Nucleon Decay}

As shown in the previous subsection, the success of GCU in 
 \msGUT\ is reproduced in the \aGUT\ scenario. 
Then, it may seem that the same problem as in \msGUT\ arises 
 in \aGUT s: the nucleon decay via dimension 5 operators tends 
 to be too rapid. 
In fact the same problem arises if we take $\tilde h_u+\tilde h_d=0$. 
This condition corresponds to the condition that the colored Higgs 
 should have a mass around $\MGUT$, and looks required for GCU.
Note that, however, the relation $\tilde h_u+\tilde h_d\sim 0$ does not 
imply $\tilde h_u+\tilde h_d=0$, 
 because there is an ambiguity involving
$\order1$ coefficients. 
As mentioned in \S\ref{Qeff}, contributions to masses 
 from higher-dimensional interactions are not suppressed 
 in contrast to the usual situation.
For instance, the VEV of an adjoint Higgs $A$, $\VEV{A}$, which breaks 
 SU(5) symmetry, contributes to the mass of $X$ and $\bar X$ 
 through higher-dimensional interactions, 
 $\lambda^{x+\bar x+na}\bar XA^n X$. 
The orders of such contributions are the same as that from
the mass term $\lambda^{x+\bar x}\bar XX$, because 
$\VEV{A}\sim \lambda^{-a}$.
Therefore the $\order1$ coefficients do not respect SU(5) symmetry 
 at all. 
This allows a non-zero value of $\tilde h_u+\tilde h_d$.
If $\tilde h_u+\tilde h_d$ is negative, 
 the nucleon decay via dimension 5 operators is suppressed. 
The suppression 
requires the effective mass of the colored Higgs 
$m_{\rm eff}^{H^c} \sim \lambda^{\tilde h_u+\tilde h_d}\Lambda$
$>$ $\order{10^{18}\GeV}$, 
and therefore $\tilde h_u+\tilde h_d\leq -3$ is needed.
Note that the physical masses of the colored Higgs are smaller than 
 $\Lambda$, although the effective mass is larger, 
 $m_{\rm eff}^{H^c}>\Lambda$, as shown in concrete models in the next 
 chapter. 

In this way, the nucleon decay via dimension 5 operators can be 
 suppressed if we take $\tilde h_u+\tilde h_d\leq -3$. 
On the other hand, the nucleon decay mediated by gauge bosons is 
 enhanced compared to the usual GUTs.
This is because the cutoff scale $\Lambda$ is required to be 
 around the usual GUT scale $\MGUT\sim2\times10^{16}\GeV$ 
 by the condition for GCU 
 (\ref{unification}), and the unification scale $\MU$ is smaller than 
 $\Lambda$ and thus than $\MGUT$. 
In fact in \aGUT s, $\MU$ is given by 
 the VEV of the Higgs $A$ that breaks SU(5) symmetry 
 as $\VEV A\sim \lambda^{-a}\Lambda\ll\Lambda\sim\MU$. 
If we choose $a=-1$ and $\lambda\sim 0.22$ as typical values,
\footnote{
Note that the slowest proton decay via dimension 6 operators
is obtained when $a=0$, and
the value must be the same as that in the usual GUT scenario. 
However, when $a=0$, generally terms of the form
$\dt A^n W_\alpha W^\alpha$ are allowed,
where $W_\alpha$ is a SUSY field strength.
This makes it impossible to realize natural GCU. 
The most
natural way to forbid these terms is to choose $a$ to be negative,
which leads to a shorter proton lifetime.
}
the proton lifetime can be roughly estimated, 
using a formula in Ref.\cite{Murayama} and a recent result provided
by a lattice calculation for the hadron matrix element parameter
$\alpha$\cite{lattice},
as
\begin{equation}
\tau_p(p\rightarrow e\pi^0)\sim 1\times 10^{34}
\left(\frac{\Lambda_A}{5\times 10^{15}\ {\rm GeV}}\right)^4
\left(\frac{0.01({\rm GeV})^3}{\alpha}\right)^2  {\rm yrs}.
\nn
\end{equation}
This value
is near the present experimental limit\cite{SKproton}.
Thus, \aGUT s predict that the proton decay $p\rightarrow e+\pi$ 
 will be observed in future experiments.

\subsection{Summary}

In this section, We have shown the success of GCU in 
 \msGUT\ is completely reproduced in the \aGUT\ scenario. 
Usually, if we adopt a simple group whose
rank is higher than that of the standard gauge group (for example, 
SO(10), \E6, SU(6), etc.), GCU can always 
be realized by  tuning the additional degrees of freedom related with 
the several scales of Higgs VEVs.
However, in the \aGUT\ scenario,
all the charges of the Higgs fields, except that of the MSSM doublet 
Higgs, are cancelled in the relations for GCU
(\ref{unification}),
and therefore we have no tuning parameters for GCU 
other than those in the minimal SU(5) SUSY-GUT. 
This result is independent of detail of models. 
In fact the assumptions of the argument are following: 
\begin{enumerate}
\item The unification group $G$ is simple.
\item The VEV relation (\ref{VEV}) holds.
\item Below a certain scale, the MSSM is realized.
\end{enumerate}
If these assumptions are realized, the argument can be applied 
 even if the scenario does not employ anomalous U(1) symmetries%
\footnote{
In this case, $o_i$'s are not charges, but merely certain numbers 
 that we assign for fields.
} and/or there are some flat directions. 
The second assumption is naturally realized in the GUT scenario 
with anomalous
\footnote{
We can use non-anomalous U(1) symmetry instead of anomalous U(1)
symmetry if 3 conditions in the Introduction are satisfied. 
However, since we don't know such models, we adopt anomalous U(1) 
symmetry in this letter.
}
 U(1) gauge symmetry as shown in \S\ref{Ansatz}.
Moreover, some of the above conditions can be weakened. 
For example, even when the gauge group
is non-simple, GCU is realized if the charge 
assignment
respects SU(5) symmetry. 

Finally, we would like to make comments on the magnitudes of 
 the gauge coupling constants, which we have not taken care of 
 in this section. 
If there appear many Higgs below $\MGUT$, the gauge couplings 
 tend to become large, and the analysis based on perturbation 
 made in this section may not be applicable.
Thus, we have to be careful that the gauge coupling at cutoff scale 
 is in the perturbative region when we construct models.


\chapter{Models}
\label{Models}

\section{SO(10) Models}

In this section, we introduce the SO(10) models discussed 
 in Ref.\cite{maekawa}.

\subsection{Realization of DW VEV}
\label{RealDWVEV}

Before examining concrete models, let us show how the 
 Dimopoulos-Wilczek\ (DW) type of VEV of an adjoint Higgs $A$, 
 $\VEV{A}\propto\tau_2\times {\rm diag.}(1,1,1,0,0)$, which is
 proportional to the generator $B-L$, 
 can be realized in \aGUT s. 
As mentioned in \S\ref{DTS}, usually the realization needs 
 fine-tuning, but it is possible to realize the DW VEV 
 as the result of an equation of motion. 
For example, this turns out to be the case 
 when we introduce an additional adjoint Higgs $A'$ 
 and assign its anomalous U(1) charge so that $(0<)-3a\leq a'<-4a$.
As mentioned in \S\ref{Ansatz}, the superpotential that is relevant 
 for determining this VEV is the part that is linear 
 in the positively charged Higgs field. 
Thus, the relevant part in this case is in general
written as
\begin{equation}
W_{A^\prime}=\lambda^{a^\prime+a} A^\prime A+\lambda^{a^\prime+3a}(
 (A^\prime A)_{\bf 1}(A^2)_{\bf 1}
+(A^\prime A)_{\bf 54}(A^2)_{\bf 54}),
\label{SO(10)WA'}
\end{equation}
where the suffixes {\bf 1} and {\bf 54} indicate the representation 
of the composite
operators under the SO(10) gauge symmetry, and we omit $\order1$ 
 coefficients. 
We can choose a gauge where the VEV of $A$ is written as
 $\VEV{A}=\tau_2\times {\rm diag.}(x_1,x_2,x_3,x_4,x_5)$. 
In this gauge, the $F$-flatness condition of the $A'$ field requires
 $x_i(\alpha\lambda^{-2a}+\beta(\sum_j x_j^2)+\gamma x_i^2)=0$, 
 where $\alpha$, $\beta$ and $\gamma$ are $\order1$ parameters. 
Here, the last term comes from the interaction 
 $(A^\prime A)_{\bf 54}(A^2)_{\bf 54}$. 
This EOM gives only two solutions $x_i^2=0$ and 
 $x_i^2=-\frac{\alpha}{\gamma+N\beta}\lambda^{-2a}$,  
 where $N$ is the number of $x_i \neq 0$ solutions and  $N=0,1,\cdots,5$.
The DW VEV is obtained when $N=3$. 
Note that the higher-dimensional terms $A^\prime A^{2L+1}$ $(L>1)$ are 
forbidden by the SUSY-zero mechanism, 
 and it is difficult to forbid them by other symmetries, such as 
 discrete symmetries, because $A^2$ should be a singlet under such 
 symmetries in order to allow both $A'A$ and $A'A^3$. 
If such terms are allowed, 
the number of possible VEVs other than the DW VEV
becomes larger, and thus it becomes less natural to obtain the DW VEV. 
This is a
crucial point in the \aGUT scenario, and the anomalous U(1) gauge symmetry
plays an essential role in forbidding the undesirable terms.

In this manner, we can realize the DW VEV without fine-tuning. 
In order to make sure that the DTS problem is indeed solved, 
 we have to examine the whole Higgs sector. 
In particular, the rank reducing VEV, \eg\ VEVs of spinor Higgs 
 $C$ and $\bar C$, should not couple to $W_{A'}$ while 
 these VEVs and the VEV of $A$ should couple with each others 
 in order to avoid the pseudo NG\ (PNG) modes. 
As shown in the next subsection, the decoupling can be realized 
 by the SUSY-zero mechanism, and the avoidance of PNG can be achieved 
 through the Barr-Raby mechanism\cite{BarrRaby}.

\subsection{Higgs Sector of SO(10) Models}

In order to break SO(10) down to $G_{\rm{SM}}$, we need at least 
 an adjoint Higgs $A({\bf45})$ and one pair of spinor Higgs 
 $C({\bf16})$ and $\bar C({\bf{\cc{16}}})$. 
The gauge singlet operators $A^2$ and $\bar CC$ must
have negative total anomalous U(1) charges
to obtain non-vanishing VEVs, 
as discussed in \S\ref{Ansatz}. 
Then, they cannot have interaction terms, especially mass terms, 
 by themselves. 
Thus, we have to introduce 
 corresponding conjugate fields possessing positive charges 
 $A'({\bf45})$, $\bar C'({\bf\cc{16}})$ and $C'({\bf16})$ 
in order to give masses to all the Higgs fields.\footnote{
Strictly speaking, since some of the Higgs fields are eaten by the 
Higgs mechanism, in principle, a smaller number of positive fields 
can give superheavy masses
to all the Higgs fields. Here we do not examine this possibility.
}
When we employ the DW mechanism to solve the DTS problem, we need two 
 vector Higgs $H({\bf10})$ and $H'({\bf10})$, one of which 
 has a vanishing self mass term, \ie\ $H^2$ is forbidden. 
This mass term is forbidden by the SUSY-zero mechanism 
 if $h<0$. 
Then, $H'$ must have positive charge to allow the interaction $HAH'$.%
\footnote{
In this case, the mass term $HH'$ cannot be forbidden by the 
 SUSY-zero mechanism and we have to introduce an additional symmetry, 
 for example $Z_2$ symmetry.
}
This is, in a sense, a minimal set of (non-singlet) Higgs content, 
 and this minimal content is enough to construct realistic models. 

An example of charge assignments of the content of the Higgs 
 sector including singlet Higgs is shown in Table\ \ref{SO(10)Content}. 
\begin{table}
\begin{center}
\begin{tabular}{|c|c|c|} 
\hline
                  &   non-vanishing VEV  & vanishing VEV \\
\hline 
{\bf 45}          &   $A(a=-1,-)$        & $A'(a'=3,-)$      \\
{\bf 16}          &   $C(c=-3,+)$        
                  & $C'(c'=2,-)$      \\
${\bf \overline{16}}$&$\bar C(\bar c=0,+)$ 
                  & $\bar C'(\bar c'=5,-)$ \\
{\bf 10}          &   $H(h=-3,+)$        & $H'(h'=4,-)$      \\
{\bf 1}           &$\Theta(\theta=-1,+)$,$Z(z=-2,-)$,
                  $\bar Z(\bar z=-2,-)$& $Z'(s=3,+)$ \\
\hline
\end{tabular}
\caption{Typical values of anomalous U(1) charges.}
\label{SO(10)Content}
\end{center}
\end{table}

\subsubsection{VEV determination} 

As mentioned in \S\ref{Ansatz}, 
the superpotential required by determination of the VEVs can be 
 written as
\begin{equation}
W=W_{A'} + W_{Z'} + W_{C'}+W_{\bar C'}+ W_{H'}
+W_{NV}.
\end{equation}
Here, $W_X$ denotes the terms linear in the positive charged field
$X$, which has vanishing VEV. 
And $W_{NV}$ contains terms consisting of only unprimed fields, 
 \ie\ fields possessing non-vanishing VEVs.
They are given as 
\beqn
  W_{A'}  &=&  \lambda^{a'+a}A' A+\lambda^{a'+3a}(
                     (A' A)_{\bf 1}(A^2)_{\bf 1}
                    +(A' A)_{\bf 54}(A^2)_{\bf 54})  \\
  W_{Z'}  &=&  \lambda^{z'+c+\bar c}Z'\left((\bar CC)+\lambda^{-(c+\bar c)}
                  +\lambda^{-(c+\bar c)+2a}A^2\right)  \\
  W_{C'}  &=&  \bar C(\lambda^{\bar c' +c+a}A
                    +\lambda^{\bar c' +c+z}Z)C' \\
  W_{\bar C'}  &=&  \bar C'(\lambda^{\bar c' +c+a} A
                         +\lambda^{\bar c' +c+z}Z)C\\
  W_{H'}  &=&  \lambda^{h+a+h'}H' AH,
\eeqn 
 for the charge assignment of Table\ \ref{SO(10)Content}.
Note that terms including two fields with vanishing VEVs like 
$\lambda^{2h^\prime}H^\prime H^\prime$
give contributions to the mass terms but not to the VEVs.
Also, we 
 ignore terms that do not include the products of only singlet 
 components under $G_{\rm{SM}}$, like
 ${\bf 16}^4$, ${\bf \overline{16}}^4$, ${\bf 10\cdot 16}^2$,
 ${\bf 10 \cdot \overline{16}}^2$ and ${\bf 1\cdot 10}^2$, 
 even if these terms are allowed by the symmetry.
This is because they can give contributions only on vacua other than 
 SM-like vacua, while 
 they have contributions on the mass spectrum.
All the terms in $W_{NV}$ contain only fields with non-vanishing 
 VEVs. 
In the typical charge assignment, 
 they do not play a significant role for VEV determination,
 because they
 do not include the products of components that are exclusively singlets 
 under $G_{\rm{SM}}$ and thus we can safely ignore them. 

$W_{A'}$ is the same as that in Eq.(\ref{SO(10)WA'}) and 
 the argument in \S\ref{RealDWVEV} can be applied. 
Note that the rank reducing Higgs $C$ and $\bar C$ do not appear 
 in $W_{A'}$ thanks to the SUSY-zero mechanism. 
Thus, we can realize the DW VEV without fine-tuning. 
The VEV $\VEV{A({\bf 45})}_{B-L}=\tau_2\times {\rm diag.}(v,v,v,0,0)$, 
 breaks SO(10) into \Ga.

The $F$-flatness condition of $Z'$ requires $\VEV{\bar CC}\sim 
\lambda^{-(c+\bar c)}$. 
The magnitude of $\VEV C$ and $\VEV{\bar C}$ are determined 
 by the $D$-flatness condition of SO(10) 
 as $\abs{\VEV{C}}=\abs{\VEV{\bar C}}\sim \lambda^{-(c+\bar c)/2}$, 
 as in \S\ref{Ansatz}.

Next, we discuss the $F$-flatness conditions of $C'$ and $\bar C'$,
which realize the alignment of the VEVs $\VEV{C}$ and $\VEV{\bar C}$
and results in masses for the PNG fields. 
This simple mechanism was proposed by Barr and Raby\cite{BarrRaby}.
The $F$-flatness conditions $F_{C'}=F_{\bar C'}=0$ give
$(\lambda^{a-z} A+Z)C=\bar C(\lambda^{a-\bar z} A+\bar Z)=0$. 
Recall that the VEV of $A$ is 
proportional to the $B-L$ generator $Q_{B-L}$ 
(precisely, $\VEV{A}=\frac{3}{2}vQ_{B-L}$), and that
 the spinor representation ${\bf 16}$ is decomposed into 
$({\bf 3},{\bf 2},{\bf 1})_{1/3}$, 
$({\bf \bar 3},{\bf 1},{\bf 2})_{-1/3}$, 
$({\bf 1},{\bf 2},{\bf 1})_{-1}$ and 
$({\bf 1},{\bf 1},{\bf 2})_{1}$ 
under \Ga.
Since $\VEV{\bar CC}\neq 0$, 
 $Z$ is fixed such that $Z\sim -\frac{3}{2}\lambda v Q_{B-L}^0$, 
where $Q_{B-L}^0$ is
the $B-L$ charge of the component of $C$ that has non-vanishing VEV. 
Once the VEV of $Z$ is determined, 
no other component fields can have non-vanishing VEVs, 
because they have different charges $Q_{B-L}$. 
If the component that obtains a non-zero VEV is 
 $({\bf 1},{\bf 1},{\bf 2})_1$ 
 (and therefore $\VEV{Z}\sim -\frac{3}{2}\lambda v$), 
the gauge group 
\Ga\ is broken down to the SM gauge group. 
Once the direction of the VEV $\VEV{C}$ is 
determined, the VEV $\VEV{\bar C}$ must be directed 
 in the same direction, 
because of the $D$-flatness
condition. Therefore, $\VEV{\bar Z}\sim -\frac{3}{2}\lambda v$.

Finally the $F$-flatness condition of $H'$ leads to vanishing VEVs of 
 the color-triplet Higgs, $\VEV{H_T}=0$. 

Now, all VEVs have been fixed as 
\beqn
  \VEV {X^+}  &=&  0,  \\
  \VEV A  &=&  \tau_2\times{\rm diag.}(v,v,v,0,0)\ 
               ,\ v=\lambda^{-a},  \\
  \VEV C  &=&  \VEV{N^c_C}\ =\ \lambda^{-{c+\bar c\over2}},  \\
  \VEV{\bar C}  &=&  \VEV{{N^c_{\bar C}}}\ =
                    \ \lambda^{-{c+\bar c\over2}},  \\
  \VEV{{D^c}_H}  &=& 0, 
\eeqn
where $X^+$ denotes all the positively charged Higgs. 
The symmetry breaking pattern is given as 
\bctr
\begin{tabular}{cclcl}
 SO(10) & $\longrightarrow$ & \Ga & \quad & at $\lambda^{-a}\Lambda$ \\
        & $\longrightarrow$ & \GSM & \quad 
        & at $\lambda^{-\half(c+\bar c)}\Lambda$. 
\end{tabular}
\ectr
The parameter that parametrizes the effective charges is written 
 as $\Delta c=\half(c-\bar c)$. 

There are several terms that must be forbidden for the stability 
of the DW mechanism. For example, $H^2$, $HZH'$ and 
$H\bar Z H'$ induce a large mass of the doublet Higgs, 
and the term $\bar CA' A C$ would destabilize the DW VEV of 
$\VEV{A}$. 
We can easily forbid these terms using the SUSY-zero mechanism.
For example, if we choose
$h<0$, then $H^2$ is forbidden, and if we choose $\bar c+c+a+a'<0$, 
then
$\bar CA' A C$ is forbidden. 
Once these dangerous terms are forbidden
by the SUSY-zero mechanism, higher-dimensional terms that could 
 also become
dangerous (for example, 
$\bar CA' A^3 C$ and $\bar CA' C\bar CA C$) are automatically
forbidden.
This is another attractive property of the \aGUT\ scenario. 
The dangerous terms which should be forbidden are
\begin{equation}
  H^2, HH',HZH', \bar CA'C, \bar CA'AC, \bar CA'ZC, A'A^4, A'A^5, 
\label{forbidden}
\end{equation}
and the terms required to realize DTS are
\begin{equation}
A'A, A'A^3, HAH',\bar C'(A+Z)C, \bar C(A+Z)C', S\bar CC.
\label{required}
\end{equation}
Here we denote both $Z$ and $\bar Z$ as ``$Z$''.
In order to forbid (\ref{forbidden}) but not (\ref{required}), 
we introduce $Z_2$ parity and assign charges like as 
 in Table\ \ref{SO(10)Content}.

Of course, the above conditions are necessary but not sufficient. 
To determine whether a given assignment actually works well,
we have to examine the mass matrices of the Higgs sector.

\subsubsection{Mass spectrum of the Higgs sector}

Here, we examine the mass matrix of the Higgs sector in 
 Table\ \ref{SO(10)Content} for each representation of the SM to 
 show that all the extra fields indeed acquire superheavy masses and 
 the MSSM is realized at low energy scale. 
For this purpose, we have to 
take into account not only terms detailed in the previous section 
but also terms that contain two fields with vanishing
VEVs. 

Under the decomposition SO(10)$\supset$SU(5)$\supset$\GSM, 
the spinor ${\bf 16}$, vector ${\bf 10}$ and adjoint ${\bf 45}$
are decomposed in terms of the representations of the SM group as 
\begin{eqnarray}
{\bf 16}&\rightarrow&
\underbrace{[Q+U^c+E^c]}_{\bf 10}+\underbrace{[D^c+L]}_{\bf \bar 5}
+\underbrace{N^c}_{\bf 1}, \nn \\
{\bf 10}&\rightarrow&
\underbrace{[D^c+L]}_{\bf \bar 5}+\underbrace{[\bar D^c+\bar L]}_{\bf 5},
\label{class}\\
{\bf 45}&\rightarrow&
\underbrace{[G+W+X+\bar X+N^c]}_{\bf 24}
+\underbrace{[Q+U^c+E^c]}_{\bf 10}
+\underbrace{[\bar Q+\bar U^c+\bar E^c]}_{\bf \overline{10}}
+\underbrace{N^c}_{\bf 1}. \nn
\end{eqnarray}

First, we examine the mass spectrum of ${\bf 5}$ and ${\bf\bar5}$ of
SU(5). 
Considering the additional terms
$\lambda^{2h'} H' H'$,
$\lambda^{c'+\bar c'}\bar C' C'$, 
$\lambda^{c'+c+h'}C'CH'$, 
$\lambda^{\bar c'+z+\bar c+h}Z\bar C'\bar CH$,
$\lambda^{\bar c'+\bar c+h'}\bar C'\bar CH'$ and
$\lambda^{2\bar c+h'}\bar C^2H'$,
the mass matrices $M_I$ ($I=D^c,L$) are given by 
\begin{equation}
M_I=\bordermatrix{
\bar I\backslash I&H & {C} & 
                  {H'}&{C'} \cr
H & 0 & 0 & \lambda^{h+h' +a}\VEV{A}  & 0 \cr
{\bar C}& 0 & 0 & \lambda^{h'+2\bar c}\VEV{\bar C} 
& \lambda^{\bar c+c'} \cr
{H'} & \lambda^{h+h' +a}\VEV{A} & 0& \lambda^{2h'} 
 & \lambda^{h'+c'+c}\VEV{C} \cr
{\bar C'} &  
\lambda^{h+\bar c'+\bar c}\VEV{\bar C} 
 & \lambda^{c+\bar c'} 
 & \lambda^{h'+\bar c'+\bar c}\VEV{\bar C}
  & \lambda^{c'+\bar c'} \cr},
\end{equation}
where the vanishing elements result from the SUSY-zero mechanism. 
Substituting the scales of non-vanishing VEVs, we can find that 
 the non-vanishing elements are written as a simple sum of the 
 effective charges of the relevant fields.
It is worthwhile examining the general structure of the 
mass matrices.
The first two columns and rows correspond to fields with 
non-vanishing VEVs that have smaller charges, and the last 
two columns and rows correspond to fields with vanishing VEVs 
that have larger charges.
Therefore, it is useful to divide the matrices into four $2\times 2$ 
matrices as
\begin{equation}
  M_I  =  \left(
    \begin{array}{cc}
      0 & A_I \\ B_I & C_I
    \end{array}
    \right).
\end{equation}
We can see 
that the ranks of $A_I$ and $B_I$ are reduced to 1
when the VEV $\VEV{A}$ vanishes.
This implies that
the rank of $M_L$ is reduced, and actually it becomes 3.
However, the ranks of $A_{D^c}$ and $B_{D^c}$ remain 2, 
because the field $A$ becomes non-zero on $D^c$.
Therefore DTS is realized.
The mass spectrum of $L$ is obtained as
$(0,\lambda^{2h'},\lambda^{\bar c+c'},\lambda^{\bar c'+c})$.
The massless modes of the doublet Higgs are estimated to be 
\begin{equation}
\bar L_H\,,\,\, L_H+\lambda^{h-c+\Delta c}L_C.
\label{mixing}
\end{equation}
The elements of the 
matrices $A_I$ and $B_I$  become generally larger than the 
elements of the matrices $C_I$ because the total effective 
 charges of the corresponding pair of fields in $A_I$ 
and $B_I$ are smaller than those in $C_I$.
Therefore, the mass spectrum of $D^c$ is essentially estimated 
by the matrices $A_{D^c}$ and $B_{D^c}$ as 
$(\lambda^{h+h'},\lambda^{h+h'},\lambda^{\bar c+c'},
\lambda^{\bar c'+c})$.
Note that in order to realize proton decay, we have to pick up 
at least one element of $C_I$. 
Because such an element is generally smaller than the mass scale 
of $D^c$, proton decay is suppressed.
In fact, the colored Higgs effective mass that appears 
 in the expression of proton lifetime is estimated as
$(\lambda^{h+h'})^2/\lambda^{2h'}=\lambda^{2h}$, 
which is larger than the cutoff scale, because $h<0$.

Next, we examine the mass matrices for the representations 
$I=Q,U^c$ and $E^c$,
which are contained in the {\bf 10} of SU(5),
where the additional terms
$\lambda^{2a'}A' A'$, 
$\lambda^{c'+\bar c'}\bar C' C'$,
$\lambda^{c'+a'+\bar c} \bar CA' C'$ and
$\lambda^{\bar c'+a'+c} \bar C' A' C$
must be taken into account.
The mass matrices are written as 
\begin{equation}
M_I=\bordermatrix{
\bar I\backslash I&A &{C}& {A'} & {C'} \cr
A &0& 0 & \lambda^{a'+a} \alpha_I  
     & \lambda^{\bar c+c'+a}\VEV{\bar C} \cr
{\bar C}&0 & 0 & 0 & \lambda^{\bar c+c'}\beta_I \cr
{A'} &\lambda^{a+a'} \alpha_I & 0& \lambda^{2a'}  
        & \lambda^{\bar c+c'+a'}\VEV{\bar C} \cr
{\bar C'} &\lambda^{c+\bar c'+a}\VEV{C} 
                  &\lambda^{c+\bar c'}\beta_I 
                  & \lambda^{c+\bar c'+a'}\VEV{C} 
                  & \lambda^{c'+\bar c'}\cr},
\label{mass10}
\end{equation}
where $\alpha_Q=\alpha_{U^c}=0$ and $\beta_{E^c}=0$, because
there are NG modes in symmetry breaking processes
SO(10)$\rightarrow$\Ga and 
 SU(2)\sub R$\times$U(1)$_{B-L}\rightarrow$U(1)$_Y$.
Defining $2\times 2$ matrices as in the $I=L,D^c$ case,
 we can easily find that the ranks of $A_I$ and $B_I$ are reduced.
Thus for each $I$, the $4\times 4$ matrices $M_I$ have one 
vanishing eigenvalue, which corresponds to the NG mode 
 that is eaten through the Higgs mechanism. 
The mass spectrum of the remaining three
modes is ($\lambda^{c+\bar c'}$, $\lambda^{c'+\bar c}$,
$\lambda^{2a'}$) for the color-triplet modes $Q$ and $U^c$, and
($\lambda^{a+a'}$, 
$\lambda^{a+a'}$,
$\lambda^{c'+\bar c'}$) for the color-singlet modes $E^c$.

Finally,  we examine the mass matrices for the representations 
$I=G,W$ and $X$, 
which are contained in the {\bf 24} of SU(5).
Considering the additional term $\lambda^{a+a'}AA'$, 
 the mass matrices $M_I(I=G,W,X)$ are given as
\begin{equation}
M_I=\bordermatrix{
\bar I\backslash I &    A       &        {A'}   \cr
A    &     0        & \alpha_I\lambda^{a+a'}  \cr
{A'} & \alpha_I\lambda^{a+a'} & \lambda^{2a'}  \cr}.
\end{equation}
Two $G$ and two $W$ acquire masses $\lambda^{a'+a}$.
Because $\alpha_X=0$, one pair of $X$ is massless and this 
 massless mode is eaten through the Higgs mechanism. 
The other pair has a rather light mass of $\lambda^{2a'}$.

In this way, all the extra fields indeed acquire superheavy masses and 
 the physical massless modes are only two doublet Higgs. 
The gauge coupling unification is realized, as mentioned 
 in \S\ref{GCU}. 
Therefore, the Higgs sector goes well. 
The next issue is about the matter sector, where fields are 
 odd under the R-parity while those of the Higgs sector are 
 assigned even R-parity. 
Such an assignment of the R-parity guarantees that the argument 
 regarding VEVs in \S\ref{Ansatz} does not change 
 if these matter fields have vanishing VEVs.

\subsection{Matter Sector of SO(10) Models}
\label{SO(10)Matter} 

In this section, we show how realistic mass matrices of 
 quarks and leptons are realized in the \aGUT\ scenario.

In that scenario, higher-dimensional terms give contributions 
 of the same order as renormalizable terms.
Thus the order of coefficient of each term 
 in the low energy effective theory 
 respects the GUT symmetry, but the precise value of the coefficient 
 does not respect the symmetry 
 at all if GUT breaking VEVs can couple to the term. 
This means that the wrong GUT relation between the down-type quarks 
 and charged leptons can be easily avoided. 
Unfortunately, it is difficult to avoid the wrong GUT relation 
 between the down-type quarks and up-type quarks
 if we employ the minimal content of the matter sector, 
 three ${\bf 16}$ representations $\Psi_i$ 
 ($\psi_1\geq\psi_2\geq\psi_3$). 
To avoid the relation, we introduce an additional matter field $T$ 
 in the ${\bf10}$ representation, which is vector-like so that 
 no exotic particles are expected at low energy 
 while it can modify the origin of each generation in 
 the ${\bf{\bar5}}$ sector. 
The ${\bf5}$ component of $T$ acquire a mass with a linear combination 
 of ${\bf{\bar5}}$ components of $\Psi_i$ and $T$ as 
\begin{eqnarray}
  &&  {\bf 5}_T ( \lambda^{t+\psi_1+c}\VEV{C}, 
                  \lambda^{t+\psi_2+c}\VEV{C}, 
                  \lambda^{t+\psi_3+c}\VEV{C}, 
                  \lambda^{2t})
      \left(
      \begin{array}{c}  {\bf \bar 5}_{\Psi_1} \\
                        {\bf \bar 5}_{\Psi_2} \\ 
                        {\bf \bar 5}_{\Psi_3} \\ 
                        {\bf \bar 5}_T
      \end{array}
      \right) \\
  &=&  {\bf 5}_T ( \lambda^{t+\psi_1+\Delta c}, 
                   \lambda^{t+\psi_2+\Delta c}, 
                   \lambda^{t+\psi_3+\Delta c}, 
                   \lambda^{2t})
       \left(
       \begin{array}{c}  {\bf \bar 5}_{\Psi_1} \\ 
                         {\bf \bar 5}_{\Psi_2} \\ 
                         {\bf \bar 5}_{\Psi_3} \\ 
                         {\bf \bar 5}_T
      \end{array}
      \right), 
\label{massT}
\end{eqnarray}
 through the interaction terms $T\Psi C$ and $T^2$. 
Thanks to the factorization property of the FN mechanism, 
 the ratio of elements of the mass matrix is determined 
 by the effective charges of ${\bf{\bar5}}$ fields.
Thus, fields possessing the smallest effective charges 
 become the main modes of the massive ${\bf{\bar5}}$. 
If $\s t>\s\psi_3$, the three light modes 
 $({\bf \bar 5}_1, {\bf \bar 5}_2, {\bf \bar 5}_3) $ 
 are written as  
 $({\bf \bar 5}_{\Psi1}, 
  {\bf \bar 5}_T+ \lambda^{\s t-\s\psi_3}{\bf \bar 5}_{\Psi3},
  {\bf \bar 5}_{\Psi_2})$. 
In this case, the origins of each generation of up-type quarks 
 and down-type quarks are different, and thus they have different 
 hierarchical structures. 
Note that in the framework of the FN mechanism, 
 the mixing angles of quarks and leptons are determined 
 by the difference in their effective charges. 
In our case, the difference in the ${\bf{\bar5}}$ sector which gives 
 lepton mixing is smaller than that in the ${\bf10}$ sector which 
 gives quark mixing. 
This means that the ${\bf{\bar5}}$ sector has milder hierarchy than 
 the ${\bf10}$ sector has. 
In addition, in that framework, this also means that the lepton mixing 
 angles are larger than the quark mixing angles. 
These properties are consistent with experiments. 

\subsubsection{Quark mass matrices}

The Dirac mass matrices for quarks and leptons are 
obtained from the interaction
\begin{equation}
\lambda^{\psi_i+\psi_j+h}\Psi_i\Psi_jH.
\label{SO(10)Yukawa}
\end{equation}
The mass matrices for the up-type quarks are independent of $T$ 
 and written as 
\begin{equation}
  M_U  =  
  \left(
  \begin{array}{ccc}
  \lambda^{2(\psi_1-\psi_3)} & \lambda^{\psi_1+\psi_2-2\psi_3} 
                             & \lambda^{\psi_1-\psi_3} \\
  \lambda^{\psi_1+\psi_2-2\psi_3} & \lambda^{2(\psi_2-\psi_3)} 
                             & \lambda^{\psi_2-\psi_3} \\
  \lambda^{\psi_1-\psi_3} & \lambda^{\psi_2-\psi_3} & 1 \\
  \end{array}
  \right) \lambda^{2\psi_3+h}\VEV{H_u}.
\label{uMass}
\end{equation} 
If we take $2\psi_3+h=0$ to reproduce the large top Yukawa coupling 
 and $\psi_1-\psi_3=3, \psi_2-\psi_3=2, 
      \lambda\sim\sin\theta_C\sim0.2$ to get correct orders of 
 the CKM matrix elements, 
\begin{equation}
  U_{\rm CKM}  =  \left(
                  \begin{array}{ccc}
                    1 & \lambda &  \lambda^3 \\
                    \lambda & 1 & \lambda^2 \\
                    \lambda^3 & \lambda^2 & 1
                  \end{array}
                  \right),  
\label{SO(10)CKM}
\end{equation}
 namely $(\psi_1,\psi_2,\psi_3,h)=(n+3,n+2,n,-2n)$, 
 the Yukawa matrix of up-type quarks is given as 
\begin{equation}
  Y_U  =  \left(
          \begin{array}{ccc}
            \lambda^6 & \lambda^5 & \lambda^3 \\
            \lambda^5 & \lambda^4 & \lambda^2 \\
            \lambda^3 & \lambda^2 & 1    
          \end{array}
          \Rs.
\label{uquark}
\end{equation}
This gives a bit too large up Yukawa coupling, and we need a fine-tuning 
 of $\order{10\%}$ to get a correct value of the coupling. 

Next, let us examine down-type quarks. 
${\bf \bar 5}_T+ \lambda^{\s t-\s\psi_3}{\bf \bar 5}_{\Psi3}$ receives 
 contributions from $\Psi\Psi H$ through ${\bf \bar 5}_{\Psi3}$ 
 and from $\Psi TC$ through ${\bf \bar 5}_T$ if there is the following 
 Higgs mixing
\bequ
  H_d=\cos\gamma L_H+\sin\gamma L_C, 
\eequ
 as in the example of Table\ \ref{SO(10)Content} (See (\ref{mixing})).
When the Higgs mixing is given as (\ref{mixing}), the magnitudes of 
 these contributions are the same, and 
 the Yukawa matrix of down-type quarks is given as 
\begin{equation}
  Y_D  =  \lambda^2\left(
          \begin{array}{ccc}
            \lambda^4 & \lambda^{\s t-\s\psi_3+1} & \lambda^3 \\
            \lambda^3 & \lambda^{\s t-\s\psi_3} & \lambda^2 \\
            \lambda   & \lambda^{\s t-\s\psi_3-2} & 1    
          \end{array}
          \right)
  \mbox{ or } 
  \lambda^{\s t-\s\psi_3}\left(
  \begin{array}{ccc}
   \lambda^{6-(\s t-\s\psi_3)} & \lambda^{5-(\s t-\s\psi_3)} & \lambda^3 \\
   \lambda^{5-(\s t-\s\psi_3)} & \lambda^{4-(\s t-\s\psi_3)} & \lambda^2 \\
   \lambda^{3-(\s t-\s\psi_3)} & \lambda^{2-(\s t-\s\psi_3)} & 1    
  \end{array}
  \right)
\label{dquark}
\end{equation}
This gives a realistic ratio $m_s/m_b$ 
 if $1\lesssim\s t-\s\psi_3\lesssim3$.

Note that if the SU(2)\sub R symmetry was exact, 
 the CKM matrix would be a unit matrix.
This is due to the cancellation 
 between the contributions from up-type quarks and down-type quarks. 
Thus, we need SU(2)\sub R breaking effects in quark mass matrices 
 in order to get a non-trivial CKM matrix. 
The additional matter field $T$ introduces such an effect, 
 but it is not sufficient. 
The simplest interaction for that purpose is $\Psi_i\Psi_jH\bar CC$. 
If such an interaction is allowed for the $(1,2)$ component, 
 which requires $c+\bar c\geq -5$, 
 the cancellation can be avoided.

Because the ratio of the top Yukawa coupling and the bottom Yukawa 
 coupling is 
 $\lambda^2$,
 $\tan \beta\equiv \VEV{H_u}/\VEV{H_d}$ is predicted to be 
 moderately large: 
 $\lambda^2m_t(\MGUT)/m_b(\MGUT)\sim5$.

\subsubsection{Lepton mass matrices}

The Yukawa matrices in the lepton sector are the transposes 
of $Y_D$, except for an overall factor $\eta$ induced by the 
renormalization group effect: 
\begin{equation}
  Y_{E,N}  =  \lambda^2
              \left(
                \begin{array}{ccc}
                  \lambda^4 & \lambda^3 & \lambda \\
                  \lambda^{\Delta+1} & \lambda^{\Delta} 
                & \lambda^{\Delta-2} \\
                  \lambda^3   & \lambda^2         & 1 
                \end{array}
              \right)
                     \eta.
\end{equation}
The right-handed neutrino masses come from the interaction
$
\lambda^{\psi_i+\psi_j+2\bar c}\Psi_i\Psi_j\bar C\bar C
$
as
\begin{equation}
M_{\rm R}=\lambda^{\psi_i+\psi_j+2\bar c}\VEV{\bar C}^2
   =\lambda^{2n-2\Delta c}\left(
\begin{array}{ccc}
\lambda^6 & \lambda^5 & \lambda^3 \\
\lambda^5 & \lambda^4 & \lambda^2 \\
\lambda^3   & \lambda^2         & 1
\end{array}
\right).
\end{equation}
Therefore the neutrino mass matrix is estimated as
\begin{equation}
M_\nu=Y_DM_{\rm R}^{-1}Y_D^T\VEV{H_u}^2
    =\lambda^{4-2n+2\Delta c}\left(
\begin{array}{ccc}
\lambda^2 & \lambda^{\s t-\s\psi_3-1} & \lambda \\
\lambda^{\s t-\s\psi_3-1} & \lambda^{2(\s t-\s\psi_3)-4} 
          & \lambda^{\s t-\s\psi_3-2} \\
\lambda   & \lambda^{\s t-\s\psi_3-2}         & 1 
\end{array}
\right)\VEV{H_u}^2\eta^2.
\end{equation}
This is the so-called Seesaw mechanism.
This neutrino mass matrix was easily calculated by using 
 effective charges without considering the right-handed neutrinos. 
For example, the (3,3) element is given by 
 ${\bf\bar5}_{\Psi_2}{\bf\bar5}_{\Psi_2}{\bf5}_H{\bf5}_H$,\footnote{
Note that this operator has negative charge and is thus forbidden 
 by the SUSY-zero mechanism. 
In consequence, we need the right-handed neutrinos in order to get this 
 effective mass term.
}whose 
 coefficient is given by 
 $\lambda^{2(\s \psi_2+\s H)}
 =\lambda^{2(n+2+\frac35\Delta c-2n+\frac25\Delta c)}$. 
From these mass matrices in the lepton sector, the MNS
matrix is obtained as
\begin{equation}
U_{\rm{MNS}}=
\left(
\begin{array}{ccc}
1 & \lambda^{1/2} &  \lambda \\
\lambda^{1/2} & 1 & \lambda^{1/2} \\
\lambda & \lambda^{1/2} & 1
\end{array}
\right) 
\end{equation}
when $\s t-\s\psi_3=5/2$, \ie
\bequ
 t=n+\half(c-\bar c+5).
\eequ 
This gives bi-large mixing angles for the neutrino sector,
because $\lambda^{1/2}\sim 0.5$. 
We then obtain the prediction 
$m_{\nu_2}/m_{\nu_3}\sim \lambda$, which is consistent with
the experimental data\cite{atmos,solar}: 
\bequ
  \begin{array}{ccccc}
    1.9\times 10^{-3} {\rm eV}^2  &\leq& \Delta m_{\rm atm}^2  
                                  &\leq& 3.6\times 10^{-3}{\rm eV}^2, \\
    7.4\times 10^{-5} {\rm eV}^2  &\leq& \Delta m_{\rm solar}^2
                                &\leq& 8.5\times 10^{-5}{\rm eV}^2. 
\label{neutrino}
\end{array}
\eequ
The relation $U_{e3}\sim \lambda$ is also an interesting 
 prediction of this matrix. 
Comparing it with the global fit to neutrino oscillations 
 which gives an upper limit $U_{e3}\leq 0.15$ 
 at $90\%$ confidence level\cite{GlobalFit}, 
 we can expect that $U_{e3}$ will be measured in near future.
Also, the normal hierarchy, $m_{\nu_1}/m_{\nu_3}\sim\lambda^2$
 is another prediction clashing any hope to observe the neutrinoless 
 double $\beta$ decay in near future. 

If we define a parameter $l$ as $4-2n+c-\bar c=-(5+l)$, 
 it is given by using the heaviest light neutrino mass $m_{\nu_3}$ 
 as 
\begin{equation}
  \lambda^l\sim\lambda^{-5}\frac{\eta^2\VEV{H_u}^2}
    {m_{\nu_3}\Lambda}.
\end{equation}
The parameter $\eta$ is roughly estimated as
\bequ
  \eta\VEV{H_u} \sim \eta\VEV{H_d}\tan\beta
              \sim m_\tau{m_t(\MGUT)\over m_b(\MGUT)}
              \sim 200\GeV.
\eequ
For the following set of parameters, 
 $l=-3,\eta\VEV{H_u}=200\GeV, \Lambda=2\times10^{16}\GeV,\lambda=0.2$, 
 we get the masses 
\beqn
  m_{\nu_3}  &\sim&  5\times10^{-2}\mbox{eV},  \\
  m_{\nu_2}  &\sim&  1\times10^{-2}\mbox{eV},  \\
  m_{\nu_1}  &\sim&  2\times10^{-3}\mbox{eV}, 
\eeqn
 which are consistent with the experimental results (\ref{neutrino}).

\section{\E6 Models} 

\subsection{\E6 Unification of the Higgs Sector}
\label{E6Higgs}

We have shown that the DTS mechanism discussed in the
 previous section can be extended to \E6 unification 
 in Refs.\cite{E6,reduced}. 
Here, we examine a simple extension of the Higgs sector of 
 the SO(10) models to \E6 models\cite{E6}. 

In order to break the \E6 gauge group into the standard gauge group,
we introduce the following Higgs content:
\begin{enumerate}
\item Higgs fields that break \E6 into SO(10):
$\Phi({\bf 27})$ and $\bar \Phi ({\bf \overline{27}})$
($\left|\VEV{\Phi({\bf 1,1})}\right|=
\left|\VEV{\bar \Phi({\bf 1,1})}\right|$).
\item An adjoint Higgs field that breaks SO(10)
into \Ga: $A({\bf 78})$
($\VEV{{\bf 45}_A}=\tau_2\times {\rm diag.}(v,v,v,0,0)$).
\item Higgs fields that break 
 SU(2)\sub R$\times$U(1)$_{B-L}$ into U(1)$_Y$:
$C({\bf 27})$ and $\bar C({\bf \overline{27}})$
($\left|\VEV{C({\bf 16,1})}\right|=
\left|\VEV{\bar C({\bf \overline{16},1})}\right|$).
\end{enumerate}
Here, $X({\bf R})$, ${\bf R_1}_X$ and $X({\bf R_1, R_2})$ denote 
 a field $X$ possessing the \E6 representation ${\bf R}$, 
 the component of $X$ possessing ${\bf R_1}$ of SO(10) and 
 the component of $X$ possessing ${\bf R_2}$ of SU(5) contained in 
     ${\bf R_1}$ of SO(10), respectively.  
Of course, 
 the anomalous U(1) charges of 
the gauge singlet operators, $\bar \Phi\Phi$, $\bar CC$ and $A^2$, 
must be negative.

Naively thinking, it seems that we would have to introduce at least 
the same number of superfields with positive charges 
  like the Higgs introduced above in order to make the superfields 
  with positive charges massive. 
We find, however, that this is not the case,
 because some of the Higgs
fields with non-vanishing VEVs are absorbed through the Higgs mechanism.
Actually, when the \E6 gauge group is broken to SO(10) 
by the non-vanishing VEV 
$\left|\VEV{\Phi}\right|=\left|\VEV{\bar \Phi}\right|$, the fields
${\bf 16}_\Phi$ and ${\bf \overline{16}}_{\bar\Phi}$ are absorbed 
 through the Higgs mechanism.%
\footnote{
Strictly speaking, a linear combination of $\Phi$, $C$ and $A$
and of $\bar\Phi$, $\bar C$ and $A$
becomes massive through the super-Higgs mechanism. The main modes are
${\bf 16}_\Phi$ and ${\bf \overline{16}}_{\bar \Phi}$, respectively.
}
Therefore, if two additional {\bf 10}'s of SO(10) in the Higgs 
sector with non-vanishing VEVs can be massive, 
we can then save one pair of ${\bf27}$ and ${\bf\cc{27}}$ 
 with positive charges.
At first glance, such a mass term may seem to be forbidden by the 
SUSY-zero mechanism.
Actually, if all fields with non-vanishing VEVs had negative 
anomalous U(1) charges, their mass term would be forbidden. 
As discussed in \S\ref{Ansatz}, however, some of the fields 
 with positive charges can have non-vanishing VEVs 
 if the total charges of $G$-singlets with non-vanishing VEVs 
 are negative. 
For example, we can set $\phi=-3$ and $\bar \phi=2$. 
Since $\bar \Phi$
has positive charge, the term $\bar \Phi^3$ is allowed, and it 
induces a mass of ${\bf 10}_{\bar \Phi}$ through the non-vanishing 
VEV $\VEV{\bar \Phi}$. If the term $\bar \Phi^2\bar C$ is allowed,
masses for the two ${\bf 10}$'s,  ${\bf 10}_{\bar \Phi}$ and 
${\bf 10}_{\bar C}$, are induced, so that we can save one pair 
 of ${\bf27}$ and ${\bf\cc{27}}$ Higgs. 

An example for the field content in the Higgs sector is given 
 in Table\ \ref{E6Content}.
The symbols $\pm$ denote the quantum numbers for a $Z_2$ parity symmetry 
 which is introduced for the same reason as in the SO(10) models. 
\begin{table}
\begin{center}
\begin{tabular}{|c|c|c|} 
\hline
                  &   non-vanishing VEV  & vanishing VEV \\
\hline 
{\bf 78}          &   $A(a=-1,-)$        & $A'(a'=4,-)$      \\
{\bf 27}          &   $\Phi(\phi=-3,+)$\  $C(c=-6,+)$ &  $C'(c'=7,-)$  \\
${\bf \overline{27}}$ & $\bar \Phi(\bar \phi=2,+)$ \  $\bar C(\bar c=-2,+)$ &
                  $\bar C'(\bar c'=8,-)$ \\
{\bf 1}           &   $Z_2(z_2=-2,-)$,$Z_5(z_5=-5,-)$,
                       $\bar Z_5(\bar z_5=-5,-)$ \   &  \\
\hline
\end{tabular}
\caption{
The typical values of anomalous U(1) charges.
}
\label{E6Content}
\end{center}
\end{table}
Here, the Higgs field $H$ of the SO(10) model is contained in $\Phi$, 
 and the $G$-singlet $\bar\Phi\Phi$ can play the same role as 
 the FN filed $\Theta$. 
This \E6
Higgs sector has the same number of superfields with non-trivial 
representations as in the SO(10) Higgs sector, in spite of the fact
that the larger group \E6 requires additional Higgs fields
to break \E6 to SO(10).

\subsubsection{DTS and alignment}

Generally, in \E6 GUT, the interactions in the superpotential
that are made of only ${\bf 27}$ and ${\bf \overline{27}}$ are written 
in terms of the units ${\bf 27}^3$, ${\bf \overline{27}27}$ and 
${\bf\overline{27}}^3$.
Note that terms like ${\bf 27}^3$ or ${\bf\overline{27}}^3$  
do not contain the product of singlet components of $G_{\rm{SM}}$. 
Therefore, we can ignore these terms when considering 
 SM-like vacua, while these terms can constrain the existence of 
 vacua other than 
 SM-like vacua. 
This point is discussed below.

The important terms in the superpotential
to determine the VEVs are 
\begin{equation}
W=W_{A^\prime} + W_{C^\prime}+W_{\bar C^\prime}+W({\bar \Phi}).
\end{equation}
Since we have a positively charged field $\bar \Phi$
 that has a non-vanishing VEV, we have to take into account only such 
$W({\bar \Phi})$'s that include only fields possessing 
 non-vanishing VEVs. 
Since 
$\bar \Phi\Phi$ and $\bar \Phi C$ have negative total charges,
the superpotential has essentially terms like $\overline{\bf{27}}^3$.
Therefore, the superpotential $W(\bar \Phi)$ constrains vacua 
other than SM-like vacua. 

Let us discuss first the VEVs of $\Phi$ and $\bar\Phi$. 
When $\phi+\bar\phi\leq0$, they have non-vanishing VEVs, 
 and the $D$-flatness condition of \E6 requires
$\VEV{\Phi}=\VEV{\bar \Phi}$, up to phases. 
The VEV of $\bar \Phi$ 
 can be rotated by the \E6 gauge transformation into the following 
form:
\begin{equation}
\VEV{\bar \Phi}=
\begin{pmatrix}
 \bar u \cr
          {\bf 0} \cr
          \bar u_1 \cr
          \bar u_2 \cr
          {\bf 0} 
\end{pmatrix}        
\begin{array}{l}
  \hbox{$\}SO(10)$ singlet (real)} \\
  \hbox{$\}SO(10)$ ${\bf \overline{16}}$}       \\
  \hbox{$\}$the first component of SO(10) {\bf 10}\ (complex)} \\
  \hbox{$\}$the second component of SO(10) {\bf 10}\ (real)} \\
  \hbox{$\}$the third to tenth components of SO(10) {\bf 10}}.
\end{array}
\end{equation}
For simplicity, we adopt a superpotential of the form 
\begin{equation}
W({\bar \Phi})=\bar \Phi^3+\bar \Phi^2\bar C.
\end{equation}
Then, the $F$-flatness conditions of ${\bf 10}_{\bar C}$ and
${\bf 1}_{\bar C}$ lead to
${\bf 1}_{\bar \Phi}{\bf 10}_{\bar \Phi}=0$ and 
${\bf 10}_{\bar \Phi}^2=0$, respectively.
Thus, two type of vacua are allowed:
$\bar u\neq 0,\bar u_1=\bar u_2=0$ and  
$\bar u=0,\bar u_1=i\bar u_2\neq 0$.
This implies that a non-vanishing VEV of
${\bf 1}_{\bar \Phi}$ requires
the vanishing of the VEV $\VEV{{\bf 10}_{\bar \Phi}}$. 
In this vacuum, \E6 is broken down to SO(10). 
Moreover, 
in this vacuum, ${\bf 10}_{\bar C}$ has vanishing VEV, because of
the $F$-flatness conditions for ${\bf 10}_{\bar \Phi}$.
Interestingly enough, a vacuum alignment occurs naturally.
In the following, for simplicity, we often write $\lambda^n$ in place 
of the operators $(\bar\Phi\Phi)^n$, though these operators are not 
always singlets.

The superpotential $W_{A^\prime}$ is in general
written as
\begin{eqnarray}
W_{A^\prime}&=&\lambda^{a^\prime+a}A^\prime A
+\lambda^{a^\prime+3a}A^\prime A^3
+\lambda^{a'+a+\bar \phi+\phi}\bar \Phi A'A\Phi \nonumber \\ 
&&+\lambda^{a'+3a+\bar \phi+\phi}\bar \Phi A'A^3\Phi, 
\end{eqnarray}
under the condition, $-3a+\bar \phi+\phi\leq a^\prime < -5a$.
Here we assume 
$c+\bar c, c+\bar \phi, \bar c+\phi<-(a'+a)$
to forbid the terms $\bar C A^\prime A C$ (which destabilizes the 
DW form of the VEV of $A$), $\bar CA'A\Phi$ and $\bar \Phi A'AC$
(which may lead to undesirable vacua in which $\VEV{\bar C}=\VEV{C}=0$
 by the $F$-flatness conditions of ${\bf16}_{A'}$ and 
 ${\bf\cc{16}}_{A'}$). 
If $A$ and $(\Phi,\bar \Phi)$ were separated in the superpotential, 
PNG fields would appear. 
Because the terms $\bar \Phi A'A\Phi$ and $\bar \Phi A'A^3\Phi$ connect 
$A'$ and $A$ with $\Phi$ and $\bar \Phi$, the PNG fields acquire non-zero 
masses.
Moreover, these terms realize the alignment between the VEVs
$\left|\VEV{\Phi}\right|=\left|\VEV{\bar \Phi}\right|$ and $\VEV{A}$. 
Note that these terms
are also important to induce the term 
$({\bf 45}_{A'}{\bf 45}_{A})_{\bf 54}({\bf 45}_{A}^2)_{\bf 54}$,
which is not included in the term $A'A^3$,
 because of a cancellation (see Appendix\ \ref{factorization}).
In terms of 
SO(10), which is not broken by the VEV 
$\left|\VEV{\Phi}\right|=\left|\VEV{\bar \Phi}\right|$,
the effective superpotential is given as 
\begin{eqnarray}
W_{A'}^{\eff}&=&
{\bf 45}_{A'}(1+{\bf 1}_{A}^2+{\bf 45}_{A}^2
+{\bf \overline{16}}_{A}{\bf 16}_{A}){\bf 45}_{A}  \nn \\
&&+{\bf \overline{16}}_{A'}(1+{\bf 1}_{A}^2+{\bf 45}_{A}^2
+{\bf \overline{16}}_{A}{\bf 16}_{A}){\bf 16}_{A} \nn \\
&&+{\bf 16}_{A'}(1+{\bf 1}_{A}^2+{\bf 45}_{A}^2
+{\bf \overline{16}}_{A}{\bf 16}_{A}){\bf \overline{16}}_{A} \nn \\
&&+{\bf 1}_{A'}{\bf 1}_{A}(1+{\bf 1}_{A}^2+{\bf 45}_{A}^2
+{\bf \overline{16}}_{A}{\bf 16}_{A}), 
\end{eqnarray}
 and the $F$-flatness conditions are written
\begin{eqnarray}
0=\frac{\partial W}{\partial {\bf 45}_{A'}}&=& (1+{\bf 1}_{A}^2
+{\bf 45}_{A}^2+{\bf \overline{16}}_{A}{\bf 16}_{A}){\bf 45}_{A}, 
\label{F45}\\
0=\frac{\partial W}{\partial {\bf \overline{16}}_{A'}}&=&(1+{\bf 1}_{A}^2
+{\bf 45}_{A}^2+{\bf \overline{16}}_{A}{\bf 16}_{A}){\bf 16}_{A}, 
\label{Fbar16} \\
0=\frac{\partial W}{\partial {\bf 16}_{A'}}&=& {\bf \overline{16}}_{A}
(1+{\bf 1}_{A}^2+{\bf 45}_{A}^2+{\bf \overline{16}}_{A}{\bf 16}_{A}), 
\label{F16}\\
0=\frac{\partial W}{\partial {\bf 1}_{A'}}&=& {\bf 1}_{A}
(1+{\bf 1}_{A}^2+{\bf 45}_{A}^2+{\bf \overline{16}}_{A}{\bf 16}_{A}).
\label{F1}
\end{eqnarray}
The terms in each parenthesis of Eqs.(\ref{F45})-(\ref{F1}) 
 look common because we omit the coefficients. 
Indeed they are common due to an \E6 relation that holds
 when $A$ and $(\Phi,\bar \Phi)$ are not coupled with each others. 
On the other hand, when they are coupled, such an \E6 relation 
 is absent, and generally, the values in the parentheses of 
 Eqs.(\ref{Fbar16}), (\ref{F16}) and (\ref{F1}) are not zero, 
 leading to $\VEV{{\bf 16}_A}=\VEV{{\bf \overline{16}}_A}=0$. 
We have two 
possibilities for the VEV of ${\bf 1}_A$: one vacuum with 
$\VEV{{\bf 1}_A}=0$ and the other vacuum with $\VEV{{\bf 1}_A}\neq 0$. 
In the latter vacuum, the DW mechanism in \E6 GUT does not work, 
because the non-vanishing VEV $\VEV{{\bf 1}_A}$
directly gives its bare mass to the doublet Higgs. 
Therefore, the former vacuum in which $\VEV{{\bf 1}_A}=0$ is 
favorable to realize DTS.
Note that if the term $\bar \Phi A'\Phi$ is allowed, the vacuum
$\VEV{{\bf 1}_A}=0$ disappears. This destroys the realization of
DTS. Here, this term is forbidden by $Z_2$ parity.
As in the SO(10) case, we have several possibilities for the VEV of 
${\bf 45}_A$, one of which is the DW VEV 
$\VEV{{\bf 45}_A}_{B-L}=\tau_2\times {\rm diag.}(v,v,v,0,0)$,
where $v\sim\lambda^{-a}$.
These VEVs break the SO(10) into \Ga. 

Next, we discuss the $F$-flatness conditions of $C^\prime$ and 
$\bar C^\prime$,
which not only determine the scale of the VEV 
$\VEV{\bar CC}\sim \lambda^{-(c+\bar c)}$ but also realize 
the alignment of the VEVs $\VEV{C}$ and $\VEV{\bar C}$.
For simplicity, we assume that 
$\VEV{{\bf 1}_C}=\VEV{{\bf 1}_{\bar C}}=0$, 
though there may be vacua in which these components have 
non-vanishing VEVs.
Then, since $\VEV{{\bf 10}_C}=\VEV{{\bf 10}_{\bar C}}=0$ 
by the above argument,
only the components ${\bf 16}_C$ and ${\bf \overline{16}}_{\bar C}$
can have non-vanishing VEVs.
The superpotential to determine these VEVs 
can be written as 
\begin{eqnarray}
W_{C'}&=&
       \lambda^{\bar\phi+c'}\bar \Phi \Ll
         \lambda^{c+\bar c+a}\bar CAC  
       + \lambda^{2c+2\bar\phi+a}\bar\Phi\bar\Phi ACC  \right.  \nn\\ 
    &&\qquad \LL
       + \lambda^{c+\bar\phi+a}\bar\Phi 
          f_1\Ls \Phi\bar\Phi, A, Z_i
         \Rs C  
      + f_2\Ls \Phi\bar\Phi, A, Z_i
       \Rs \Rl C'  \nn\\
    &&+\lambda^{\bar c+c'}\bar C 
         f_3\Ls \Phi\bar\Phi, A, Z_i
       \Rs  C',  \\
W_{\bar C'}&=&
       \lambda^{\bar c'+\phi}\bar C' 
      f_4\Ls \Phi\bar\Phi, A, Z_i
       \Rs \Phi  +\lambda^{\bar c'+c}\bar C' 
         f_5\Ls \Phi\bar\Phi, A, Z_2
       \Rs  C.
\end{eqnarray}
Here, $f_i$'s are certain functions whose forms are easily found 
 and $Z_i$ represents $Z_2$, $Z_5$ and $\bar Z_5$. 
Note that these give common values for each multiplet of \Ga\  
 and generally give different values for different multiplet, 
 because these are functions of $A$, $\Phi$, $\bar\Phi$ and 
 singlet fields. 
The vacua are $\VEV{\bar CC}=0$ and $\VEV{\bar CC}\neq 0$.
In the desired vacuum $\VEV{\bar CC}\neq0$, the $F$-flatness conditions 
 of ${\bf 16}_{C'}$ and ${\bf \cc{16}}_{\bar C'}$ give 
 non-trivial conditions, which cause
the alignment of the VEVs $\VEV{A}$ and $\VEV{C}(\VEV{\bar C})$, as in 
the SO(10) case. Then, the above four $F$-flatness conditions with
respect to  ${\bf 1}_{C'}$, ${\bf 1}_{\bar C'}$, ${\bf 16}_{C'}$ and 
${\bf \cc{16}}_{\bar C'}$
determine the scale of
the four VEVs
$\VEV{\bar CC}\sim \lambda^{-(c+\bar c)}$,
$\VEV{Z_i}\sim \lambda^{-z_i}(i=2,5)$ and 
$\VEV{\bar Z_5}\sim \lambda^{-\bar z_5}$.
The VEVs $\left|\VEV{C}\right|=\left|\VEV{\bar C}\right|
\sim \lambda^{-(\bar c+c)}$ 
break SU(2)\sub R$\times$U(1)$_{B-L}$ into U(1)$_Y$. 

Now, all the VEVs are determined as 
\beqn
  \VEV \Phi  &=&  \VEV{\Phi({\bf1},{\bf1})}\ 
                 =\ \lambda^{-{\phi+\bar \phi\over2}},  \\
  \VEV{\bar\Phi}  &=&  \VEV{\bar\Phi({\bf1},{\bf1})}\ =
                    \ \lambda^{-{c+\bar c\over2}},  \\
  \VEV{\bf45}_A  &=&  \tau_2\times{\rm diag.}(v,v,v,0,0)\ 
               ,\ v=\lambda^{-a},  \\
  \VEV C  &=&  \VEV{C({\bf16},{\bf1})}\ =\ \lambda^{-{c+\bar c\over2}},  \\
  \VEV{\bar C}  &=&  \VEV{\bar C(\bf\cc{16},{\bf1})}\ =
                    \ \lambda^{-{c+\bar c\over2}}  ,
\eeqn
 and all the other VEVs are zero. 
The symmetry breaking pattern is given by 
\bctr
\begin{tabular}{cclcl}
 \E6    & $\longrightarrow$ & SO(10) &
          & at $\lambda^{-{\phi+\bar \phi\over2}}\Lambda$ \\ 
        & $\longrightarrow$ & \Ga & \quad & at $\lambda^{-a}\Lambda$ \\
        & $\longrightarrow$ & \GSM & \quad 
          & at $\lambda^{-\half(c+\bar c)}\Lambda$. 
\end{tabular}
\ectr
The parameters that parametrize the effect of the additional 
 FN mechanism are written 
 as $\Delta\phi=\half(\phi-\bar\phi)$ for U(1)$_{V'}$ 
 and $\Delta c=\half(c-\bar c)$ for U(1)$_V$.

\subsubsection{Mass Spectrum of the Higgs Sector}

Since all the VEVs are fixed, we can derive the mass spectrum
of the Higgs sector.

\E6 representations are decomposed in terms of 
 SO(10)$\times$U(1)$_{V'}$ as
\begin{eqnarray}
{\bf 27}&=& {\bf 16}_1+{\bf 10}_{-2}+{\bf 1}_4, \\
{\bf 78}&=& {\bf 45}_0+{\bf 16}_{-3}+{\bf \cc{16}}_3+{\bf 1}_0,
\end{eqnarray}
which are further decomposed into SU(5) representations 
as Eqs.(\ref{class}).

In the following, we study how the mass matrices of the above fields
are determined by anomalous U(1) charges.
Note that for the mass terms, we must
take into account not only the terms given in the previous argument 
but also the terms that contain two fields with vanishing
VEVs (see Appendix \ref{E6HiggsSpectrum}).

Before going into details, it is worthwhile examining the NG modes
that are eaten through the Higgs mechanism, because in some cases,  
it is not obvious that there are vanishing eigenvalues 
 in the mass matrices. 
There appear the following NG modes:
\begin{enumerate}
\item ${\bf 16}+{\bf \cc{16}}+1$ of SO(10) (namely,
$Q+U^c+E^c+h.c.+N^c$) 
in the breaking \E6$\rightarrow$ SO(10).
\item $Q+U^c+X+h.c.$ in the breaking 
 SO(10)$\rightarrow$\Ga.
\item $E^c+h.c.+N^c$ in the breaking
SU(2)\sub R$\times$U(1)$_{B-L}\rightarrow$U(1)$_Y$.
\end{enumerate}
Namely, there are NG modes possessing 
 $2\times({\bf10},{\bf\cc{10}})$, 
 $({\bf5},{\bf\bar5})$ and $4\times{\bf1}$ of SU(5) 
 and $(X,\bar X)$.

First, we examine the mass matrices of ${\bf 24}$ in SU(5).
Considering the additional term $A'^2$,
we get the following mass matrices $M_I$, $I=G,W,X$:
\begin{equation}
M_I=\bordermatrix{
 I\backslash \bar I  &   {\bf{45}}_A &  {\bf{45}}_{A'}            \cr
  {\bf{45}}_A         &     0      & \alpha_I\lambda^{a'+a} \cr
  {\bf{45}}_{A'}    & \alpha_I\lambda^{a'+a}& \lambda^{2a'}    \cr
},
\end{equation}
where $\alpha_X=0$ and $\alpha_I\neq 0$ for $I=G,W$. 
One pair of $X$ is massless and is eaten through the Higgs mechanism.
The mass spectra are $(0, \lambda^{2a'})$ for $I=X$ and 
$(\lambda^{a'+a},\lambda^{a'+a})$ for $I=G,W$.

Next, we examine the mass matrices for the representations $I=Q, U^c$ 
and $E^c$, which are contained in ${\bf 10}$ of SU(5). The mass 
matrices $M_I$ are written as
\begin{equation}
\bordermatrix{
I\backslash \bar I  &{\bf{\cc{16}}}_{\bar \Phi} & {\bf{\cc{16}}}_{\bar C}&
                     {\bf{\cc{16}}}_{A} &{\bf{45}}_{A} 
                     &{\bf{\cc{16}}}_{\bar C'}   & {\bf{\cc {16}}}_{ A'} 
                     & {\bf{45}}_{A'}  \cr
{\bf{16}}_\Phi & 0 & 0 & 0& 0& \lambda^{\bar c'+\phi}  
             & \lambda^{\phi+a'-\Delta\phi}
             & 0 \cr
{\bf{16}}_C & 0 & 0 &0 & 0& \beta_I\lambda^{\bar c'+c} &  0  & 0 \cr
{\bf{16}}_A & 0 & 0 & 0& 0& \lambda^{\bar c'+a+\Delta\phi}   
          & \lambda^{a'+a} & 0 \cr
{\bf{45}}_{A} & 0 & 0 & 0 & 0& \lambda^{\bar c'+a+\Delta c}  & 0 & 
       \alpha_I\lambda^{a+a'}  \cr
{\bf{16}}_{C'} & \lambda^{c'+\bar\phi} & \beta_I\lambda^{c'+\bar c} &
        \lambda^{a+c'-\Delta\phi} & \lambda^{a+c'-\Delta c} & 
        \lambda^{c'+\bar c'} & \lambda^{a'+c'-\Delta\phi} & 
        \lambda^{a'+c'-\Delta c}  \cr
{\bf{16}}_{A'} & \lambda^{\bar \phi+a'+\Delta\phi} & 0 & 
        \lambda^{a+a'} & 0&\lambda^{\bar c'+a'+\Delta\phi} &  
        \lambda^{2a'}& \lambda^{2a'+\Delta\phi-\Delta c}  \cr
{\bf{45}}_{A'} & 0 & 0 &0 & \alpha_I\lambda^{a'+a}
        & \lambda^{\bar c'+a'+\Delta c} &  
        \lambda^{2a'-\Delta\phi+\Delta c} 
        & \lambda^{2a'} \cr
},
\end{equation}
where we have used the relations
$\lambda^\phi\VEV{\Phi}\sim (\lambda^{\bar \phi}\VEV{\bar \Phi})^{-1}
\sim\lambda^{\Delta\phi}$ and
$\lambda^c\VEV{C}\sim(\lambda^{\bar c}\VEV{\bar C})^{-1}
\sim\lambda^{\Delta c}$ ($\Delta\phi=\half(\phi-\bar\phi)$, 
$\Delta c=\half(c-\bar c)$).
Because one pair of ${\bf{10}}$ and ${\bf\cc{10}}$ (whose main 
modes are ${\bf{16}}_\Phi$ and ${\bf\cc{16}}_{\bar\Phi}$) 
is eaten through the Higgs mechanism in the process of breaking \E6 
to SO(10), we can simply omit ${\bf 16}_\Phi$ and 
${\bf \cc{16}}_{\bar \Phi}$
 during the derivation of the mass spectrum. 
Then, the mass matrices can be written in the form of four 
$3\times 3$ matrices as
\begin{equation}
  M_I  =  \left(
    \begin{array}{cc}
      0 & A_I \\ B_I & C_I
    \end{array}
    \right)
\end{equation}
as in the SO(10) case.
We can find that the ranks of $A_I$ and $B_I$ reduce to two 
 because
$(\alpha_I=0, \beta_I\neq 0)$ for $I=Q,U^c$ and
$(\alpha_I\neq 0, \beta_I=0)$ for $I=E^c$, where the vanishing values 
 are due to the NG theorem. 
The mass spectra become 
$(0,0,\lambda^{a'+a}, \lambda^{a'+a},\lambda^{c'+\bar c}, 
\lambda^{\bar c'+c}, \lambda^{2a'})$ for $I=Q,U^c$ and 
$(0,0,\lambda^{a'+a},\lambda^{a'+a},\lambda^{a'+a},
\lambda^{a'+a},\lambda^{\bar c'+c'})$ for $I=E^c$.

Finally, we examine the mass matrices of ${\bf 5}$ and ${\bf \bar 5}$
in SU(5) and show how DTS is realized.
Considering the additional terms,
we write the mass matrices $M_I$ for the representations 
$I=D^c,L$ and their conjugates as
\begin{equation}
M_I=\left(\begin{matrix} 0 & 0 & A_I \cr  B_I & C_I & D_I \cr 
                         E_I & F_I & G_I \cr\end{matrix}\right),
\end{equation}
\begin{equation}
A_I=\bordermatrix{
 I\backslash \bar I   &  {\bf10}_{C'} 
                    & {\bf10}_{\bar C'} &
                    {\bf\cc{16}}_{\bar C'} &
                    {\bf\cc{16}}_{A'}  \cr
{\bf10}_\Phi & S_I\lambda^{c'+\phi+\Delta\phi} & \lambda^{\bar c'+\phi} & 
0  & 0 \cr
{\bf10}_C  & 0 & \lambda^{\bar c'+c} & 0 & 0 \cr
{\bf16}_C & 0 & 0 & \lambda^{\bar c'+c}
   & 0 \cr
{\bf16}_A  & 0 & \lambda^{\bar c'+a+\Delta c}& \lambda^{\bar c'+a+\Delta\phi}
   & \lambda^{a'+a} \cr
}, 
\end{equation}
\begin{equation}
B_I=
\bordermatrix{
 I\backslash \bar I  &{\bf10}_\Phi & {\bf10}_C& 
                    {\bf\cc{16}}_{\bar C} &{\bf\cc{16}}_{A}   \cr
{\bf10}_{\bar C} &
0 & 0& 0 & 0
 \cr
{\bf16}_\Phi & 0 & 0 &  0 & 0  \cr
{\bf10}_{\bar \Phi} & 
0 & 0& 0 & 
\lambda^{\bar\phi+a-\Delta\phi-\Delta c}   \cr
},
\end{equation}
\begin{equation}
C_I=\bordermatrix{
 I\backslash \bar I   &  {\bf10}_{\bar \Phi} 
                    & {\bf10}_{\bar C} &
                    {\bf\cc{16}}_{\bar \Phi} \cr
{\bf10}_{\bar C}  & \lambda^{\bar \phi+\bar c-\Delta\phi} & 0 & 0 \cr
{\bf16}_\Phi  & 0 & 0 & 0 \cr
{\bf10}_{\bar \Phi}  &
\lambda^{2\bar\phi-\Delta\phi} & \lambda^{\bar \phi+\bar c-\Delta\phi}&
\lambda^{2\bar\phi-\Delta c} \cr
}, 
\end{equation}
\begin{equation}
D_I=\bordermatrix{
 I\backslash \bar I   &  {\bf10}_{C'} 
                    & {\bf10}_{\bar C'} &
                    {\bf\cc{16}}_{\bar C'} &
                    {\bf\cc{16}}_{A'}  \cr
{\bf10}_{\bar C}  & \lambda^{\bar c+c'} &
\lambda^{\bar c'+\bar c-\Delta\phi}& \lambda^{\bar c'+\bar c-\Delta c} & 
\lambda^{\bar c+a'-\Delta\phi-\Delta c} \cr
{\bf16}_\Phi  & 0 & \lambda^{\bar c'+\phi-\Delta\phi+\Delta c}& 
   \lambda^{\bar c'+\phi}
   & \lambda^{a'+\phi-\Delta\phi} \cr
{\bf10}_{\bar \Phi}  &
\lambda^{c'+\bar\phi} & \lambda^{\bar c'+\bar\phi-\Delta\phi}&
\lambda^{\bar c'+\bar\phi-\Delta c} & 
 \lambda^{\bar\phi+a'-\Delta\phi-\Delta c} \cr
}, 
\end{equation}
\begin{equation}
E_I=\bordermatrix{
 I\backslash \bar I  & {\bf10}_\Phi&{\bf10}_C &{\bf\cc{16}}_{\bar C}&{\bf\cc{16}}_{A}\cr
{\bf10}_{C'} &S_I\lambda^{c'+\phi+\Delta\phi} & 0 & 0 &\lambda^{c'+a-\Delta c} \cr
{\bf10}_{\bar C'} &\lambda^{\bar c'+\phi} &\lambda^{\bar c'+c} &
\lambda^{\bar c'+\bar c-\Delta c} & \lambda^{\bar c'+a-\Delta\phi-\Delta c} \cr
{\bf16}_{C'} & 0 & 0 &\lambda^{c'+\bar c} &\lambda^{a+c'-\Delta\phi} \cr
{\bf16}_{A'}& 0 & 0 & 0 &\lambda^{a'+a} \cr
}, 
\end{equation}
\begin{equation}
F_I=\bordermatrix{
 I\backslash \bar I  & {\bf10}_{\bar \Phi} &{\bf10}_{\bar C} &{\bf\cc{16}}_{\bar \Phi} \cr
{\bf10}_{C'} &\lambda^{c'+\bar\phi} &\lambda^{c'+\bar c} & 
   \lambda^{c'+\bar\phi+\Delta\phi-\Delta c} \cr
{\bf10}_{\bar C'} &\lambda^{\bar c'+\bar\phi-\Delta\phi} & 
       \lambda^{\bar c'+\bar c-\Delta\phi} & 
       \lambda^{\bar c'+\bar\phi-\Delta c} \cr
{\bf16}_{C'} & \lambda^{c'+\bar\phi-\Delta\phi+\Delta c} &
      \lambda^{c'+\bar c-\Delta\phi+\Delta c} & \lambda^{c'+\bar\phi} \cr
{\bf16}_{A'} & 0 & 0 & \lambda^{\bar\phi+a'+\Delta\phi} \cr
}, 
\end{equation}
\begin{equation}
G_I=\bordermatrix{
 I\backslash \bar I  & {\bf10}_{C'} 
                    & {\bf10}_{\bar C'}  &
                    {\bf\cc{16}}_{\bar C'} &
                    {\bf\cc{16}}_{A'}  \cr
{\bf10}_{C'}  & \lambda^{2c'+\Delta\phi} & 
  \lambda^{\bar c'+c'} &
  \lambda^{\bar c'+c'+\Delta\phi-\Delta c}  & \lambda^{c'+a'-\Delta c} \cr
{\bf10}_{\bar C'}  &\lambda^{\bar c'+c'} & 
\lambda^{2\bar c'-\Delta\phi} & \lambda^{2\bar c'-\Delta c} & 
\lambda^{\bar c'+a'-\Delta\phi-\Delta c} \cr
{\bf16}_{C'} & \lambda^{2c'+\Delta c} & 
  \lambda^{\bar c'+c'-\Delta\phi+\Delta c} &\lambda^{\bar c'+c'} &  
\lambda^{a'+c'-\Delta\phi} \cr
{\bf16}_{A'}& \lambda^{c'+a'+\Delta\phi+\Delta c} & 
\lambda^{\bar c'+a'+\Delta c} & \lambda^{\bar c'+a'+\Delta\phi} &  
\lambda^{2a'} \cr
},
\end{equation}
where $S_{D^c}\neq 0$ and $S_L=0$.
We can see that the rank of $A_L$ is three, which is smaller than
the rank of $A_{D^c}$. This implies that the rank of $M_L$ is smaller
than the rank of $M_{D^c}$, and 
the rank of the matrix $M_I$ is actually 10 for $I=D^c$
and 9 for $I=L$. One pair of massless fields, 
${\bf\bar5}$ and ${\bf5}$ (whose main modes are 
${\bf 16}_{\Phi}$ and ${\bf \cc{16}}_{\bar \Phi}$), 
gives the NG mode, which is eaten through the Higgs mechanism during the 
breaking from \E6 to SO(10). 
The other massless mode for $I=L$ is identified as the so-called 
 MSSM doublet Higgs. 
The massless mode is given as 
\begin{eqnarray}
H_u&\sim&\bar L({\bf 10}_\Phi)+\lambda^{\phi-c}\bar L({\bf 10}_C), 
\label{HuMixing}\\
H_d&\sim&L({\bf 10}_\Phi)+\lambda^{\phi-c}L({\bf 10}_C).
\label{HdMixing}
\end{eqnarray}
As noted above, ${\bf 16}_{\Phi}$ and ${\bf \cc{16}}_{\bar \Phi}$
are eaten through the Higgs mechanism, and ${\bf 10}_{\bar \Phi}$ and
${\bf 10}_{\bar C}$ can become massive through the matrix $C_I$, 
whose elements are generally larger than the elements of $B_I$, 
$D_I$ and $F_I$.
Thus their masses can be estimated as 
$(\lambda^{\bar \phi+\bar c-\Delta\phi},
\lambda^{\bar \phi+\bar c-\Delta\phi})$.
With this observation, we ignore the matrices 
 $B_I$, $C_I$, $D_I$ and $F_I$ and consider only 
 $A_I$, $E_I$ and $G_I$ in the following argument.
Because the elements of $A_I$ and $E_I$ are generally larger than those
of $G_I$, we can estimate the mass spectrum of the other modes of
$D^c$ from $A_{D^c}$ and $B_{D^c}$
as $(\lambda^{c'+\phi+\Delta\phi},\lambda^{c'+\phi+\Delta\phi},
\lambda^{\bar c'+c},\lambda^{\bar c'+c},\lambda^{ \bar c'+ c},
\lambda^{ c'+\bar c},\lambda^{a'+a},\lambda^{a'+a})$, and
that of $L$ as
$(0,\lambda^{\bar c'+c},\lambda^{\bar c'+c},\lambda^{ \bar c'+ c},
\lambda^{ c'+\bar c},\lambda^{a'+a},\lambda^{a'+a},
\lambda^{2c'+\Delta\Phi})$.
As in the SO(10) models, in order to realize proton decay, 
 we have to pick up at least one element of $C_I$, 
 which is generally smaller than the mass scales of $D^c$'s, 
 leading to suppressed proton decay via dimension 5 operators. 
The effective mass of the colored Higgs is estimated as
$(\lambda^{c'+\phi+\Delta\phi})^2/\lambda^{2c'+\Delta\phi}=
\lambda^{2\phi+\Delta\phi}$, 
which is usually larger than the cutoff scale.
For example, for the typical charge assignment in Table\ \ref{E6Content}, 
$2\phi+\Delta\phi=-17/2$.

It is worthwhile to summarize the required terms and the undesirable terms. 
There are several terms which must be forbidden in order to realize 
DTS:
\begin{enumerate}
\item $\Phi^3$, $\Phi^2C$, $\Phi^2C'$, $\Phi^2C'Z$ induce a large mass 
of the doublet Higgs.
\item $\bar CA'C$,$\bar CA'AC$,$\bar \Phi A'\Phi$ would destabilize the
DW form of $\VEV{A}$. 
\item $\bar \Phi A'C$, $\bar CA'\Phi$, $\bar \Phi A'AC$, $\bar CA'A\Phi$, 
$\bar \Phi A'ZC$, $\bar C A'Z\Phi$ lead to the undesirable VEV 
$\VEV{{\bf 16}_C}=0$, 
unless  another singlet field is introduced. 
\item $A'A^n(n\geq 4)$ make it less natural to obtain a DW VEV.
\end{enumerate}
In contrast, the following terms are necessary:
\begin{enumerate}
\item $A'A$, $\bar \Phi A'A^3\Phi$ to obtain a DW VEV $\VEV{A}$.
\item $\Phi^2AC'$ for DTS.
\item $\bar C'(A+Z)C$, $\bar C(A+Z)C'$ to achieve alignment between 
the VEVs $\VEV{A}$ and $\VEV{C}$ and to give superheavy masses 
to the PNGs.
\item $\bar\Phi A'A\Phi$ to realize alignment between 
the VEVs $\VEV{A}$ and $\VEV{\Phi}$ and to give superheavy masses 
to the PNGs.
\item $\bar \Phi^3$,$\bar \Phi^2\bar C$ to give superheavy masses to 
two ${\bf 10}$ of SO(10). 
\end{enumerate}

\begin{figure}[htb]
\begin{center}
\leavevmode
\put(300,50){{\large $\bf{\log \mu (GeV)}$}}
\put(0,260){{\Large $\bf{\alpha^{-1}}$}}
\put(29,240){$\alpha_1^{-1}$}
\put(31,150){$\alpha_2^{-1}$}
\put(31,90){$\alpha_3^{-1}$}
\includegraphics[width=11cm]{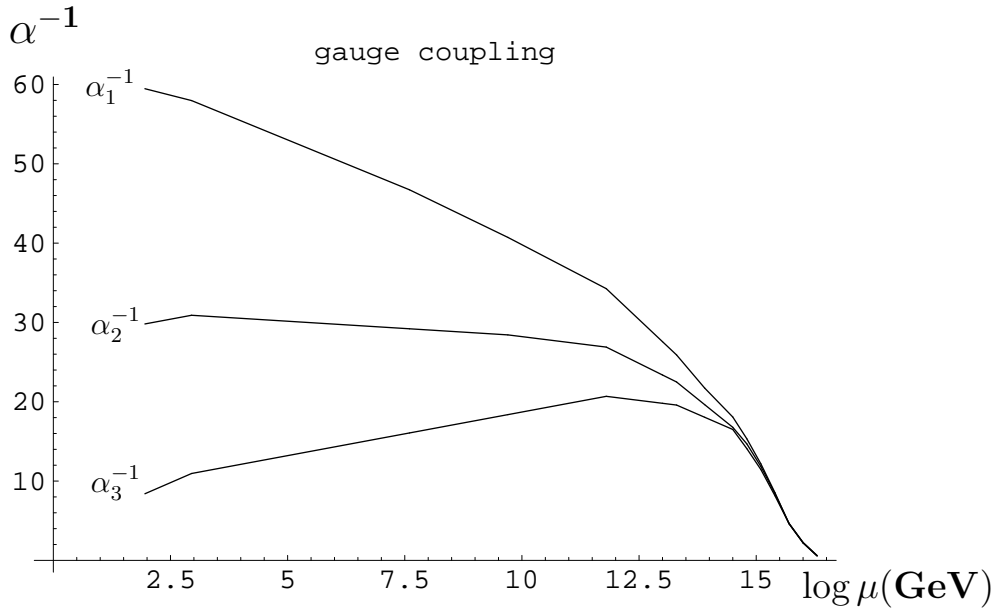}
\vspace{-2cm}
\caption{
An example of the gauge coupling flows:
Here we adopt $\lambda=0.25$, 
$\alpha_1^{-1}(M_Z)=59.47$, $\alpha_2^{-1}(M_Z)=29.81$,
$\alpha_3^{-1}(M_Z)=8.40$, and the SUSY breaking scale 
$\MSB\sim 1\TeV$.
We also use the anomalous U(1) charges shown in
 Table\ \ref{E6Content} and $(\psi_1,\psi_2,\psi_3)=(9/2,7/2,3/2)$. 
We use the ambiguities in the $\order1$ coefficients.}
\label{GCUaGUT}
\end{center}
\end{figure}

In this way, all the extra fields other than one pair of doublet Higgs 
 indeed acquire superheavy masses. 
GCU is realized, as mentioned in \S\ref{GCU}. 
For this Higgs sector, however, 
 the unified gauge coupling at the cutoff scale tends to become large, 
 even when the matter sector is the minimal one, 
 \ie\ $\Psi_i({\bf 27})\ (i=1,2,3)$. 
An example of the gauge coupling flows is shown in Fig.\ref{GCUaGUT}. 
The value of the unified gauge coupling strongly depends 
 on the actual charge assignment, and if all anomalous U(1) charges 
 become smaller, the unified gauge coupling 
at the unified scale $\lambda^{-a}$ becomes smaller.\footnote{
This means a large $a'+a$ is disfavored by this fact. 
Note that 
 $\bar\Phi A'A\Phi$ is required to give large masses 
 to would-be PNG modes, leading to $a+a'\geq-(\phi+\bar\phi)$.
Thus, we cannot take $\phi+\bar\phi$ so small, and therefore 
 $\VEV\Phi$ and $\VEV{\bar\Phi}$ cannot be so small.
} 
For example, we can adopt half-integer charges as
$a=-1/2,a'=5/2,\phi=-3,\bar\phi=2,c=-5,\bar c=-1,c'=13/2,\bar c'=13/2,
z_i=-i/2(i=3,7,11)$, where the half integer charges
play the same role as the $Z_2$ parity, 
and $\psi_1=9/2,\psi_2=7/2,\psi_3=3/2$ with odd $R$-parity 
 in the matter sector. 
For this charge assignment, the unified gauge coupling 
 at the cutoff scale is smaller than in the previous model. 
However, because the unification scale $\lambda^{-a}$ is larger than 
 that of the previous model, the model predicts 
 a longer lifetime of the nucleon, 
 which is roughly estimated as
\begin{equation}
\tau_p(p\rightarrow e^+\pi^0)\sim 1\times 10^{35}\left(\frac{\Lambda_U}
{10^{16}\ {\rm GeV}}\right)^4
\left(\frac{0.01\ {\rm GeV}^3}{\alpha}\right)^2  {\rm yrs}.
\end{equation}
This predicted value is significantly longer than the present 
experimental lower bound.

Another way to maintain the unified gauge coupling in the 
 perturbative region is to reduce the number of Higgs. 
This is the topic of the next subsection. 

Before examining this possibility, let us mention how theses models 
 can be compatible with the matter sector, which is discussed 
 in \S\ref{E6Matter}. 
The relevant parameters are essentially in a number of two. 
The first one is $l$, which parametrizes the neutrino mass scale as 
\bequ
  m_{\nu_3} \sim \lambda^{-(l+5)} \frac{\VEV{H_u}^2\eta^2}
                                   {\Lambda}, 
\label{l}
\eequ
 and is also introduced in SO(10) models in \S\ref{SO(10)Matter}. 
The other is the one that parametrizes the lepton mixing, $r$, 
 which is defined as 
\begin{equation}
\lambda^r\equiv 
 \frac{\lambda^{c}\VEV{C({\bf16})}}{\lambda^\phi\VEV{\Phi({\bf1})}}. 
\label{r}
\end{equation}
This $r$ corresponds to $3-(\s t-\s\psi_3)$ of SO(10) models 
 in \S\ref{SO(10)Matter}. 
For $\lambda\sim0.2$, allowed values of these parameters are 
\beqn
 -1<&l&<-4, \\
 0<&r&<3/2.
\eeqn
In this case, they are given as
\beqn
 l&=&-2\Ls\s\psi_2+\Delta c+\phi\Rs-5, 
\label{E6l}\\
 r&=&\Delta{c}-\Delta{\phi{}}, 
\label{E6r}
\eeqn
where $\Delta\phi=\half\Ls\phi-\bar\phi\Rs$ and
 $\Delta c=\half\Ls c-\bar c\Rs$.
The charge assignment shown in Table\ \ref{E6Content} and 
 $(\psi_1,\psi_2,\psi_3)=(9/2,7/2,3/2)$ yield 
 $r=\half$ and $l=-2$. 
Thus, it can be consistent with the matter sector.

\subsection{Simpler \E6 Higgs Sector}
\label{SimpleE6}

In the previous \E6 model, $C({\bf 16})$ and $\bar C({\bf\cc{16}})$ of 
the SO(10) model are embedded into the ${\bf 27}$ field and 
the ${\bf \cc{27}}$ field, respectively. 
However, they may also be embedded into the ${\bf 78}$ field, 
resulting in simpler \E6 models\cite{reduced}. 
Here, we examine this alternative embedding.

Since we introduce two adjoint Higgs $A'$ and $A$, we have two kinds of 
possibilities for reducing the Higgs sector.
\begin{enumerate}
\item
  The VEV $\VEV{{\bf 16}_{A'}}$ or $\VEV{{\bf \cc{16}}_{A'}}$ 
  is non-vanishing.
\item
  The VEV $\VEV{{\bf 16}_{A}}$ or $\VEV{{\bf \cc{16}}_{A}}$ 
  is non-vanishing.
\end{enumerate}
Note that it must be forbidden that ${\bf 16}$ and 
${\bf \cc{16}}$ have non-vanishing VEVs simultaneously, 
which destabilizes the DW form of VEVs. 
For example, if the VEVs $\VEV{\bf 16_{A'}}$ and 
$\VEV{\bf \cc{16}_{A'}}$ are non-zero, 
the interactions $A'^n$ destabilize the DW form of VEVs 
because $F_{{\bf 45}_{A'}}$ includes the VEVs $\VEV{\bf 16_{A'}}$ and 
$\VEV{\bf \cc{16}_{A'}}$. 
At first glance, such an asymmetric VEV structure would be forbidden 
by the $D$-flatness conditions.
But it is shown below that such an interesting VEV can satisfy
the $D$-flatness conditions.

\subsubsection{Possibility 1: $\VEV{{\bf 16}_{A'}}\neq 0$}

The typical Higgs content is represented 
 in Table\ \ref{SimpleE6ContentA'}.

\begin{table}
\begin{center}
\begin{tabular}{|c|c|} 
\hline 
{\bf 78}          &   $A(a=-1)$\ $A'(a'=5)$       \\
{\bf 27}          &   $\Phi(\phi=-5)$\  $C'(c'=7)$  \\
${\bf \cc{27}}$ & $\bar C(\bar c=-6)$ \ $\bar \Phi'(\bar \phi'=6)$   \\
{\bf 1}           &   $\Theta(\theta=-1)$\,\, $Z_i(z_i=-1)$ \ ($i=$1-5)\,\, 
                      $Z'(z'=6)$   \\
\hline
\end{tabular}
\caption{
Typical values of anomalous U(1) charges.
}
\label{SimpleE6ContentA'} 
\end{center}
\end{table}
Suppose that among the above Higgs fields, only ${\bf 45}_A$, 
${\bf 1}_\Phi$, ${\bf \cc{16}}_{\bar C}$ and
${\bf 16}_{A'}$ have non-vanishing VEVs such as
\begin{eqnarray}
  \VEV{{\bf 45}_A}&=&\tau_2\times{\rm diag.} (v,v,v,0,0)
    \quad (v\sim \lambda^{-a}),\\
  \abs{\VEV{{\bf 1}_\Phi}}&=&\abs{\VEV{\bf \cc{16}}_{\bar C}}
                     \ =\ \abs{\VEV{\bf 16}_{A'}}
                       \sim \lambda^{-\frac{1}{3}(\bar c+a'+\phi)}.
\label{VEVA'}
\end{eqnarray}
As mentioned above, if $\phi+a'+\bar c<0$, the $G$-singlet 
 ${\bar CA'\Phi}$ can have a non-vanishing VEV, 
 which means that $A'$ has a non-vanishing VEV. 
Actually, this vacuum satisfies the relations
$\VEV{\tr A'^n}=0$ and 
$\VEV{\bar CA'\Phi}\sim\lambda^{-(\bar c+a'+\phi)}$, which are 
consistent with the VEV relation (\ref{VEV}).
And this vacuum satisfies not only the $D$-flatness conditions
for SO(10) but also that of U(1)$_{V'}$ 
\begin{equation} 
D_{V'}\ :\  4\abs{{\bf 1}_{\Phi}}^2-3\abs{{\bf 16}_{A'}}^2
               -\abs{{\bf \cc{16}}_{\bar C}}^2=0.
\end{equation}
Therefore, this vacuum satisfies all the \E6 $D$-flatness conditions.

Next we discuss the $F$-flatness conditions to know how such 
a vacuum can be obtained. 
For simplicity, we assume that any component fields other than 
${\bf 45}_A$, ${\bf 1}_\Phi$, ${\bf \cc{16}}_{\bar C}$ and
${\bf 16}_{A'}$ have vanishing VEVs. To determine the VEV of ${\bf 45}_A$,
it is sufficient to consider the superpotential
\begin{equation}
W_{A'}=A'A+A'A^3+A'A^4+A'A^5.
\end{equation}
Here, for simplicity, singlet fields $Z_i$'s and coefficients are not 
 written explicitly.
The $F$-flatness condition of 
${\bf 45}_{A'}$ leads to the DW VEV,
$\VEV{{\bf 45}_A}\sim \tau_2\times{\rm diag.} (v,v,v,0,0)$. 
(Here $A'A^5$ is needed to avoid the ``factorization problem'', 
 as shown in Appendix\ \ref{factorization}.)
Because the positively charged field $A'$ has a non-vanishing VEV 
$\VEV{{\bf 16}_{A'}}\neq 0$, the $F$-flatness conditions of the
negatively charged fields may become non-trivial conditions. 
Fortunately, in this model, there is no such non-trivial condition.
For example, $F_{{\bf \cc{16}}_A}=0$ is trivial because 
${\bf \cc{16}}_A$ is a NG mode in the superpotential $W_{A'}$.

The $F$-flatness condition of $Z'$, which is obtained from the 
superpotential
\begin{equation}
W_{Z'}=Z'(1+\bar CA'\Phi+f_Z(A,Z_i)),
\end{equation}
 where $f_Z$ is a certain function of $A$ and $Z_i$'s, 
leads to
\begin{equation}
\VEV{\bar CA'\Phi}\sim \lambda^{-(\bar c+a'+\phi)}.
\end{equation}
The $D$-flatness conditions of SO(10) and U(1)$_{V'}$ lead to
\begin{equation}
  |\VEV{{\bf 1}_\Phi}|=|\VEV{{\bf \cc{16}}_{\bar C}}|
                      =|\VEV{{\bf 16}_{A'}}|
                    \sim \lambda^{-\frac{1}{3}(\bar c+a'+\phi)},
\end{equation}
which correspond to the desired vacuum shown in Eq. (\ref{VEVA'}).

The $F$-flatness conditions of $C'$, which are obtained from the
superpotential
\begin{equation}
W_{C'}=\bar C(1+Z_i+A+A'(f_C(A, Z_i)+\bar CA'\Phi)C',
\end{equation}
 where $f_C$ is another function of $A$ and $Z_i$'s, 
are written as
\begin{eqnarray}
F_{{\bf 16}_{C'}}&=&(1+Z_i+A){\bf\cc{16}}_{\bar C}=0, \\
F_{{\bf 1}_{C'}}&=&
(f_C(A, Z_i)+\bar CA'\Phi){\bf\cc{16}}_{\bar C}{\bf16}_{A'}=0.
\label{alignment}
\end{eqnarray}
These conditions realize an alignment between the VEVs 
$\VEV{{\bf 45}_A}$, $\VEV{{\bf \cc{16}}_{\bar C}}$ and 
$\VEV{{\bf 16}_{A'}}$ by shifting the VEVs of the singlet fields $Z_i$, 
and as a result, the PNG fields become massive.
The $F$-flatness condition of ${\bf \cc{16}}_{A'}$, which is obtained 
from the superpotential
\begin{equation}
W_{A'A'}=A'(f_A(A,Z_i)+\bar CA'\Phi)A',
\end{equation}
 where $f_A$ is another function of $A$ and $Z_i$'s, 
also realizes an alignment between $\VEV{{\bf 45}_A}$
 and $\VEV{{\bf 16}_{A'}}$.

It is interesting to note that in this model, the 
 ``generalized sliding singlet mechanism'' in \S\ref{Sliding} 
 is naturally realized.
The $F$-flatness conditions of $\bar \Phi'$, which are obtained from the
superpotential
\begin{equation}
W_{\bar\Phi'}=\bar \Phi'(1+Z_i+A+A'(f_\Phi(A, Z_i)+\bar CA'\Phi))\Phi,
\end{equation}
 where $f_\Phi$ is another function of $A$ and $Z_i$'s, 
 are written as
\begin{eqnarray}
F_{{\bf 1}_{\bar \Phi'}}&=&(1+Z_i){\bf1}_\Phi=0, 
\label{sliding} \\
F_{{\bf \cc{16}}_{\bar \Phi'}}&=&
(f_\Phi(A, Z_i)+\bar CA'\Phi){\bf1}_\Phi {\bf16}_{A'}=0.
\end{eqnarray}
At first glance, the component field ${\bf 10}_{\Phi}$, 
which includes a pair of doublet Higgs, 
seems to have a mass term from the superpotential 
$\bar \Phi'(1+Z_i+A)\Phi$. 
However, this mass term is a function of $A$ and singlets, and 
 the doublet Higgs has the same quantum number under 
 the generator $\VEV{A}$ as the component field ${\bf 1}_\Phi$ 
 which has non-vanishing VEV. 
This is the condition that the generalized sliding singlet mechanism 
 can take place, and the $F$-flatness condition Eq.(\ref{sliding}) 
 ensures that the masses of the doublets also vanish. 
Note that the other components of $\Phi$ have different charges, 
 and have superheavy masses. 
As a result, DTS is realized. 
In this mechanism, we need not introduce the $Z_2$ symmetry 
 that is required in the DW mechanism. 

In the above model, for intelligibility, we introduced a positively 
charged singlet $Z'$ in order to fix the VEV 
$\VEV{\bar CA'\Phi}\sim \lambda^{-(\bar c+a'+\phi)}$. However, 
one of the non-trivial $F$-flatness conditions of ${\bf 1}_{C'}$,
${\bf \cc{16}}_{A'}$ and ${\bf \cc{16}}_{\bar \Phi'}$
can play the same role as $Z'$. If we do not introduce the field $Z'$,
the number of the negatively charged singlet fields $Z_i$'s becomes four.

It is worthwhile to note how to determine the anomalous U(1) charges. 
In order to realize DTS, the terms
\begin{equation}
A'A^5,\bar \Phi'A\Phi, \bar C(A+Z)C'
\end{equation}
must be allowed, and the term
\begin{equation}
\bar CA'^2\Phi
\end{equation}
must be forbidden. 
These requirements can be rewritten as inequalities.
We determined the charges in order to satisfy those inequalities.

Unfortunately, we have not found any realistic matter sector with such a 
Higgs sector. Actually, the mixing parameter $r$ of (\ref{r}), 
 which is obtained as
\begin{equation}
\lambda^r\equiv\frac{\lambda^{a'+\phi}\VEV{{\bf16}_{A'}{\bf1}_\Phi}}
                    {\lambda^{\phi}\VEV{{\bf1}_\Phi}}
               =\lambda^{a'}\VEV{{\bf16}_{A'}}
               =\lambda^{\frac{1}{3}(2a'-\bar c-\phi)}
\end{equation}
 in this case, 
 must be around 1/2 in order to obtain bi-large neutrino mixings, but 
 it looks difficult to realize, because $2a'-\bar c-\phi\gg 1$.

\subsubsection{Possibility 2: $\VEV{{\bf16}_A}\neq0$}

Here, we consider another possibility in which 
the $C({\bf16})$ of the SO(10) model is embedded into the negatively 
charged adjoint Higgs $A({\bf78})$. This possibility is more promising,
because the condition for a realistic matter sector, 
$2a-\bar c-\phi\sim 1$, can be realized.
The content of the Higgs sector is the same as 
in the previous possibility, except for the charges and 
the number of singlets.

To begin with, we examine the $D$-flatness conditions.
Because $\VEV{{\bf45}_A}\neq0$ and 
$\VEV{{\bf16}_A}\neq0$, 
the $D$-flatness condition in the ${\bf\cc{16}}$ 
direction gives a non-trivial condition.
In order to compensate the contribution from $A$ in the
condition, $\Phi$ and/or 
$\bar C$ must have non-zero VEV in both {\bf1} and {\bf16}
(${\bf\cc{16}}$) components.
Therefore, non-trivial $D$-flatness conditions are 
\beqn
  D_{V+V'} &\quad:\quad& \abs{{{\bf1}_\Phi}}^2  =  
      \abs{{{\bf16}_A}}^2 + \abs{{{\bf1}_{\bar C}}}^2, 
\label{DVVp} \\
  D_V &\quad:\quad& \abs{{{\bf\cc{16}}_{\bar C}}}^2 = 
      \abs{{{\bf16}_A}}^2 + \abs{{{\bf16}_\Phi}}^2, 
\label{DV} \\
  D_{\cc{\bf16}} &\quad:\quad& {{\bf45}_A}^*{{\bf16}_A}
        = {{\bf1}_\Phi}^*{{\bf16}_\Phi}
         -{{\bf\cc{16}}_{\bar C}}^*{{\bf1}_{\bar C}}. 
\label{D16} 
\eeqn
In addition, we suppose that the VEVs
\beqn
  \VEV{{\bf\cc{16}}_{\bar C}}\VEV{{\bf16}_A}\VEV{{\bf1}_\Phi} \nn
  &\sim& 
  \VEV{{\bf\cc{16}}_{\bar C}}\VEV{{\bf45}_A}\VEV{{\bf16}_\Phi} \nn\\
  &\sim& 
  \VEV{{\bf{1}}_{\bar C}}\VEV{{\bf45}_A}\VEV{{\bf1}_\Phi} \nn\\
  &\sim& 
  \lambda^{-(\bar c+a+\phi)}
  \,\,\equiv\,\, \lambda^{-3k}, 
\label{3Fcondi}\\
  \VEV{{\bf45}_A} &\sim& \lambda^{-a} 
\label{AFcondi}
\eeqn
are obtained from $F$-flatness conditions as is generally expected.%
\footnote{
Strictly speaking, if three conditions in Eq.(\ref{3Fcondi}) were 
determined by $F$-flatness conditions, the $F$-flatness and $D$-flatness 
conditions would become over-determined. 
Therefore, only two of the three conditions are
determined by $F$-flatness conditions. 
Then, another solution,
\begin{equation}
\VEV{{\bf 1}_A}\sim \VEV{{\bf 16}_A}\sim \lambda^{-a}\ll
\VEV{{\bf 1}_\Phi}\sim \VEV{{\bf 16}_\Phi}\sim \VEV{{\bf 1}_{\bar C}}
\sim \VEV{{\bf 16}_{\bar C}},
\nn
\end{equation}
may be allowed, through which the natural gauge coupling unification 
will not be realized.
Though the $\order1$ coefficients determine which vacuum is realized,
the desired vacuum is obtained in some (finite) region of the parameter space
for the $\order1$ coefficients.
}
From these conditions except for one $D$-flatness condition 
 Eq.(\ref{D16}), 
the orders of VEVs are determined 
as follows: 
\beqn
  &\VEV{{\bf\cc{16}}_{\bar C}} 
    \sim \VEV{{\bf16}_A}
    \sim \VEV{{\bf1}_\Phi}
    \sim \lambda^{-k} 
    \equiv \lambda^{-a}\lambda^r, & \label{VEV1}\\
  &\VEV{{\bf{1}}_{\bar C}}
    \sim \VEV{{\bf16}_\Phi}
    \sim \lambda^{a-2k} 
    \sim \lambda^{-a}\lambda^{2r}, & \label{VEV2}\\
  &\VEV{{\bf45}_A} \sim \lambda^{-a},& \label{VEV3}
\eeqn
for $\lambda^{-a}\gg\lambda^{-k}$.
Here, $r=a-k$ is the mixing parameter, 
 introduced in \S\ref{E6Matter}. 
For these VEVs, the effective charges can be defined and therefore 
the natural gauge coupling unification is realized. 
Taking into account Eq.(\ref{D16}), it may appear that this condition 
requires $r=0$.
However, since $r$ should be small ($\sim1/2$) to explain 
 the bi-large neutrino mixings 
and there is an ambiguity due to order one coefficients,
Eq.(\ref{D16}) can be satisfied even if $r>0$.
To be more precise, Eq.(\ref{D16}) has the form 
$\lambda^{-2a+r} = \lambda^{-2a+3r}+\lambda^{-2a+3r}$, 
and the r.h.s can become 
$2\lambda^{-2a+3r}\sim\lambda^{-2a+r}\lambda^{2r-1/2}$, 
allowing $r=1/4$.
And the ambiguities in $\order1$ coefficients leaves room for 
a larger $r$. 

Next, we examine $F$-flatness conditions.
The typical charge assignment of the Higgs sector is represented 
in Table\ \ref{SimpleE6ContentA}.
\begin{table}
\begin{center}
\begin{tabular}{|c|c|} 
\hline 
{\bf 78}          &   $A(a=-1,+)$\ $A'(a'=5,+)$       \\
{\bf 27}          &   $\Phi(\phi=-3,+)$\  $C'(c'=6,-)$  \\
${\bf \cc{27}}$ &$\bar \Phi'(\bar \phi'=5,+)$ \ $\bar C(\bar c=0,-)$  \\
{\bf 1}           &   $\Theta(\theta=-1,+)$\ $Z_i(z_i=-1,+)$\ $(i=1,2)$   \\
\hline
\end{tabular}
\caption{
Typical values of anomalous U(1) charges.
}
\label{SimpleE6ContentA}
\end{center}
\end{table}
Here the VEVs are again determined by 
\begin{equation}
  W=W_{A^\prime} + W_{\bar\Phi^\prime}+W_{C^\prime},
\end{equation}
where 
\beqn
  W_{A'} &=& A'(A+A^3+A^4+A^5), \\
  W_{\bar\Phi'} &=& \bar\Phi'(1+A+Z_i+A^2+AZ_i+Z_i^2)\Phi, 
\label{Wphi}\\
  W_{C'} &=& \bar C(1+A+Z_i+\cdots+(\bar C\Phi)^2)C'. 
\eeqn
As in the previous model, the $F$-flatness condition of 
${\bf 45}_{A'}$ leads to the DW VEV,
$\VEV{{\bf 45}_A}\sim \tau_2\times{\rm diag.} (v,v,v,0,0)$. 
The $F$-flatness condition of ${\bf1}_{\bar\Phi'}$ makes the \E6 singlet 
part in the parenthesis of Eq.(\ref{Wphi}) vanish, leading to vanishing 
doublet mass terms by the generalized sliding singlet mechanism.
The $F$-flatness condition of ${\bf\cc{16}}_{\bar\Phi'}$ gives a factored 
equation
\bequ
  (1+A+Z_i)\Ll {\bf45}_A{\bf16}_\Phi + {\bf16}_A{\bf1}_\Phi\Rl
    = 0, 
\eequ
which can be checked by an explicit calculation based on 
\E6 group theory. 
The above two $F$-flatness conditions are satisfied by shifting
the VEVs of two singlets $Z_i$. 
The two $F$-flatness conditions of ${\bf 1}_{C'}$ and ${\bf 16}_{C'}$
and the three $D$-flatness conditions in Eqs. (\ref{DVVp})-(\ref{D16}) 
determine the five VEVs ${\bf 16}_A$, ${\bf 1}_{\Phi}$,
${\bf 16}_\Phi$, ${\bf 1}_{\bar C}$ and ${\bf \cc{16}}_{\bar C}$.
It is straightforward to analyse the mass matrices of Higgs 
to check that all modes are superheavy except for one pair of 
 doublet Higgs contained in ${\bf10}_\Phi$.\footnote{
We would emphasize in this model that all the singlet fields also 
become superheavy, while
in the models treated in \S\ref{E6Higgs}, 
one massless singlet field appears.}

Now, we examine if the conditions are compatible with the matter 
sector, in which we introduced the same three superfields 
 as in \S\ref{E6Matter}. 
Applying the same discussion to this case, the 
parameters $r$ and $l$ will be given as 
\beqn
  &\lambda^r \sim \frac{\lambda^c \VEV{{\bf{16}}_C}}
                       {\lambda^\phi \VEV{{\bf1}_\Phi}}
             \sim \frac{\lambda^{a+\phi} \VEV{{\bf{16}}_A}
                                         \VEV{{\bf1}_\Phi}}
                       {\lambda^\phi\VEV{{\bf1}_\Phi}}
             = \lambda^{a-k}, & \\
  &\lambda^{-(5+l)} \sim \lambda^{4+\phi-2\bar c}
                         \VEV{{\bf\cc{16}}_{\bar C}}^{-2}
                    \sim \lambda^{4+\phi-2\bar c +2k}.&
\eeqn
For example, a set of charges $(a,\phi, \bar c)=(-1, -3, 0)$ 
 in Table\ \ref{SimpleE6ContentA} gives 
$(r, l)=(1/3, -10/3)$, which is allowed, as shown in \S\ref{E6Matter}. 

In this model, we need
\bequ
 A'A^5,\,\bar\Phi'AZ\Phi,\,\bar C\bar C\Phi C'
\eequ 
and have to forbid 
\bequ
 \bar CA'\Phi,\,\bar\Phi'\bar C\Phi\Phi.
\eequ
However, it is difficult to forbid $\bar CA'\Phi$ while allowing 
$A'A^5$ by the SUSY-zero mechanism for small $r (=\frac{1}{3}
(\bar c+\phi-2a))$.
Therefore we need another mechanism to forbid $\bar CA'\Phi$, 
{\it e.g.} we have to introduce an additional $Z_N$ symmetry. 
In $W_{\bar{\Phi}'}$, the simplest superpotential that one could imagine, 
$W_{\bar{\Phi}'}=\bar{\Phi}'A\Phi$, is not consistent with the 
$D$-flatness conditions for these VEVs (\ref{VEV1})-(\ref{VEV3}).
This is again a characteristic of the \E6 group, and therefore, 
\E6 breaking effects such as $\bar{\Phi}'(AZ+A^2)\Phi$ are needed.

As in the models discussed in \S\ref{E6Higgs},  
the half-integer charges of matter supermultiplets play
the same role as R-parity in this model.
Another charge assignment
$(a,\phi, \bar c, z_i, a', \bar \phi', c')=$ 
$(-1, -3, 1/3, -1, 5, 5, 23/3)$  gives another example of a 
consistent model, which requires no additional $Z_N$ symmetries.

\subsection{Matter Sector}
\label{E6Matter}

In this subsection, we review the matter sector of \E6 models 
 discussed in Ref.\cite{BM}.

\subsubsection{E-Twisting Mechanism}
Let us first  recall the so-called E-twisting mechanism,
which is naturally realized in \E6 unification models\cite{ETwist}. 
In the case of \E6, the ${\bf 16}$ and ${\bf 10}$ representations 
 of SO(10) are 
 automatically included in a fundamental multiplet {\bf 27} of \E6, 
 which is decomposed under \E6$\supset$SO(10)$\supset$SU(5) as 
\begin{equation}
{\bf27} \rightarrow \underbrace{[( {\bf 16,10}) +({\bf 16,\bar 5})
+({\bf 16,1})]}_{\bf 16}
+\underbrace{[({\bf 10,\bar 5})+({\bf 10,5})]}_{\bf 10}
+ \underbrace{[({\bf 1,1})]}_{{\bf 1}},
\label{27}
\end{equation}
where the representations of SO(10) and SU(5) are explicited above. 
Thus, even with the minimal matter content, 
 \ie\ $\Psi_i({\bf 27})\ (i=1,2,3)$, there appear extra fields: 
 three ${\bf10}$'s and three ${\bf1}$'s 
 in addition to the three ${\bf16}$'s. 
Because these extra fields are vector-like, 
 they do not change the number of light fields, 
 but they can change the family structure, 
 as in the SO(10) models in \S\ref{SO(10)Matter}.
Note that in SO(10) models, we have to introduce an additional 
 ${\bf10}$ field to realize a twisting family structure. 

In terms of SU(5), there are three ${\bf5}$ and six ${\bf\bar5}$. 
Among them, three pairs of $({\bf 5,\bar 5})$ become heavy.\footnote{
The possible right-handed neutrino modes
 $\Psi_i({\bf 16,1})$ and 
$\Psi_i({\bf 1,1})$ also acquire large masses, 
but here we concentrate on the family structure of ${\bf
 \bar 5}$.
}
Indeed, the Higgs fields $\Phi$ and $C$ can yield such masses. 
The superpotential that gives large masses for
 $({\bf 5},{\bf \bar 5})$ pairs are written as 
\begin{equation}
 W = \lambda^{\psi_i+\psi_j+c}\Psi_i\Psi_jC
   + \lambda^{\psi_i+\psi_j+\phi}\Psi_i\Psi_j\Phi.
\end{equation}
The VEV $\VEV{\Phi({\bf 1,1})}$ gives the $3\times 3$ mass matrix  of
$\Psi_i({\bf 10,5})\Psi_j({\bf 10,\bar 5})$ pairs as 
\begin{equation}
M_\Phi=\bordermatrix{
&\Psi_1({\bf 10, \bar 5})&\Psi_2({\bf 10, \bar5})
&\Psi_3({\bf 10,\bar 5})\cr
\Psi_1({\bf 10,5})
&\lambda^{2\psi_1}&\lambda^{\psi_1+\psi_2}
&\lambda^{\psi_1+\psi_3}  \cr
\Psi_2({\bf 10,5})& \lambda^{\psi_1+\psi_2}   &  \lambda^{2\psi_2}
&\lambda^{\psi_2+\psi_3} \cr
\Psi_3({\bf 10,5})
 &\lambda^{\psi_1+\psi_3} & \lambda^{\psi_2+\psi_3}
 & \lambda^{2\psi_3}  \cr}
 \lambda^{\phi}\VEV{\Phi({\bf 1,1})},
\end{equation}
while the VEV $\VEV{C({\bf 16,1})}$ gives the mass terms
of $\Psi_i({\bf 16,\bar 5})$ and $\Psi_j({\bf 10,5})$ as
\begin{equation}
M_C=
\bordermatrix{
&\Psi_1({\bf 16,\bar 5})&\Psi_2({\bf 16,\bar 5})
&\Psi_3({\bf 16,\bar5})\cr
\Psi_1({\bf 10,5})&\lambda^{2\psi_1}&\lambda^{\psi_1+\psi_2}
&\lambda^{\psi_1+\psi_3} \cr
\Psi_2({\bf 10,5})&\lambda^{\psi_1+\psi_2}&\lambda^{2\psi_2}
&\lambda^{\psi_2+\psi_3} \cr
\Psi_3({\bf 10,5})
&\lambda^{\psi_1+\psi_3}&\lambda^{\psi_2+\psi_3}
&\lambda^{2\psi_3} \cr}
\lambda^{c} \VEV{C({\bf 16,1})}.
\end{equation}
Then, the full mass matrix is proportional to
\begin{equation}
\bordermatrix{
&\Psi_1({\bf 16,\bar 5})&\Psi_2({\bf 16,\bar 5})
&\Psi_3({\bf 16,\bar5})
&\Psi_1({\bf 10, \bar 5})&\Psi_2({\bf 10, \bar5})
&\Psi_3({\bf 10,\bar 5})\cr
\Psi_1({\bf 10,5})&\lambda^{2\psi_1+r}&\lambda^{\psi_1+\psi_2+r}
&\lambda^{\psi_1+\psi_3+r}
&\lambda^{2\psi_1}&\lambda^{\psi_1+\psi_2}
&\lambda^{\psi_1+\psi_3}  \cr
\Psi_2({\bf 10,5})&\lambda^{\psi_1+\psi_2+r}&\lambda^{2\psi_2+r}
&\lambda^{\psi_2+\psi_3+r}
& \lambda^{\psi_1+\psi_2}   &  \lambda^{2\psi_2}
&\lambda^{\psi_2+\psi_3} \cr
\Psi_3({\bf 10,5}) &\lambda^{\psi_1+\psi_3+r}&\lambda^{\psi_2+\psi_3+r}
&\lambda^{2\psi_3+r}
 &\lambda^{\psi_1+\psi_3} & \lambda^{\psi_2+\psi_3}
 & \lambda^{2\psi_3}  \cr}, 
 \label{full}
\end{equation}
where we have defined the parameter $r$ as 
\begin{equation}
\lambda^r\equiv \frac{\lambda^c\VEV{C({\bf 16,1})}}
                     {\lambda^\phi \VEV{\Phi({\bf 1,1})}},
\end{equation}
which we use frequently in \E6 models. 
Note that some of the matrix elements may vanish 
 by the SUSY-zero mechanism,
 but for the moment, we assume that no such zeros appear 
 in the superpotential.
In general, there are three  massless modes among the 
six ${\bf \bar 5}$ fields
by solving the above  $3\times 6$ matrix. 
However, since the matrix has hierarchical structure, 
 we can easily find which ${\bf\bar5}$'s remain massless. 
It is determined by their effective charges, therefore by 
 the parameter $r$, so that fields possessing smaller 
 effective charges become massive. 
We can classify all the cases as follows: 
\begin{enumerate}
\item $\psi_1-\psi_3<r$ : 
  $({\bf16}_{\Psi_1},{\bf16}_{\Psi_2},{\bf16}_{\Psi_3})$ type.
\item  $0<r<\psi_1-\psi_3$ : 
  $({\bf16}_{\Psi_1},{\bf16}_{\Psi_2},{\bf10}_{\Psi_1})$ type.
\item  $\psi_3-\psi_1<r<0$ : 
  $({\bf16}_{\Psi_1},{\bf10}_{\Psi_1},{\bf10}_{\Psi_2})$ type.
\item  $r<\psi_3-\psi_1$ : 
  $({\bf10}_{\Psi_1},{\bf10}_{\Psi_2},{\bf10}_{\Psi_3})$ type.
\end{enumerate}
When $0<r<\psi_1-\psi_3$, we can realize different family 
 structures for up-type quarks and down-type quarks 
 so that we can reproduce realistic Yukawa matrices 
 as in SO(10) models where $\s t>\s\psi_3$. 
Now, we consider this case, 
where we can write the three massless modes 
 $({\bf \bar 5}_1,{\bf \bar 5}_2,{\bf \bar 5}_3)$ as 
\begin{eqnarray}
{\bf \bar 5}_1 &=& {\bf 16}_{\Psi_1}
+\lambda^{\psi_1-\psi_3}{\bf 16}_{\Psi_3}
+\lambda^{\psi_1-\psi_2+r}{\bf 10}_{\Psi_2}
+\lambda^{\psi_1-\psi_3+r}{\bf 10}_{\Psi_3}, 
\label{51} \\
{\bf \bar 5}_2 &=& {\bf 10}_{\Psi_1}
+\lambda^{\psi_1-\psi_3-r}{\bf 16}_{\Psi_3}
+\lambda^{\psi_1-\psi_2}{\bf 10}_{\Psi_2}
+\lambda^{\psi_1-\psi_3}{\bf 10}_{\Psi_3}, 
\label{52} \\
{\bf \bar 5}_3 &=& {\bf 16}_{\Psi_2}
+\lambda^{\psi_2-\psi_3}{\bf 16}_{\Psi_3}
+\lambda^{r}{\bf 10}_{\Psi_2}
+\lambda^{\psi_2-\psi_3+r}{\bf 10}_{\Psi_3},
\label{53}
\end{eqnarray}
where the first terms on the right-hand sides are the main components of
these massless modes, and the other terms are mixing terms with
heavy states, ${\bf 16}_{\Psi_3}$,
 ${\bf 10}_{\Psi_2}$ and ${\bf 10}_{\Psi_3}$.

If we use SUSY-zero coefficients, various types of massless modes can
be realized.
For example,
if $\psi_1+\psi_3+\phi<0$, SUSY zeros appear, and
the Yukawa terms $\Psi_3\Psi_i\Phi$ $(i=1,2,3)$
are forbidden. 
Hence, when $2\psi_2+\phi>0$ the mass matrix $M_\Phi$ becomes
\begin{equation}
M_\Phi= \bordermatrix{
&\Psi_1({\bf 10, \bar 5})&\Psi_2({\bf 10, \bar5})
&\Psi_3({\bf 10,\bar5})\cr
\Psi_1({\bf 10,5})&\lambda^{2\psi_1}&\lambda^{\psi_1+\psi_2}&0  \cr
\Psi_2({\bf 10,5})& \lambda^{\psi_1+\psi_2}   &  
                    \lambda^{2\psi_2}  &0 \cr
\Psi_3({\bf 10,5})  & 0 & 0   & 0  \cr}\lambda^{\phi}v,
\end{equation}
and the massless mode ${\bf 10}_{\Psi_3}$ does not mix 
through  non-diagonal mass
matrix elements with any other ${\bf \bar5}$ field. We call 
such a massless field an ``isolated" field. There are various 
different patterns of massless modes containing the ``isolated" fields.
For example, if the conditions $2\psi_2+\phi\geq 0$, $2\psi_3+c\geq 0$ and
$\lambda^{2\psi_1+\phi}\VEV{{\bf1}_\Phi} >
 \lambda^{\psi_1+\psi_2+c}\VEV{{\bf16}_C}$ are satisfied
in addition
to the above condition $\psi_1+\psi_3+\phi<0$, we have the pattern
$({\bf 16}_{\Psi_1},{\bf 16}_{\Psi_2},{\bf 10}_{\Psi_3})$, i.e.,
\begin{equation}
\left(\begin{array}{c}
      {\bf \bar 5}_1 \\
      {\bf \bar 5}_2 \\
      {\bf \bar 5}_3
      \end{array} \right)=
\left(\begin{array}{c}
      \Psi_1({\bf 16},{\bf \bar 5})+\cdots \\
      \Psi_2({\bf 16},{\bf \bar 5})+\cdots \\
      \Psi_3({\bf 10},{\bf \bar 5})
      \end{array} \right),
\end{equation}
which has been adopted in Ref.\cite{ETwist}.
Note that ${\bf\bar5}_3$ has no components from 
 $({\bf 16},{\bf \bar 5})$ and therefore 
 we need the Higgs mixing (\ref{10-16mixing}) 
 to make ${\bf\bar5}_3$ massive.
Here, we do not consider such isolated fields and thus assume 
\beqn
  0  &\leq&  \psi_1+\psi_3+c,  \\
  0  &\leq&  \psi_1+\psi_3+\phi,
\eeqn
 for simplicity.

\subsubsection{Quark mass matrices}

Because the E-twisting mechanism does not change the structure 
 of the ${\bf10}$ sector of SU(5), 
 the mass matrix for up-type quarks 
 are given in essentially the same way as in SO(10) models. 
Their Yukawa interactions are obtained from the interaction 
\bequ
 \lambda^{\psi_i+\psi_j+\phi}\Psi_i\Psi_j\Phi 
\eequ
and $\lambda^{\psi_i+\psi_j+c}\Psi_i\Psi_j C$ if there are Higgs 
 mixing 
\bequ
 H_u\sim{\bf10}_\Phi+\lambda^{c-\phi}{\bf10}_C, 
\label{10-10mixing}
\eequ
as (\ref{HuMixing}). 
They give contributions of the form 
 ${\bf 16}_\Psi{\bf 16}_\Psi{\bf 10}_\Phi$. 
These contributions are of the same order and can be written 
 like (\ref{uMass}).
Thus, we take $(\psi_1,\psi_2,\psi_3,\phi)=(n+3,n+2,n,-2n)$
 for the same reason as explained in \S\ref{SO(10)Matter}. 
Then, we get the Yukawa matrix of up-type quarks as 
\begin{equation}
  Y_U  =  \left(
          \begin{array}{ccc}
            \lambda^6 & \lambda^5 & \lambda^3 \\
            \lambda^5 & \lambda^4 & \lambda^2 \\
            \lambda^3 & \lambda^2 & 1    
          \end{array}
          \Rs.
\end{equation}

Down-type quarks have contributions from the same interactions 
 as up-type quarks, and when another Higgs mixing 
\bequ
  H_d=\cos\gamma{\bf L}_{{\bf 10}}+\sin\gamma{\bf L}_{{\bf 16}_C}
\label{10-16mixing}
\eequ
 exists, the interaction $\Psi\Psi C$ gives the other condition
 that gives a contribution of the form 
 ${\bf 16}_\Psi{\bf 10}_\Psi{\bf 16}_C$. 
When $\sin\gamma \sim \lambda^{\phi-c-r}$, these contributions 
 are of the same order.
Because we assume $0<r<\psi_1-\psi_3=3$, the massless modes of 
 the ${\bf\bar5}$ sector have essentially the same structure as in 
 SO(10) models from \S\ref{SO(10)Matter}. 
And, from the mixing (\ref{52}), 
 we can find $\s t-\s\psi_3\leftrightarrow\psi_1-\psi_3-r$. 
Thus, the Yukawa matrix of down-type quarks is given as 
\begin{equation}
  Y_D  =  \lambda^2\left(
          \begin{array}{ccc}
            \lambda^4 & \lambda^{4-r} & \lambda^3 \\
            \lambda^3 & \lambda^{3-r} & \lambda^2 \\
            \lambda   & \lambda^{1-r} & 1    
          \end{array}
          \right).  
\end{equation}

As mentioned in \S\ref{SO(10)Matter}, we need SU(2)\sub R breaking 
 effect in order to get non-trivial CKM matrix. 
In the SO(10) models, we have to allow a higher-dimensional 
 interaction such as $\Psi_i\Psi_jH\bar CC$. 
On the other hand, in \E6 models, the Higgs mixing 
 (\ref{10-10mixing}) (if the mixing breaks SU(2)\sub R) 
 or (\ref{10-16mixing}) is enough to make CKM matrix non-trivial, 
 and we get the correct orders for the CKM matrix elements: 
\begin{equation}
  U_{\rm CKM}  =  \left(
                  \begin{array}{ccc}
                    1 & \lambda &  \lambda^3 \\
                    \lambda & 1 & \lambda^2 \\
                    \lambda^3 & \lambda^2 & 1
                  \end{array}
                  \right).  
\end{equation}

\subsubsection{Lepton mass matrices} 

The Yukawa matrix of charged leptons is again the transpose 
of $Y_D$, except for an overall factor $\eta$ induced by the 
renormalization group effect: 
\begin{equation}
  Y_{E}  =  \lambda^2
              \left(
                \begin{array}{ccc}
                  \lambda^4 & \lambda^3 & \lambda \\
                  \lambda^{\Delta+1} & \lambda^{\Delta} 
                & \lambda^{\Delta-2} \\
                  \lambda^3   & \lambda^2         & 1 
                \end{array}
              \right)
                     \eta.
\end{equation}

As for the neutrino sector, we have a $3\times6$ Dirac mass matrix 
 and a $6\times6$ Majorana mass matrix, because there are six 
 right-handed neutrinos.
Thus we have to make calculations using large matrices 
 in order to get a contribution to the Seesaw mechanism. 
However, the discussion in \S\ref{Qeff} ensures that 
 such a contribution is of the same order as that of the higher-dimensional 
 interaction ${\bf\bar5}_2{\bf\bar5}_2{\bf5}_\Phi{\bf5}_\Phi$ 
 and can be estimated by the simple sum of the effective charges as 
\begin{equation}
  M_\nu  =  \lambda^{4-2n+2\Delta c}
            \left(
            \begin{array}{ccc}
              \begin{array}{ccc}
                \lambda^2 & \lambda^{2-r} & \lambda \\
                \lambda^{2-r} & \lambda^{2-2r} & \lambda^{1-r} \\
                \lambda   & \lambda^{1-r}         & 1
              \end{array}
            \end{array}
            \right)\VEV{H_u}^2\eta^2, 
\end{equation}
which leads to the following MNS matrix:
\begin{equation}
U_{\rm{MNS}}=
\left(
\begin{array}{ccc}
1 & \lambda^{r} &  \lambda \\
\lambda^{r} & 1 & \lambda^{1-r} \\
\lambda & \lambda^{1-r} & 1
\end{array}
\right). 
\end{equation}

In this way, we get the same matrices as in \S\ref{SO(10)Matter} 
 by the exchange $\s t-\s\psi_3\rightarrow3-r$, 
 thanks to the effective charge. 
This means that the same discussion for neutrino given 
 in \S\ref{SO(10)Matter} can be applied here. 
Namely, we can reproduce bi-large mixing angle 
 when $r=3-(\s t-\s\psi_3)$, that is 
\begin{equation}
  c-\bar c=\phi-\bar \phi+1, 
\end{equation}
 and if we define the parameter $l$ as 
\begin{equation}
  \phi-\bar\phi=2n-9-2r-l, 
\end{equation}
 $l$ is expressed by using the heaviest light neutrino mass 
 $m_{\nu_3}$ as 
\begin{equation}
  \lambda^l\sim\lambda^{-5}\frac{\eta^2\VEV{H_u}^2}
    {m_{\nu_3}\Lambda}.
\end{equation}
These parameters should have values within the following range: 
\beqn
 -1<&l&<-4, \\
 0<&r&<3/2.
\eeqn

\subsubsection{Suppression of FCNC processes} 

The crucial difference between SO(10) models and \E6 models are 
 that in \E6 model the non-diagonal elements of the sfermion 
 mass-squared matrix can be suppressed to some extent. 
This is because the first and second generation of the ${\bf\bar5}$ 
 sector is contained in a single multiplet $\Psi_1$, and thus 
 they are degenerate in the symmetric limit. 
To be more precise, they behave as doublets under an SU(2) 
 subgroup of \E6. 
We call the subgroup as SU(2)\sub E. 
This subgroup is broken by VEVs, $\VEV A$, $\VEV\Phi$, 
 $\VEV{\bar\Phi}$ and so on. 
This means that the rates of FCNC processes are proportional to 
 these VEVs. 
Unfortunately, these VEVs are usually too large to suppress 
 the FCNC processes so that they cannot be observed 
 in the present experiments. 
Thus, the SUSY-flavor problem can be softened 
 but cannot be solved in the \E6 models. 
A sufficient suppression can be obtained in models which 
 employ a horizontal symmetry, as discussed in the next chapter.

\chapter{Models with Horizontal Symmetry}
\label{Modelsw/Horiz}

We have several reasons for introducing a horizontal symmetry $G_H$.
One of them is to understand the origin of the flavor violation
in Yukawa couplings of quarks and leptons. Actually many studies have
been made along this direction\cite{discrete}-\cite{nonabel}.

The second reason is to unify quarks and leptons in fewer multiplets,
though it is strongly related with the first motivation.
Usual GUTs with SU(5), SO(10) or \E6 gauge group can 
 make unification of quarks and leptons within one generation. 
In order to unify all the quarks and leptons into a single multiplet,
a larger gauge group such as SO($12+2n$), \E7, or \E8 is required,
though these unified groups cannot realize the chiral matter 
in 4D theories. 
However, actually it is possible in higher dimensional field
theories, and in that cases, a horizontal symmetry may appear in
the effective 4D theories.

The third reason is to solve the SUSY-flavor problem\cite{FCNC}. 
A non-abelian flavor (horizontal) symmetry can potentially 
 solve this problem. 
If the first
two generation fields become  a doublet under the flavor symmetry, 
$\Psi_a$($a=1,2$), the sfermion masses of the first two 
generation fields become
universal, unless the flavor symmetry is broken. 
This is important in solving the SUSY-flavor problem.

\section{Horizontal Symmetry for SUSY-Flavor Problem}
\label{HorizFCNC}

In this section, we examine the idea that the SUSY-flavor problem 
 can be solved when a horizontal symmetry is
 introduced. 
We follow the argument given in Ref.\cite{horiz}. 
Of course, in order to obtain realistic hierarchical structures 
 of Yukawa couplings, 
 the flavor symmetry must be broken, for example 
 by a VEV $\VEV{F_a}$. 
Then, generally the universal sfermion
masses are lifted by the breaking effect. 
Various models in which the breaking 
effects can be controlled have been considered 
in the literature\cite{IR,Sdiscrete,Sabel,Snonabel}.
However, in GUT models with bi-large neutrino mixings, 
the universality of sfermion
masses of the first two generations 
is not sufficient to solve the SUSY-flavor problem. 
This is because the large mixings and the $\order1$ discrepancy 
 between the sfermion masses of the third generation fields and 
 those of the first two generation fields lead to 
 too rapid FCNC processes. 
The \E6 unification can naturally solve this problem\cite{horiz}.
The essential point is that in the \E6 unification 
 all the three light modes of ${\bf \bar 5}$
 fields come from the first two generation fields 
 $\Psi_a({\bf 27})$ as shown in \S\ref{E6Matter}. 
Therefore, all the three light modes of ${\bf \bar 5}$ 
have universal sfermion masses, which are important 
in solving the SUSY-flavor problem.

\subsection{U(2) Models}
\label{U(2)}

Let us show how the large mixings and the $\order1$ discrepancy 
 lead to too rapid FCNC processes. 
For simplicity,  we consider a simple model with 
a global horizontal symmetry U(2), under which the three
generations of quarks and leptons,
$\Psi_i=(\Psi_a,\Psi_3)$ ($a=1,2$), transform as ${\bf 2+1}$, while 
the Higgs field $H$ is a singlet.
Then only the Yukawa interaction $\Psi_3\Psi_3H$ is allowed 
by the horizontal 
symmetry, which accounts for the large top Yukawa coupling. 
Suppose that the U(2) horizontal symmetry is broken by the 
VEVs of a doublet $\VEV{\bar F^a}=\delta^a_2V$ and of an anti-symmetric
tensor $\VEV{A^{ab}}=\epsilon^{ab}v$ ($\epsilon^{12}=-\epsilon^{21}=1$) as
\begin{equation}
  {\rm U(2)}_H \mylimit{V}{\rm U(1)}_H \mylimit{v} {\rm nothing}.
\end{equation}
The ratios of the VEVs to the cutoff scale $\Lambda$, 
 $\epsilon\equiv V/\Lambda\gg\epsilon'\equiv v/\Lambda$, 
 give the following hierarchical structure of the Yukawa couplings as 
\begin{equation}
Y_{U,D,E}\sim\left( 
 \begin{matrix}
  0 & \epsilon' & 0 \\
  \epsilon' & \epsilon^2 & \epsilon \\
  0 & \epsilon & 1
 \end{matrix}
\right).
\end{equation}
Moreover,
the U(2)$_H$ symmetric interaction
 $\dtf \Psi^{\dagger a}\Psi_a S^\dagger S$, where $S$ 
 has a non-vanishing VEV $\VEV{S}\sim \theta^2\tilde m^2$ 
 and should not be confused with the dilaton field discussed 
 in \S\ref{U(1)A},
leads to approximate universality of the first and second generation 
sfermion masses:
\begin{equation}
\tilde m^2_{U,D,E}\sim \tilde m^2\left(
 \begin{matrix}
  1 & 0 & 0 \\
  0 & 1+\epsilon^2 & \epsilon \\
  0 & \epsilon & 1+R_{U,D,E} 
 \end{matrix}
\right).
\label{U(2)sfermion} 
\end{equation}
Here $R_{U,D,E}$ is $\order1$, because $\Psi_3$ has nothing to do 
 with $\Psi_a$, $\epsilon^2$ comes from higher dimensional interactions, 
 such as 
\bequ
 \dtf(\Psi_a\bar F^a)^\dagger \Psi_b\bar F^bS^\dagger S,
\label{OffDiag} 
\eequ
through a non-vanishing VEV $\VEV{\bar F}$. 
We have neglected the contributions from $\epsilon'$. 
The important parameters which are constrained 
by the FCNC processes are defined as 
\begin{equation}
\delta_{f_\chi}\equiv 
 V_{f_\chi}^\dagger\frac{\tilde m^2_{f_\chi}-\tilde m^2}{\tilde m^2}
 V_{f_\chi},
\end{equation}
where $V_{f_\chi}$ ($f=U,D,E$, $\chi=L,R$) is a diagonalizing matrix 
 for fermions\cite{GGMS}.
The constraints are given as, for example,
\begin{eqnarray}
\sqrt{|{\rm Im}(\delta_{D_{\rm L}})_{12}(\delta_{D_{\rm R}})_{12})|}&\leq&
 2\times 10^{-4}\left(\frac{\tilde m_Q}{500\ {\rm GeV}}\right), \nonumber \\
|{\rm Im}(\delta_{D_{\rm R}})_{12}| & \leq & 1.5\times 10^{-3}
\left(\frac{\tilde m_Q}{500\ {\rm GeV}}\right),
\label{K}
\end{eqnarray}
at the weak scale from $\epsilon_K$ in the $K$ meson mixing,
and 
\begin{equation}
|(\delta_{E_{\rm L}})_{12}|\leq 4\times 10^{-3}\left(
\frac{\tilde m_{\rm L}}{100\ {\rm GeV}}\right)^2
\label{mu}
\end{equation}
from the $\mu\rightarrow e\gamma$ process.

As shown above, the U(2)$_H$ symmetry indeed realizes not only 
hierarchical Yukawa couplings 
but also approximately universal sfermion masses 
of the first two generation fields.
These mass matrices lead to the relations
\begin{equation}
\frac{\tilde m_2^2-\tilde m_1^2}{\tilde m^2}
\sim \frac{m_{F2}}{m_{F3}},
\label{Mass}
\end{equation}
where $m_{Fi}$ and $\tilde m_i$ are the masses of 
 the $i$-th generation fermions
 and of the $i$-th generation sfermions, respectively.
Unfortunately, these predictions of this simple model are 
 too large for the ${\bf\bar5}$ sector which has milder 
 hierarchy in fermion masses, leading to too rapid FCNC processes. 
Furthermore, even if we can manage to make the $\epsilon$ 
 contributions in (\ref{U(2)sfermion}) harmless, 
 \eg\ by forbidding the higher dimensional interactions 
 (\ref{OffDiag}) by hand, 
 the bi-large mixings in the lepton sector lead to too rapid FCNC 
 processes through the $\order1$ contribution $R_{\bf\bar5}$. 
To be more precise, even if we can realize 
 $\Delta_{f_\chi}\equiv
   \frac{\tilde m^2_{f_\chi}-\tilde m^2}{\tilde m^2}
  =\diag(0,0,R_{f_\chi})$, the large mixings in the ${\bf\bar5}$ sector, 
 such as
\begin{equation}
V_{\bf\bar5}=
\left(
\begin{array}{ccc}
1 & \lambda^{r} &  \lambda \\
\lambda^{r} & 1 & \lambda^{1-r} \\
\lambda & \lambda^{1-r} & 1
\end{array}
\right), 
\end{equation}
 where $r\sim\half$, induce a large $(\delta_{f_\chi})_{12}$ as 
\begin{equation}
\delta_{\bf\bar5}=R_{\bf\bar5}\Ls
 \begin{matrix}
                \lambda^2 & \lambda^{2-r} & \lambda \\
                \lambda^{2-r} & \lambda^{2-2r} & \lambda^{1-r} \\
                \lambda   & \lambda^{1-r}         & 1
 \end{matrix}
\right).
\label{U(2)del5}
\end{equation}
Comparing with the experimental constraints 
 (\ref{K}) and (\ref{mu}), which suggest 
 $(\delta_{f_\chi})_{12}\lesssim\lambda^4$ for 
 $\s m_Q\sim500\GeV,\,\s m_{\rm L}\sim100\GeV$, 
 we find that the $(1,2)$ component of (\ref{U(2)del5}) 
 is too large.

In addition, the hierarchical
Yukawa couplings predicted by this simple model are similar for the 
up-quark sector, the down-quark sector, and the charged lepton sector, 
 which is inconsistent with the experimental results.
Moreover,  in the neutrino sector, 
it seems to be difficult to obtain 
the large neutrino mixing angles.

\subsection{\E6$\times$SU(2)\sub H$\times$U(1)\sub A Models}

Note that all the problems mentioned in \S\ref{U(2)} 
 arise from the ${\bf\bar5}$ sector, 
 and also that the ${\bf\bar5}$ sector has twisted family structure 
 in the models discussed 
 in \S\ref{SO(10)Matter} and \S\ref{E6Matter}. 
By this twist, we can get milder hierarchies and thus 
 larger mixings in the ${\bf\bar5}$ sector, 
 even though the original hierarchy is similar to that 
 in the ${\bf10}$ sector. 
Another consequence of the twist is that 
 a smaller SU(2)\sub H breaking can reproduce 
 milder hierarchies of Yukawa couplings in the ${\bf\bar5}$ sector 
 than in ${\bf10}$ sector. 
Because the lift from the degeneracy is given 
 by the SU(2)\sub H breaking, 
 we can avoid the disfavored relation (\ref{Mass}).
Furthermore, in \E6 models, all the generation of ${\bf\bar5}$ 
 come from $\Psi_1({\bf27})$ and $\Psi_2({\bf27})$ which 
 behave as doublets under SU(2)\sub H in \E6$\times$SU(2)\sub H
 models. 
This means that, in these models, all the ${\bf\bar5}$'s 
 are degenerate in the SU(2)\sub H symmetric limit, and 
 it is expected  $R_{\bf\bar5}\ll1$.

Let us illustrate this by using an example of \E6 models employing 
 anomalous U(1) symmetry and SU(2) horizontal (gauge) symmetry shown 
 in Table\ \ref{E6SU(2)Content}.
\begin{table}
\begin{center}
\begin{tabular}{|c|cc|cccccccccc|} 
\hline
   & $\Psi_a$ & $\Psi_3$ 
   & $A$ & $A'$ 
     & $\Phi$ & $\bar\Phi$ & $C$ & $\bar C$ 
     & $C'$ & $\bar C'$ 
   & $F_a$ & $\bar F^a$  \\
\hline 
 \E6 
   & ${\bf 27}$ & ${\bf 27}$ 
   & ${\bf78}$ & ${\bf78}$ 
     & ${\bf27}$ & ${\bf\cc{27}}$ & ${\bf27}$ & ${\bf\cc{27}}$ 
     & ${\bf{27}}$ & ${\bf\cc{27}}$ 
   & ${\bf 1}$ & ${\bf 1}$ \\
 SU(2)\sub H 
   & {\bf 2} &{\bf 1} 
   & {\bf 1} &{\bf 1} &{\bf 1} &{\bf 1} &{\bf 1}
     &{\bf 1} &{\bf 1} &{\bf 1}
   &{\bf 2} &${\bf\bar2}$ \\ 
  U(1)\sub A 
   & $5$ & $2$ 
   & $-1$ & $5$ 
     & $-4$ & $2$ & $-5$ & $-2$ 
     & $9$ & $7$ 
   & $-2$ & $-3$ \\
 \hline
\end{tabular}
\caption{
Typical values of anomalous U(1) charges of non-singlet fields.
}
\label{E6SU(2)Content}
\end{center}
\end{table}
Here we omit singlet fields and additional $Z_N$ symmetries 
 for simplicity. 
Note that all the \E6 charged Higgs are singlets under SU(2)\sub H. 
This means that the discussion in \S\ref{E6Higgs} can be applied, 
 and the effect of SU(2)\sub H appears only on $\Psi_a$ as 
\bequ
 \s\psi_1=\psi_a+\Delta f,\,\, 
 \s\psi_2=\psi_a-\Delta f, 
\eequ
 where $\Delta f\equiv \half(f-\bar f)=\half$.
Thus, from this charge assignment together with Eqs.(\ref{E6l}) 
 and (\ref{E6r}), we can find that $l=-3$, $r=3/2$ and 
\bequ
   Y_U  =  \left(
          \begin{array}{ccc}
            \lambda^7 & \lambda^6 & \lambda^{3.5} \\
            \lambda^6 & \lambda^5 & \lambda^{2.5} \\
            \lambda^{3.5} & \lambda^{2.5} & 1    
          \end{array}
          \Rs,\,\,
  Y_D  =  \lambda^{2}\left(
          \begin{array}{ccc}
            \lambda^5 & \lambda^4 & \lambda^{3.5} \\
            \lambda^4 & \lambda^3 & \lambda^{2.5} \\
            \lambda^{1.5} & \lambda^{0.5} & 1    
          \end{array}
          \Rs, 
\eequ 
 which lead to 
\begin{equation}
U_{\rm{CKM}}=V_{\bf{10}}=
\left(
\begin{array}{ccc}
1 & \lambda &  \lambda^{3.5} \\
\lambda & 1 & \lambda^{2.5} \\
\lambda^{3.5} & \lambda^{2.5} & 1
\end{array}
\right),\,\, 
U_{\rm{MNS}}=V_{\bf{\bar5}}=
\left(
\begin{array}{ccc}
1 & \lambda &  \lambda^{1.5} \\
\lambda & 1 & \lambda^{0.5} \\
\lambda^{1.5} & \lambda^{0.5} & 1
\end{array}
\right). 
\end{equation}
Note that the main modes of the three generations in the ${\bf{\bar5}}$ 
 sector are given by 
 $({\bf \bar 5}_1,{\bf \bar 5}_2,{\bf \bar 5}_3)\sim
 ({\bf16}_{\Psi_1},{\bf16}_{\Psi_2},{\bf10}_{\Psi_1})$

The sfermion mass-squared matrices are written as 
\begin{equation}
\tilde m_f^2=\left(
 \begin{matrix}
   \tilde m_{f_{\rm L}}^2 & A_f^\dagger \cr
                           A_f & \tilde m_{f_{\rm R}}^2 
 \end{matrix}
 \right),
\end{equation}
where $A_f$ is the $A$-term matrix. 
In the following discussion, we restrict our consideration 
on the mass mixings through $\tilde m_{f_P}^2$,
because a reasonable assumption, \eg\ SUSY breaking in the hidden 
sector, leads to an $A_f$ that is proportional to 
 the Yukawa matrix $Y_f$\cite{SW}. 
The corrections $\Delta_{f_\chi}$ 
in this model are approximately given as 
\begin{equation}
\Delta_{\bf 10}=\left(
 \begin{matrix}
  \lambda^5 & \lambda^6 & \lambda^{3.5}\cr
  \lambda^6 &\lambda^5 &  \lambda^{2.5} \cr
  \lambda^{3.5} & \lambda^{2.5}& R_{\bf 10} \cr 
 \end{matrix}
\right),\,\,
\Delta_{\bf \bar 5}=\left(
 \begin{matrix}
  \lambda^5 & \lambda^6 & \lambda^{5.5} \cr
  \lambda^6 & \lambda^5 & \lambda^{4.5} \cr
  \lambda^{5.5} &  \lambda^{4.5}& R_{\bf \bar 5} \cr 
 \end{matrix}
\right), 
\end{equation}
where, for example, $(\Delta_{\bf \bar 5})_{12}$ is derived 
 by using the interaction 
$\dtf\lambda^{|f-\bar f|}
 (\Psi\bar F)^\dagger (\Psi F)S^\dagger S$.
Note that 
$\frac{\tilde m_{D_{\rm R}2}^2-\tilde m_{D_{\rm R}1}^2}{\tilde m_{D_{\rm R}}^2}
 \sim\lambda^5\ll \frac{m_s}{m_b}$.
Namely, the Yukawa hierarchy in the superpotential becomes milder,
improving the undesirable relations (\ref{Mass}).
The constrained parameters $\delta_{f_\chi}$ are approximated as  
\begin{equation}
\delta_{\bf 10}=R_{\bf10}\left(
 \begin{matrix}
   \lambda^5 & \lambda^6 & \lambda^{3.5} \cr
   \lambda^6 & \lambda^5 & \lambda^{2.5} \cr
   \lambda^{3.5} & \lambda^{2.5} & 1
 \end{matrix}
             \right), \,\,
\delta_{\bf \bar 5}=R_{\bf \bar 5}
          \left(
 \begin{matrix}
  \lambda^3 & \lambda^2 & \lambda^{1.5} \cr
  \lambda^2 & \lambda & \lambda^{0.5} \cr
  \lambda^{1.5} & \lambda^{0.5} & 1 
 \end{matrix}
 \right)
\label{delta}
\end{equation}
at the GUT scale. 
As discussed above, all the sfermion masses for ${\bf \bar 5}$ become
equal at the leading order in this model.
$R_{\bf\bar5}$ is given by SU(2)\sub E breaking effects, such as 
 $\VEV{A}$, $\VEV{\Phi}$ and $\VEV{\bar\Phi}$, 
 through interactions such as
 $\dtf \Psi^\dagger \Phi^\dagger \Psi\Phi S^\dagger S$. 
$R_{\bf\bar5}$ is $\order{\lambda^2}$ while 
 $R_{\bf 10}$ is $\order1$ in this model. 
This reduces the lower limits
for the scalar quark masses to satisfy the FCNC constraints 
 to an acceptable level, $250\GeV$.

\subsubsection{Extension to SU(3)\sub H}

It is interesting that the SU(2)\sub H horizontal symmetry in the 
 previous model can be extended to SU(3)\sub H.
In such models, the three generations of quarks and leptons
can be unified into a single multiplet, $\Psi({\bf 27},{\bf 3})$.
Assuming that the horizontal gauge symmetry SU(3)\sub H is broken 
by the VEVs of two pairs of Higgs fields $F_i({\bf 1},{\bf 3})$ and 
$\bar F_i({\bf 1},{\bf \bar 3})$ $(i=2,3)$ as
\begin{equation}
|\VEV{F_{ia}}|=|\VEV{\bar F_i^a}|\sim 
\delta_i^a\lambda^{-\frac{1}{2}(f_i+\bar f_i)},
\end{equation}
the effect of SU(3)\sub H can be parameterized by using 
 the following two parameters, 
\beqn
 \Delta{f_3} &=& \half(f_3-\bar f_3), \\
 \Delta{f_2} &=& \half(f_2-\bar f_2)
\eeqn
as
\begin{equation}
\tilde \psi_i\equiv \psi-\Delta f_i,\quad
\tilde \psi_1\equiv \psi+\Delta f_2+\Delta f_3, 
\end{equation}
 and thus $\s\psi=\psi+\half\Delta f_3$.
This parametrization corresponds to that 
 in the previous model as 
\beqn
 \Psi_3 &\leftrightarrow& \Psi\bar F_3, \\
 \Psi\bar F &\leftrightarrow& \Psi\bar F_2, \\
 \Psi F &\leftrightarrow& \Psi F_2F_3, \\ 
 \Delta{f} &\leftrightarrow& \half(2\Delta{f_2}+\Delta{f_3}).
\eeqn
Note that in order to realize an $\order1$ top Yukawa coupling, 
 which is given by $\Psi\Psi\Phi\VEV{\bar F_3}^2$, 
 SU(3)\sub H must be broken at the cutoff scale, namely, 
 $\VEV{F_3}\sim1$ which is realized when $f_3+\bar f_3=0$.
To obtain the same mass matrices of quarks and leptons as in the previous
 \E6$\times$SU(2)\sub H model, the effective charges must be taken as
 $(\tilde \psi_1,\tilde \psi_2,\tilde \psi_3)=(11/2,9/2,2)$.
For example, a set of charges $(f_3,\bar f_3,f_2,\bar f_2)=(2,-2,-3,-2)$ 
and $\psi=4$ satisfies the above conditions.

\subsection{Discussions}

For both \E6$\times$SU(2)\sub H and \E6$\times$SU(3)\sub H models, 
 we have introduced a Higgs sector that breaks \E6 to $G_{\rm SM}$ 
 where all the fields have trivial quantum numbers for the horizontal 
 symmetry. 
This means, as mentioned above, that the discussion in \S\ref{E6Higgs} 
 can be applied to this case. 
Thus, it is possible to obtain complete
 \E6$\times$SU(2)\sub H and \E6$\times$SU(3)\sub H \aGUT s, where 
 the degeneracy of the sfermion masses of ${\bf\bar 5}$ fields
 is naturally obtained.
Note that, additional fields that are 
not singlets under the horizontal gauge symmetry are required for anomaly
cancellation. 
It is interesting to introduce
non-singlet Higgs fields under the horizontal gauge symmetry. 
This is the subject of \S\ref{HorizHiggs}.

Before examining that possibility, let us make comments on 
 phenomenology of these models.
Because the SU(3)\sub H symmetry must be broken at the cutoff scale 
to realize an $\order1$ top Yukawa coupling, the degeneracy of the sfermion
masses between the third generation fields $\Psi_3$ and the first
and second generation fields $\Psi_a$ $(a=1,2)$ is not guaranteed.
Therefore, the \E6$\times$SU(3)\sub H models gives the same predictions
for the structure of sfermion masses as \E6$\times$SU(2)\sub H models. 
Roughly speaking,
 all the sfermion fields have nearly equal masses, 
 except the third generation fields included in {\bf 10} of SU(5).
It must be an interesting subject to study the predictions on 
FCNC processes (for example, $B$-physics\cite{Okada}) from such a 
special structure of the sfermion masses.
More precisely, this degeneracy is lifted by $D$-term 
contributions of SU(3)\sub H and \E6. 
Though the contributions strongly depend on the concrete
models for SUSY breaking and on GUT models and 
 some of them must be small in order to suppress the 
FCNC processes, it is important
to test these GUT models
with precisely measured masses of sfermions, as discussed 
in Ref.\cite{KMY}.

\section{Horizontal Symmetry in \E6 Higgs Sectors}
\label{HorizHiggs}

In this section, we investigate \E6 models with an anomalous
U(1) gauge symmetry whose Higgs 
 sectors have non-trivial quantum numbers for the horizontal 
 symmetry, SU(2)\sub H or SU(3)\sub H\cite{HorizHiggs}.
Because \E6 contains SU(2)\sub E, these models may realize 
 well-suppressed FCNC processes as suggested in \S\ref{HorizFCNC}.
In this sence, \E6 models seem more promising than 
 SO(10) models which are examined in detail in Ref.\cite{HorizHiggs}. 
Unfortunately, however, if both \E6 and the horizontal symmetry
are simultaneously broken, it is difficult to obtain 
realistic models in which the FCNC processes are sufficiently 
suppressed. 
The point is as follows.
In order to sufficiently suppress the FCNC processes with 
the horizontal symmetry, 
the scale at which the horizontal symmetry is broken should be smaller
than $\lambda^2$. (In this section, we take 
$\lambda\sim \sin\theta_C\sim 0.22$, and we do not fix the anomalous U(1)
charge of the FN field to $-1$ but $\VEV{\Theta}\sim \lambda^\theta$.)
Generally, in \aGUT s, it is difficult to obtain 
 a smaller \E6 breaking scale than $\lambda^2$, 
 as mentioned in \S\ref{E6Higgs}. 
Therefore, if both \E6 and the horizontal symmetry are broken by a VEV
of a single field, that is, both the symmetries are broken at the same scale,
then the suppression of the FCNC processes does not become sufficient, 
 although SU(2)\sub E help to ameliorate the SUSY-flavor problem. 

Nevertheless, such a possibility still deserves to be examined, 
 by the second reason discussed at the beginning of this chapter. 
In particular, if we assume that \E8 is the unified group of a more 
 fundamental theory, 
 it seems natural that some of ${\bf27}$(${\bf\cc{27}}$) Higgs fields 
 also have non-trivial quantum numbers for the horizontal symmetry.

\subsection{\E6$\times$SU(2)\sub H$\times$U(1)\sub A Models}
\label{e6su2}

Motivated by the decomposition
 \E8$\supset$\E6$\times$SU(3)\sub H$\supset$\E6$\times$SU(2)\sub H, 
 under which {\bf248} of \E8 is decomposed as 
\beqn
 {\bf248} &\rightarrow&
   ({\bf78},{\bf1})+({\bf1},{\bf8})
  +({\bf27},{\bf3}) + ({\bf{\cc{27}}},{\bf\bar{3}}) \\
 &\rightarrow&
   ({\bf78},{\bf1})+({\bf1},{\bf3}+{\bf2}+{\bf2}+{\bf1})
  +({\bf27},{\bf2}+{\bf1}) + ({\bf{\cc{27}}},{\bf2}+{\bf1}), 
\eeqn
we assign non-trivial representations of the horizontal symmetry 
 only to {\bf27}, ${\bf\cc{27}}$ and/or {\bf1} Higgs fields.
Note that, in the matter sector, 
 $\Psi_1({\bf27})$ and $\Psi_2({\bf27})$ are treated as a doublet, 
 and the difference of their effective charges should correspond 
 to the Cabibbo angle, 
 $\s\psi_1-\s\psi_2\sim1$. 
This means that the difference of the effective charges of the two 
 components of doublets should be also around 1 
 as far as the effective charge is well-defined. 
In Table \ref{SimpleE6ContentA}, 
 we have introduced two {\bf27} ($\Phi, C'$)
 and two ${\bf\cc{27}}$ ($\bar C, \bar C'$), where 
 the difference of anomalous U(1) charges of 
 two fields each is much larger than 1. 
Thus, it is difficult to unify the Higgs sector of 
 the model, and we concentrate on the Higgs sector of 
 Table \ref{E6Content}, which contain
\bequ
\begin{array}{cccc}
  {\bf 78} &:&   A,  & A'      \\
  {\bf 27} &:&   \Phi,\ C, &  C'  \\
  {\bf\cc{27}} &:& \bar \Phi,\  \bar C, & \bar C'
\end{array}
\eequ
and some singlets.
From the same reason 
 as for the previous models (Table\ \ref{SimpleE6ContentA}), 
 it is difficult to embed 
 the primed fields into a doublet if we aim to suppress 
 the FCNC processes 
 not assuming the universal soft mass.
If we take $(\Phi,C)$ as a doublet under SU(2)\sub H, the Yukawa interaction
for the top quark, 
$\Psi_3\Psi_3\Phi$, is forbidden by the horizontal symmetry 
 so that it is difficult to realize an $\order1$ top Yukawa.
Thus the remaining possibility is to embed $\bar \Phi$ 
 and $\bar C$ into a doublet as $\bar C_a =(\bar\Phi,\bar C)$.

The Higgs content we consider below is summarized 
 in Table \ref{HiggsContentE6SU2}.
\begin{table}[th]
\[
\begin{array}{|c|c|c|} 
\hline
           &   \mbox{non-vanishing VEV}  & \mbox{vanishing VEV} \\
\hline 
  {\bf 78} &   A\Ls \VEV{{\bf45}_A}\sim\lambda^{-a}\Rs  & A'      \\
  {\bf 27} &   \Phi\Ls \VEV{{\bf1}_\Phi}
                   \sim\lambda^{-(\phi-\Delta\phi)}\Rs,\ 
               C\Ls \VEV{{\bf16}_C}\sim\lambda^{-(c-\Delta c)}\Rs 
           &  C'  \\
  {\bf\cc{27}} & \bar C_a\Ls 
                    \VEV{{\bf{1}}_{\bar C_1}}
               \sim \lambda^{-(\bar c+\Delta\phi+\Delta f)},\ 
                    \VEV{{\bf\cc{16}}_{\bar C_2}}
               \sim \lambda^{-(\bar c+\Delta c-\Delta f)} \Rs 
               & \bar C'\\
  {\bf1} & \bar F_a\Ls \VEV{\bar F_1}
                    \sim\lambda^{-(\bar f+\Delta f)}\Rs,\ 
           F_a \Ls \VEV{F_2}\sim\lambda^{-(f-\Delta f)}\Rs& \\
\hline
\end{array}
\]
\caption{
The Higgs content of \E6$\times$SU(2)\sub H$\times$U(1)\sub A models 
 except for singlets:
Here SU(2)\sub H doublets are denoted by the index $a$.
One or more discreate symmetries are introduced when needed.
}
\label{HiggsContentE6SU2}
\end{table}
All the non-vanishing VEVs are shown, and their magnitudes 
 are expressed with parameters $\Delta\phi$ etc..
For simplicity, we assume that any component fields other than 
 those shown in Table \ref{HiggsContentE6SU2} have vanishing VEVs.
We also assume that the following three $G$-singlets have 
 non-vanishing VEVs as in the VEV relations (\ref{VEV}) 
 from three $F$-flatness conditions: 
\beqn
  \Phi \bar C F &\sim& \lambda^{-(\phi+\bar c+f)} 
                   \ \equiv \lambda^{-3k},
\label{Fflat1}\\
  C \bar C\bar F &\sim& \lambda^{-(c+\bar c+\bar f)}, 
\label{Fflat2}\\
  F\bar F &\sim& \lambda^{-(f+\bar f)}, 
\label{Fflat3}
\eeqn
where the parameter $k$ should not be confused with $k$ 
 given in \S\ref{SimpleE6}. 
In addition to the three relations, 
 three $D$-flatness conditions
\beqn
  \abs{{\bf1}_\phi}^2 &=& \abs{{\bf1}_{\bar C_1}}^2, \\
  \abs{{\bf16}_C}^2 &=& \abs{{\bf\cc{16}}_{\bar C_2}}^2, \\
  \abs{{\bf1}_{\bar C_1}}^2 + \abs{\bar F_1}^2
 &=& \abs{{\bf\cc{16}}_{\bar C_2}}^2 + \abs{F_2}^2
\label{DcondiSU2}
\eeqn
determine three parameters, $\Delta\phi$, $\Delta c$ and $\Delta f$,
in terms of the anomalous U(1) charges.
Roughly speaking, there are four possible cases as follows:
\beqn
 1.&& {\bf1}_{\bar C_1} \sim {\bf\cc{16}}_{\bar C_2}
                          \geq \bar F_1,\,F_2 .
\label{vacuum1}\\
 2.&& \bar F_1 \sim {\bf\cc{16}}_{\bar C_2} 
           \geq {\bf1}_{\bar C_1},\,F_2 .
\label{vacuum2}\\
 3.&& \bar F_1 \sim F_2 
           \geq {\bf1}_{\bar C_1},\,{\bf\cc{16}}_{\bar C_2} .
\label{vacuum3}\\
 4.&& {\bf1}_{\bar C_1} \sim F_2 
                         \geq \bar F_1,\,{\bf\cc{16}}_{\bar C_2} .
\label{vacuum4} 
\eeqn
As for the second and third cases, 
 the horizontal symmetry breaking scale is larger than
 the GUT breaking scale $\VEV{\bf 1}$. 
As discussed in \S\ref{E6Higgs},
$\VEV{\bf 1}$ does not seem so small as $\lambda^2$. Therefore, 
 an SU(2)\sub H breaking scale larger than $\VEV{\bf1}$ is not sufficient
 for the suppression of the FCNC processes.
For simplicity,  we concentrate on the fourth case in the following discussion,
but a similar discussion can be applied to the other cases.
In the fourth case, we get
\beqn
  {\bf1}_\Phi = {\bf1}_{\bar C_1} \sim F_2 &\sim& \lambda^{-k} ,\\
  \bar F_1 &\sim& \lambda^{-(f+\bar f)+k} , \\
  {\bf16}_C = {\bf\cc{16}}_{\bar C_2} 
            &\sim& \lambda^{\half[-(c+\bar c)+f-k]}, 
\eeqn
in other words,
\beqn
 \Delta f &=& \frac{2f-\phi-\bar c}{3} = f-k ,
\label{deltaf}\\
 \Delta \phi &=& \frac{2\phi -f-\bar c}{3} = \phi-k ,\\
 \Delta c &=& \frac{c-\bar c +\Delta f}{2}.
\eeqn
The conditions for the fourth case ($F_2 
 \geq \bar F_1,\,{\bf\cc{16}}_{\bar C_2} $) to be realized are given by 
\bequ
 0<-k \leq -\half(f+\bar f)\,,\,\, -c-\bar c+f,
\label{4thCase}
\eequ
which are also written as 
\bequ
 f< \Delta f \leq \half(f-\bar f)\,,\,\, -c-\bar c+2f.
\label{4thCase'}
\eequ

In addition, as shown in \S\ref{E6Higgs}, the following 
 conditions are required phenomenologically:
\bit
 \item The parameter $r$ for the neutrino mixings should be 
       around $1/2$--$3/2$. 
 \item The parameter $l$ for the neutrino mass scale should be 
       around $-2$--$-3$.
 \item In order to realize the DTS, 
       $C'A\Phi\Phi$ must be allowed, 
       while $C'AC\Phi$ must be forbidden.
 \item $\bar C F\bar C F\bar C\bar F$, which corresponds to 
       $\bar\Phi\bar\Phi\bar C$ in \S\ref{E6Higgs}, must be 
       allowed in order to avoid undesirable massless modes.
 \item In order to give mass to would-be PNG modes, 
       $A'A\Phi\bar C F$ must be allowed. 
 \item For the gauge coupling unification, a smaller 
       effective mass of the colored Higgs, 
       $m^\eff_{\rm C}\sim\lambda^{2\phi+\Delta\phi}$, is preferred.
       In the model displayed in Table \ref{E6Content}, 
       $m^\eff_{\rm C}\sim\lambda^{-8.5}$.
 \item In order to reproduce the realistic quark mass matrices,
       the SU(2)\sub R symmetry must be broken in the Yukawa couplings.
       SU(2)\sub R breaking VEVs $\VEV{C}=\VEV{\bar C}$ can be 
       picked up through the SM Higgs mixing 
       ($\bar C'\bar C\bar FA\Phi^2$ is required), 
       or through higher dimensional interactions 
       (for example, $\Psi\bar CC\Psi\bar F\Phi$).        
\eit
These conditions are rewritten 
 in terms of the anomalous $U(1)$ charges as
\begin{eqnarray}
&&\half
         \lesssim r=\half(c-\phi)+\Delta f
         \lesssim \frac{3}{2} 
\label{1st}
\\
&&-2\gtrsim l=-5-2(\psi-\Delta f-\psi_3)+\phi+2\Delta c
                   \gtrsim -3
\label{2nd}\\
&&c<\phi \label{3rd}\\
&&\bar f \geq -3\bar c-2f \label{4th}\\
&&0\leq a'+a+\phi+\bar c+f\geq 0 \label{5th}\\
&&2\phi+\Delta\phi\gtrsim -8.5 \label{6th}\\
&&2\psi+\phi+c+\bar c+\bar f\geq 0\quad {\rm or}
 \quad \bar c'\geq -2\phi-\bar c-\bar f-a,
\hspace{5cm}
\label{7th}
\end{eqnarray}
Note that the first condition is not consistent with the third conditions
if $\tilde \psi_1-\tilde \psi_2=1$, that is, $\Delta f=\half$ to reproduce
the suitable value of the Cabibbo angle.
There are three ways to avoid this inconsistency: 
\begin{enumerate}
 \item To relax the first requirement. \\
       For example, $r=\frac{1}{4}$ is not an unacceptable choice, 
       although rather large ambiguity of 
       $\order1$ coefficients are needed
       to reproduce the large atmospheric neutrino oscillation.
 \item To set $c\geq\phi$ and introduce an additional discrete symmetry 
       to forbid $C'AC\Phi$.\\
       If $c=\phi$ is taken, the relation 
       $r=\half$ is obtained.
 \item To give up the effective charge.\\
       This strategy is examined in detail in Ref.\cite{HorizHiggs}. 
\end{enumerate}
Here, we construct realistic models, along 
 the first and second strategies. 
In the following, we consider only the case with $\Delta f=\half$, 
 for simplicity.

\subsubsection{Strategy 1: $c<\phi$ ($r<\half$)}
\label{strategy1}

The relation $r=\half(c-\phi+1)$ indicates that 
smaller $\phi-c(>0)$ leads to
larger $r$ bounded from above by 1/2. Therefore, if $\phi-c$ is taken 
as the minimum unit of 
U(1)\sub A charge, then $r$ acquires the closest value to $\half$.
Therefore, the smaller unit leads to the closer value 
of $r$ to $\half$.
Here, we introduce half integer U(1)\sub A charges 
and take $\theta=-\half$, 
which gives $r=\frac{1}{4}$.

In the fourth vacuum (\ref{vacuum4}), 
the SU(2)\sub H breaking scale is the
same as the \E6 breaking scale, because the VEVs 
$\VEV{{\bf 1}_\Phi}=\VEV{{\bf1}_{\bar C_1}}
               \sim \VEV{F_2}\sim \lambda^{-k}$ 
break simultaneously SU(2)\sub H and \E6. 
In order to suppress 
the FCNC processes, a smaller SU(2)\sub H breaking scale  is preferable, 
while a smaller \E6 breaking scale leads to 
a larger effective mass of the colored Higgs,
which may spoil the success of the gauge coupling unification 
and/or result in gauge couplings in the non-perturbative region, 
 as noted in \S\ref{E6Higgs}.
Taking account of the above conflict,  we take $k=-1$ here. 
Thus, the relation $k=f-\Delta f$ leads to $f=-\half$.

Then, the condition for the vacuum structure (\ref{vacuum4}), 
 Eq.(\ref{4thCase}), and the condition (\ref{4th}) give a relation 
\bequ
 2k-f \geq -3\bar c-2f, 
\label{cbar}
\eequ
that is, $\bar c\geq\frac{5}{6}$.
Because $3k=\bar c+f+\phi$, 
larger $\bar c$ with $k$ and $f$ fixed leads to smaller $\phi$
and thus a larger mass of the colored Higgs, 
 which leads to less natural
explanation for the success of the gauge coupling unification. 
Therefore, we adopt 
$\bar c=1$, which leads to 
 $\phi=-\frac{7}{2}$ and $c=-4$.

Now, Eq.(\ref{4thCase}) and $\bar f \geq -3\bar c-2f$ lead to 
 $-2\leq \bar f\leq-\frac{3}{2}$.
And we take $\bar f=-2$.

As for $a$, both $a=-1/2$ and $a=-1$ are possible.
The former yields relatively large FCNC processes 
because $\VEV{A}$ breaks the SU(2)\sub E symmetry 
 which guarantees the universality of masses of three 
 ${\bf \bar 5}$ sfermion fields. 
Therefore, we take $a=-1$, though the gauge couplings may become in 
the non-perturbative region.

Table \ref{ContentStrategy12} shows an example (and those of the 
 strategy 2).
\begin{table}
\[
\begin{array}{|c|c|c|} 
\hline
                  &   \mbox{non-vanishing VEV}  
                  &   \mbox{vanishing VEV} \\
\hline 
{\bf 78}          &   A(a=-1;-)        & A'(a'=5;-)  \\
{\bf 27}          &   \Phi(\phi=-7/2;+),\,  C(c=-4,-7/2,-3,-5/2;+) 
                  &   C'(c'=8;-)  \\
                 &&   \Psi_3 (\psi_3=7/4;+),\,\Psi_a (\psi=17/4;+)\\
{\bf \cc{27}}     &   \bar C_a(\bar c=1;+) 
                  &   \bar C'(\bar c'=11/2;-) \\
{\bf 1}           &   \bar F_a (\bar f=-2;+),\,F_a(f=-1/2;+) &\\
                  &   \Theta (\theta=-1/2;+),\, Z_i(z_i=-3/2;-) &  \\
                  &   Z_C(z_C=\mbox{-},-1/2,-1,-3/2;+) &  \\
\hline
\end{array}
\]
\caption{
Examples of the charge assignments for the first and second strategies : 
This charge assignment yields $r=\half+\frac{c-\phi}{2}$ and 
 $l=-5-c$.
When $c\geq\phi$, we impose an additional $Z_2$ symmetry 
 and introduce a singlet field $Z_C$. 
}
\label{ContentStrategy12}
\end{table}
The sign $\pm$ denotes the parity under the additional $Z_2$ symmetry 
 that plays the same role as does the $Z_2$ symmetry 
 introduced in Table \ref{E6Content}. 

$(a',c',\bar c')$ are determined by the smallest values that 
 allow $A'A^5$, $A'\Phi\bar CF$, $C'A\Phi\Phi$ and 
 $\bar C'ZC$.
We set $z$ to be the largest value which forbids $C'Z\Phi\Phi$.
Here, the matter fields ($\Psi_3$, $\Psi_a$) are also shown.
From their charges, we can find that $l=-1$ and 
 $\Psi\Psi\Phi C\bar C\bar F$ is allowed, which introduces SU(2)\sub R 
 breaking in the Yukawa couplings.
Note that only the matter fields have odd quarter integer charges, 
 and therefore they always appear in pairs, 
 which guarantees the $R$-parity to be automatically conserved.
The effective mass of the colored Higgs is given 
 as $\lambda^{-19/2}$. This value is not so much different from that 
 of the model in Table \ref{E6Content}.
And the parameter $\delta_{\bf10}$ and $\delta_{\bf{\bar5}}$ 
 are estimated as 
\bequ
\delta_{\bf10} = 
 \begin{pmatrix}
   \lambda^2 & \lambda^3 & \lambda^3 \\
   \lambda^3 & \lambda^2 & \lambda^2 \\
   \lambda^3 & \lambda^2 & 1 
 \end{pmatrix}\,,\quad
\delta_{\bf{\bar5}} = 
 \begin{pmatrix}
   \lambda^2     & \lambda^{2+r} & \lambda^3 \\
   \lambda^{2+r} & \lambda^2     & \lambda^{3-r} \\
   \lambda^3     & \lambda^{3-r} & \lambda^2 
 \end{pmatrix},
\label{e6su2delta}
\eequ
which are obtained from the non-renormalizable interactions, for example,
\begin{equation}
\dtf S^\dagger S[|\Psi F|^2+|\Psi F^\dagger|^2
+\Psi^\dagger A^2\Psi].
\end{equation}
Off diagonal elements of $\delta$ are indeed small, 
 but are still too large to suppress
 the FCNC processes sufficiently,  
 and we must require other mechanisms that 
suppresses the above non-renormalizable interactions with the spurion 
field $S$ or that gives universal sfermion masses.

\subsubsection{Strategy 2: $\phi\leq c$ $(r\geq \half)$}
\label{strategy2}

Next, let us examine the second strategy.
With the aid of an additional discrete symmetry, we can 
 forbit the interaction $C'AC\Phi$ while we allow $C'A\Phi\Phi$ 
 even when $\phi\leq c$ which always leads to $r\geq\half$.
For example, consider another $Z_2$ symmetry that only $C$ and 
$Z_C$ have odd parity. Here $z_C<\phi-c$ is required to forbid
$C'AC\Phi Z_C$ and to allow $C'A\Phi\Phi$.

In this analysis, we also introduce half-integer charges.
Then, as in the previous strategy, we set 
 $(k, f, \bar c,\phi,\bar f,a)=(-1,-\half,1,-\frac{7}{2},-2,-1)$.
For these charges, Eq.(\ref{4thCase}) requires 
 $c\leq k-\bar c+f=-\frac{5}{2}$, 
 which leads to $c=-\frac{7}{2},-3,-\frac52$.
$(a',c',z)$ are also determined as in the previous 
 strategy.
We set $z_C$ as the largest negative value satisfying $\phi>c+z_C$, and 
 $\bar c'$ is determined to allow $\bar C'ZCZ_C$.%
\footnote{
Another choice is to assign odd parity to $\bar C$ and 
 determine $\bar c'$ that $\bar C'ZC$ is allowed. 
This choice is convenient for embedding the model into 
 \E6$\times$SU(3)\sub H model, 
 and we consider this possibility later.
}
Table \ref{ContentStrategy12} summarizes the charge assignments.

Here, the matter fields ($\Psi_3$, $\Psi_a$) are also shown.
From their charges, 
 we can find that $l=-3/2,-2,-5/2$ and that 
 $\Psi\Psi\Phi CZ_C\bar C\bar F$ is allowed, which is
 important to introduce SU(2)\sub R breaking in Yukawa couplings.
Again, the $R$-parity is automatically conserved.
The effective mass of the colored Higgs is given 
 as $\lambda^{-19/2}$.
And the parameter $\delta_{\bf10}$ and $\delta_{\bf{\bar5}}$ 
 are given by the same expression as in Eqs.(\ref{e6su2delta}). 

\subsection{\E6$\times$SU(3)\sub H$\times$U(1)\sub A Models}
\label{e6su3}

In this subsection, we consider \E6$\times$SU(3)\sub H model, 
 where three $\Psi$'s 
 and three ${\bf{\cc{27}}}$'s ($\bar C,\bar\Phi,\bar C'$) from 
 a triplet and an anti-triplet of SU(3)\sub H, respectively.
In this case, the anomaly of SU(3)\sub H of the matter 
 sector is cancelled by that of the three ${\bf{\cc{27}}}$'s, 
 in contrast to the case of \S\ref{HorizHiggs} where some additional
fields must be introduced for the anomaly cancellation.

In order to yield the large top Yukawa coupling, SU(3)\sub H should be 
 broken near the cutoff scale.
Suppose that SU(3)\sub H is broken to SU(2)\sub H at the cutoff 
scale by developing the VEVs
$\VEV{E}=\VEV{\bar E}\sim \lambda^{-\frac{1}{2}(e+\bar e)}=1$, 
 \ie\ $e+\bar e=0$. 
Then it can be shown that the effective theory with
SU(2)\sub H can be identified with a certain SU(2)\sub H model 
that have the same U(1)\sub A charges as the effective charges 
in the effective SU(2)\sub H model. 
The essential point is that all the interactions in the SU(2)\sub H
model can be induced from the interactions in the SU(3)\sub H model.
This is a characteristic feature of models where a symmetry 
 is broken at the cutoff scale. 
For example, $\lambda^{2\psi_3+\phi}\Psi_3\Psi_3\Phi$ in SU(2)\sub H model
can be obtained
from the interaction 
$\lambda^{2\psi+2\bar e+\phi}\Psi \bar E\Psi \bar E\Phi$
by developing the VEV $\VEV{\bar E}\sim 1$.
Note that the coefficient of the effective interaction 
is determined by the effective charges, that is, 
$\lambda^{2\psi+2\bar e+\phi}\VEV{\bar E}^2
 \sim \lambda^{2\tilde \psi_3+\phi}$,
where $\tilde \psi_3$ is the effective charge of $\Psi_3$ of 
the effective SU(2)\sub H model.
Therefore, the total charge of an interaction 
in the SU(3)\sub H models is nothing but 
the total effective charge of the corresponding interaction
in the effective SU(2)\sub H model because 
SU(3)\sub H is broken at the cutoff scale. 
Thus, if a term is forbidden 
by the SUSY-zero mechanism in the SU(3)\sub H model,
the corresponding term in the SU(2)\sub H model 
is also forbidden by the SUSY-zero mechanism. 
Hence, 
the effective SU(2)\sub H model can be described
 by the SU(2)\sub H model.
Conversely, if an SU(2)\sub H model is found 
in which the U(1)\sub A charges are 
the same as the effective charges of an SU(3)\sub H model, 
then an SU(3)\sub H model can be found straightforwardly. 
Note that for SU(2)\sub H models, the arguments 
in the previous section can be
applied, which makes the discussion much simpler.

\subsubsection{SU(2)\sub H models for SU(3)\sub H models}
\label{Su2ForSU3}

In order to extend the horizontal symmetry to SU(3)\sub H, the difference
$m=\psi-\psi_3$ is required to be the same as $\bar m\equiv \bar c'-\bar c$.
The charge assignments shown 
 in Table \ref{ContentStrategy12} 
 have discrepancy between  
 $m = \psi-\psi_3 = \frac{5}{2}$ and 
 $\bar m=\bar c'-\bar c = \frac{9}{2}$.
Note that phenomenologically viable value of $m$ is around 
 $\frac{5}{2}$-$3$.
Thus, models with smaller $\bar m$ is needed. 
Since $(f,z_i,z_C,\bar c')$ are set as 
 $(k+\Delta f,a-\half,\phi-c-\half,-c-z_C-z_i)$ in \S\ref{e6su2},
  $\bar m$ is written as 
\bequ
 \bar m = \Ls\half-\phi-\Ls a-\half\Rs\Rs-\bar c 
        = 2\times\half-3k+f-a
        = 2\times\half+\Delta f-2k-a. 
\label{mbar}
\eequ
This means that in order to obtain a smaller $\bar m$, 
 larger $a$ and $k$ are required. 
We can construct such models (see Table \ref{ContentSU2forSU3-1}), 
 although the suppression of the FCNC processes becomes milder: 
\bequ
\delta_{\bf10} \sim 
 \begin{pmatrix}
   \lambda & \lambda^{2} & \lambda^3 \\
   \lambda^{2} & \lambda & \lambda^2 \\
   \lambda^3     & \lambda^2     & 1 
 \end{pmatrix}\,,\quad
\delta_{\bf{\bar5}} \sim 
 \begin{pmatrix}
   \lambda & \lambda^{1+r} & \lambda^2 \\
   \lambda^{1+r} & \lambda       & \lambda^{2-r} \\
   \lambda^2     & \lambda^{2-r} & \lambda 
 \end{pmatrix}.
\eequ

\begin{table}[ht]
\[
\begin{array}{|c|c|c|} 
\hline
                  &   \mbox{non-vanishing VEV}  
                  &   \mbox{vanishing VEV} \\
\hline 
{\bf 78}          &   A(a=-1/2;-)        & A'(a'=3;-)  \\
{\bf 27}          &   \Phi(\phi=-7/2;+),\,  C(c=-4,-7/2;+) 
                  &   C'(c'=15/2;-)  \\
                 &&   \Psi_3 (\psi_3=7/4;+),\,\Psi_a (\psi=19/4;+)\\
{\bf \cc{27}}     &   \bar C_a(\bar c=2;+) 
                  &   \bar C'(\bar c'=5;-) \\
{\bf 1}           &   \bar F_a (\bar f=-2;+),\,F_a(f=0;+) &\\
                  &   \Theta (\theta=-1/2;+),\, Z_i(z_i=-1;-) &  \\
                  &   Z_C(z_C=\mbox{-},-1/2;+) &  \\
\hline
\end{array}
\]
\caption{
Examples of the charge assignments of SU(2)\sub H models that 
 can be embedded into SU(3)\sub H models : 
When $c\geq\phi$, we impose an additional $Z_2$ symmetry 
 and introduce a singlet field $Z_C$. 
}
\label{ContentSU2forSU3-1}
\end{table}

In order to improve the suppression of the FCNC processes, we have to 
change some assumptions.
If we employ the other choice of $Z_2$ parity introduced in \S\ref{strategy2}
for $\bar c'$ as in the footnote there, 
 we can set $\bar c'=-c-z_i$ instead of 
 $\bar c'=-c-z_C-z_i$, so that 
 $\bar C'\Ls A+Z\Rs C$ is allowed. 
This can reduce $\bar m$.  
We thus can construct a model that can be embedded 
 into an SU(3)\sub H model with suppression of 
 the FCNC process to the same level as in the models introduced 
 in \S\ref{strategy2} (See Table \ref{SU2ForSU3-2}). 
Actually, the parameters $\delta_{\bf 10}$ and $\delta_{\bf \bar 5}$
have the same expression as in the Eqs.(\ref{e6su2delta}).
\begin{table}
\[
\begin{array}{|c|c|c|} 
\hline
                  &   \mbox{non-vanishing VEV}  
                  &   \mbox{vanishing VEV} \\
\hline 
{\bf 78}          &   A(a=-1;-)        & A'(a'=5;-)  \\
{\bf 27}          &   \Phi(\phi=-7/2;+),\,  C(c=-5/2;+) 
                  &   C'(c'=8;-)  \\
                 &&   \Psi_3 (\psi_3=7/4;+),\,\Psi_a (\psi=19/4;+)\\
{\bf \cc{27}}     &   \bar C_a(\bar c=1;+) 
                  &   \bar C'(\bar c'=4;-) \\
{\bf 1}           &   \bar F_a (\bar f=-2;+),\,F_a(f=-1/2;+) &\\
                  &   \Theta (\theta=-1/2;+),\, Z_i(z_i=-3/2;-) &  \\
                  &   Z_C(z_C=-3/2;+) &  \\
\hline
\end{array}
\]
\caption{
Another example of the charge assignments of SU(2)\sub H models that 
 can be embedded into SU(3)\sub H models : 
This charge assignment yields $r=1$ and 
 $l=-5/2$.
}
\label{SU2ForSU3-2}
\end{table}

\subsubsection{SU(3)\sub H models}

Now, we treat SU(3)\sub H models. 
The Higgs content is summarized in Table \ref{HiggsContentE6SU3}.
\begin{table}[th]
\[
\begin{array}{|c|c|c|} 
\hline
           &   \mbox{non-vanishing VEV}  & \mbox{vanishing VEV} \\
\hline 
  {\bf 78} &   A\Ls \VEV{{\bf45}_A}\sim\lambda^{-a}\Rs  & A'      \\
  {\bf 27} &   \Phi\Ls \VEV{{\bf1}_\Phi}
                   \sim\lambda^{-(\phi-\Delta\phi)}\Rs,\ 
               C\Ls \VEV{{\bf16}_C}\sim\lambda^{-(c-\Delta c)}\Rs 
           &  C'  \\
  {\bf\cc{27}} & \bar C^\alpha\Ls 
                    \VEV{{\bf\cc{16}}_{\bar C_1}}
               \sim \lambda^{-(\bar c+\Delta c-\Delta f-\Delta e/2)},\ 
                    \VEV{{\bf\cc{1}}_{\bar C_2}}
               \sim \lambda^{-(\bar c+\Delta\phi+\Delta f-\Delta e/2)} \Rs 
               & \\
  {\bf1} & F_\alpha\Ls \VEV{F_2}\sim\lambda^{-(f-\Delta f+\Delta e/2)}\Rs,\ 
           \bar F^\alpha \Ls \VEV{\bar F_2}
                    \sim\lambda^{-(\bar f+\Delta f-\Delta e/2)}\Rs& \\
         & E_\alpha\Ls \VEV{E_3}\sim\lambda^{-(e-\Delta e)}\Rs,\ 
           \bar E^\alpha \Ls \VEV{\bar E_3}
                    \sim\lambda^{-(\bar e+\Delta e)}\Rs& \\
\hline
\end{array}
\]
\caption{
The Higgs content of $E_6\times SU(3)_H\times U(1)_A$ models 
 expect for singlets:
Here SU(3)\sub H triplets and anti-triplets are denoted 
 by the lower and upper index $\alpha$, respectively. 
One or more discreate symmetries are introduced when needed.
}
\label{HiggsContentE6SU3}
\end{table}
Each component of the triplet $\Psi_\alpha$ and 
 the anti-triplet $\bar C^\alpha$
 can be chosen as 
\beqn
 (\Psi_1, \Psi_2, \Psi_3) &\sim& (\Psi EF, \Psi\bar F, \Psi E) 
\label{triplet}\\
 (\bar C_1, \bar C_2, \bar C_3) &\sim& 
    (\bar C\bar E\bar F, \bar CF, \bar CE)
\label{antitriplet},
\eeqn
and the effective charge of each element is given as
\beqn
 \s\psi &=& (\psi+\Delta f+\Delta e/2, \psi-\Delta f+\Delta e/2, 
           \psi-\Delta e) \\
 \s{\bar c} &=& (\bar c-\Delta f-\Delta e/2, \bar c+\Delta f-\Delta e/2, 
                 \bar c+\Delta e) .
\eeqn
This means that, providing $e=-\bar e=\Delta e$ and 
 integrating out $E$ and $\bar E$, 
 we get an SU(2)\sub H model where 
 $(\psi, \psi_3, \bar c, \bar c', \bar f, f)$ are given as 
 $(\psi+e/2, \psi_3-e, \bar c-e/2, \bar c'+e, \bar f-e/2, f+e/2)$ 
 in terms of the charges in the SU(3)\sub H model.%
\footnote{
As for the $Z_2$-parities discussed below, 
 we can find those of each component 
 from Eqs.(\ref{triplet}) and (\ref{antitriplet}). 
In addition, for example,  $\bar C\bar E\bar C\Phi$ and 
 $\bar C\bar E\bar FZ_C$ (whose charges are usually smaller 
 than that of $\bar C\bar E\bar F$) 
 may pick up $C_1$ component with 
 the opposite parity. 
} 
Conversely, we can construct an SU(3)\sub H model with $e=-\bar e=2$ 
 as shown in Table \ref{ContentSU3} from an SU(2)\sub H model in 
 Table \ref{SU2ForSU3-2}. 
\begin{table}[ht]
\[
\begin{array}{|c|c|c|} 
\hline
                  &   \mbox{non-vanishing VEV}  
                  &   \mbox{vanishing VEV} \\
\hline 
{\bf 78}          &   A(a=-1;-)        & A'(a'=5;-)  \\
{\bf 27}          &   \Phi(\phi=-7/2;+),\,  C(c=-5/2;+) 
                  &   C'(c'=8;-)  \\
                 &&   \Psi_\alpha (\psi=15/4;-) \\
{\bf \cc{27}}     &   \bar C^\alpha(\bar c=2;+) 
                  & \\
{\bf 1}           &   F_\alpha (f=-3/2;+),\,
                      \bar F^\alpha(\bar f=-1;-) &\\
                  &   E_\alpha (e=2;-),\,
                      \bar E^\alpha(\bar e=-2;-) &\\
                  &   \Theta (\theta=-1/2;+),\, Z_i(z_i=-3/2;-) &  \\
                  &   Z_C(z_C=-3/2;+) &  \\
\hline
\end{array}
\]
\caption{
An example of the charge assignments of SU(3)\sub H models. 
}
\label{ContentSU3}
\end{table}
Here, parity assignment of the additional $Z_2$ symmetry 
 for (anti)triplet fields 
 $(\Psi,\bar C,F,\bar F,E,\bar E)$ is 
 $(-,+,+,-,-,-)$, 
 so that $\bar C_a\,(a=1,2)$ and $\Psi_\alpha\,(\alpha=1,2,3)$ 
 have even parity while $\bar C_3$ has odd parity, 
 and the others have the same parity as in the SU(2)\sub H model. 
This parity plays essentially the same role 
 as that in the SU(2)\sub H model in Table \ref{SU2ForSU3-2}. 
The FCNC processes are suppressed to the same level as in models 
 in Table \ref{ContentStrategy12}.
This charge assignment yields $r=1$ and 
 $l=-5/2$.
Odd quarter integer charge of the matter field ($\Psi_\alpha$) 
 guarantees that the $R$-parity is automatically conserved.

\subsection{Summary and Discussion} 

Here, 
we have investigated \E6 SUSY-GUTs with 
 an anomalous U(1) symmetry and an SU(2)\sub H or SU(3)\sub H 
 horizontal symmetry, where some of GUT-breaking Higgs fields belong to 
 non-trivial representations of the horizontal symmetry. 
We have found it possible to unify the Higgs sectors for the GUT symmetry and
the horizontal symmetry. 
It is interesting that for SU(3)\sub H models, SU(3)\sub H
gauge anomaly is cancelled between the triplet matter $\Psi_\alpha$ and the 
anti-triplet Higgs $\bar C^\alpha$.

Unfortunately, the unification of the Higgs sectors of the GUT symmetry and
the horizontal symmetry yields in too rapid FCNC processes. 
This is because in the \aGUT\ scenario, 
 \E6 breaking scale is difficult to be smaller than 
 $\lambda^2$, which
 is the sufficient value for suppressing the FCNC processes. 
This fact may mean that another mechanism is required to realize the 
universality of sfermion masses, or that the fields in the Higgs sector of
the GUT symmetry do not have non-trivial quantum numbers under the horizontal 
symmetry. 
However, we hope that the arguments in this section give a hint 
to find out a realistic \E8 unification.

\chapter{Summary}
\label{Summary}

In this thesis, we have introduced a very interesting scenario. 
This is a kind of SUSY-GUT scenario that employs an anomalous 
 U(1) gauge symmetry, whose anomaly is assumed to be canceled via 
 the Green-Schwarz mechanism. 
With the aid of this U(1) symmetry, almost all the problems of 
 usual SUSY-GUTs can be solved simultaneously: 
\bit
 \item The Doublet-Triplet Splitting problem can be solved 
       via the Dimopoulos-Wilczek type of VEV, with no fine-tuning. 
 \item The success of the gauge coupling unification 
       in \msGUT\ is naturally explained. 
 \item The nucleon decay via dimension 5 operators can be suppressed, 
       while that via dimension 6 operators is predicted to be 
       enhanced compared to usual SUSY-GUTs. 
 \item Realistic fermion Yukawa matrices can be reproduced. 
       In particular, the neutrino bi-large mixings can be realized 
       in the (almost) minimal matter content.
\eit
Surprisingly, these consequences are led from natural assumptions:
\bit
 \item We introduce the ``generic interaction''. 
       Namely, we introduce all the possible interaction terms that 
       respect the symmetry of the model, including 
       non-renormalizable terms. 
       In addition, their coupling constants are $\order1$ 
       in the unit of the cutoff scale $\Lambda$, and we do not 
       introduce hierarchical structures. 
 \item We assume one of the vacua shown in (\ref{VEV}) is selected 
       as the vacuum of the model.
\eit
This means that the definition 
 of a model is given, except for a few parameters, by the 
 definition of a symmetry: 
 a symmetry group, matter content and representations 
 of the matter fields under the symmetry. 
And thus, the parameters of the models are essentially the anomalous 
 U(1) charges, whose number is the same as that of the superfields
 ($\sim\order{10}$). 
We have illustrated these characters by using concrete models 
 based on SO(10) or \E6.

Also We have examined the role of the horizontal symmetry 
 in \aGUT\ scenario. 
One of the motivations to introduce a horizontal symmetry 
 is to solve the SUSY-flavor problem. 
If we construct models so that all the Higgs that is charged 
 in \E6 are singlets under the horizontal symmetry, 
 the SUSY-flavor problem may be solved in \E6 models. 
For this purpose, however, we have to assume the $D$ terms of the 
 horizontal symmetry are very suppressed compared to $F$ terms. 
If we aim to construct models so that some ${\bf27}$ (${\bf\cc{27}}$) 
 Higgs also have nontrivial representations under the horizontal 
 symmetry, the SUSY-flavor problem is not solved sufficiently 
 but ameliorated. 
Nevertheless, it is still worthwhile to introduce a horizontal 
 symmetry, because it can realize a further unification of matter fields. 
We hope that the study introduced in \S\ref{HorizHiggs}
 gives a hint in finding a realistic \E8 unification model.

\section*{Acknowledgment}

I am very grateful to my adviser N. Maekawa for his instruction, 
 stimulating discussions, collaborations and reading 
 this thesis. 
This thesis is based on the works in collaboration with him. 

I would like to thank H. Kogetsu, I. Sonobe and S. Teraguchi 
 for their encouragements and continuous support. 
Also I would like to thank H. Hata, M. Fukuma, T. Kobayashi 
 and M. Bagnoud for their kind comments on our English.

Finally, I would like to thank everyone who have supported me.

\appendix

\chapter{Factorization}
\label{factorization}

  As mentioned in \S\ref{E6Higgs}, the naive extension
of DTS in the SO(10) models into the \E6 models does not work.
  In the SO(10) DTS, the interaction
$(A^\prime A)_{\bf 54}(A^2)_{\bf 54}$ plays an essential role.
  In the \E6 models, however, the term $A^\prime A^3$
does not include the interaction 
$({\bf 45}_{A'}{\bf 45}_A)_{\bf 54}({\bf 45}_A^2)_{\bf 54}$.
  Therefore the superpotential
\begin{equation}
W_{A^\prime} =  \lambda^{a^\prime+a}A^\prime A
              + \lambda^{a^\prime+3a} \Ls(A^\prime A)_{\bf 1}
                                             (A^2)_{\bf 1}
                                      + (A^\prime A)_{\bf 650}
                                              (A^2)_{\bf 650}\Rs
\end{equation}
does not realize the DW VEV naturally.
  Here, we show that the term $A^\prime A^3$ of \E6 actually does not
include the interaction $({\bf 45}_{A'}{\bf 45}_A)_{\bf 54}
({\bf 45}_A^2)_{\bf 54}$ of SO(10).

  The VEV of SO(10) adjoint Higgs can be represented as 
$\VEV{A}=\tau_2\times {\diag}(x_1,x_2,x_3,x_4,x_5)$, thanks to the 
SO(10) rotation and D-flatness condition.
  In this gauge,
\beqn
  A^\prime A &=& 2 \sum_i x^\prime_i x_i ,\\
  (A^\prime A)_{\bf 54}(A^2)_{\bf 54} &=& 2 \sum_i x^\prime_i {x_i}^3 
  -\frac{2}{5}\left(\sum_i x'_ix_i\right)\left(\sum_jx_j^2\right).
\eeqn
  In the same manner, the VEV of \E6 adjoint Higgs can be
represented in the form $\VEV{{\bf 1}_A}=y$, $\VEV{{\bf 16}_A}=
\VEV{{\bf \overline{16}}_A}=0$, $\VEV{{\bf 45}_A}=\tau_2\times
{\rm diag.}(x_1,x_2,x_3,x_4,x_5)$.
  In this gauge, the VEV $\VEV A$ can be represented
as $27 \times 27$ matrix as 
\begin{equation}
  \VEV{A} = \left(
  \begin{array}{ccc}
    {2\over \sqrt3}y & 0 & 0 \\
    0 & \theta^{MN}{T_{16}}^{MN}
        + {1\over2\sqrt3}y{\bf 1}_{16} & 0 \\
    0 & 0 & \theta^{MN}{T_{10}}^{MN}
            - {1\over\sqrt3}y{\bf 1}_{10} \\
\end{array}
\right)\label{A-matrix}.
\end{equation}
  Here, ${T_{i}}^{MN}$ is the $i \times i$ matrix representation of
SO(10) generators and the summation of the indices $M$ and $ N$ is 
understood from 1 to 10 with  $M>N$. 
  Also, ${\bf 1}_{i}$ is the $i \times i$ unit matrix. 
  Explicitly, we have 
\beqn
  ({T_{10}}^{MN})_{KL} &=& -i( \delta^M_K\delta^N_L
                             - \delta^M_L\delta^N_K ), \\
  ({T_{16}}^{MN})_{\alpha\beta}
                       &=& {1\over2}(\sigma^{MN})_{\alpha\beta} \nn\\
                       &=& {1\over4i}([\gamma^M , \gamma^N]P_{\rm R})
                                     _{\alpha\beta}, \\
  \theta^{MN} &=& \Lm
    \begin{array}{cl}
      x_n & M+1=N=2n,\ (n=1,\cdots,5) \\
      0   & {\rm otherwise},
    \end{array}
  \right.
\eeqn
where the $\gamma^M$ are SO(10) $\gamma$-matrices and $P_{\rm R}$ is the
right-handed projector, which can be written as 
\beqn
  \gamma^1 &=& \tau_1 \otimes {\bf 1} \otimes {\bf 1}
                      \otimes {\bf 1} \otimes {\bf 1},  \\
  \gamma^2 &=& \tau_3 \otimes {\bf 1} \otimes {\bf 1}
                      \otimes {\bf 1} \otimes {\bf 1},  \\
  \gamma^3 &=& \tau_2 \otimes \tau_1 \otimes {\bf 1}
                      \otimes {\bf 1} \otimes {\bf 1},  \\
  \gamma^4 &=& \tau_2 \otimes \tau_3 \otimes {\bf 1}
                      \otimes {\bf 1} \otimes {\bf 1},  \\
  \gamma^5 &=& \tau_2 \otimes \tau_2 \otimes \tau_1
                      \otimes {\bf 1} \otimes {\bf 1},  \\
  \gamma^6 &=& \tau_2 \otimes \tau_2 \otimes \tau_3
                      \otimes {\bf 1} \otimes {\bf 1},  \\
  \gamma^7 &=& \tau_2 \otimes \tau_2 \otimes \tau_2
                      \otimes \tau_1 \otimes {\bf 1},  \\
  \gamma^8 &=& \tau_2 \otimes \tau_2 \otimes \tau_2
                      \otimes \tau_3 \otimes {\bf 1},  \\
  \gamma^9 &=& \tau_2 \otimes \tau_2 \otimes \tau_2
                      \otimes \tau_2 \otimes \tau_1,   \\
  \gamma^{10} &=& \tau_2 \otimes \tau_2 \otimes \tau_2
                         \otimes \tau_2 \otimes \tau_3,\\
  \gamma^{11} &=& i\gamma^1\gamma^2\gamma^3\gamma^4\gamma^5
                  \gamma^6\gamma^7\gamma^8\gamma^9\gamma^{10}\nn\\
              &=& \tau_2 \otimes \tau_2 \otimes \tau_2
                      \otimes \tau_2 \otimes \tau_2,   \\
  P_{\rm R} &=& {1+\gamma^{11}\over2}.
\eeqn
In this basis, we have 
\beqn
  \theta^{MN}{T_{16}}^{MN} &=& -{1\over2}(
                    x_1 \tau_2 \otimes {\bf 1} \otimes {\bf 1}
                      \otimes {\bf 1} \otimes {\bf 1}  \nn\\
                &&\hspace{2mm}+ x_2 {\bf 1} \otimes \tau_2 \otimes {\bf 1}
                      \otimes {\bf 1} \otimes {\bf 1}  \nn\\
                &&\hspace{2mm}+ x_3 {\bf 1} \otimes {\bf 1} \otimes \tau_2
                      \otimes {\bf 1} \otimes {\bf 1}  \nn\\
                &&\hspace{2mm}+ x_4 {\bf 1} \otimes {\bf 1} \otimes {\bf 1}
                      \otimes \tau_2  \otimes {\bf 1}  \nn\\
                &&\hspace{2mm}+ x_5 {\bf 1} \otimes {\bf 1} \otimes {\bf 1}
                      \otimes {\bf 1} \otimes \tau_2
                                        ) P_{\rm R}, \\
                         &\equiv&  B, \\
  \theta^{MN}{T_{10}}^{MN} &=& \tau_2 \otimes {\rm diag.}
                               (x_1,x_2,x_3,x_4,x_5)  \\
                         &\equiv&  C.
\eeqn

  Before beginning the calculation, we should determine what coupling
can occur in the term $A^\prime A^3$ of \E6.
  Because 
${\bf 78}\times{\bf78} = {\bf1}_{\rm s}+{\bf78}_{\rm a}
                       + {\bf650}_{\rm s}+{\bf2430}_{\rm s}
                       + {\bf2925}_{\rm a}$,
$A^\prime A^3 \ni (A^\prime A)_{\bf 1}(A^2)_{\bf 1},
                  (A^\prime A)_{\bf 650}(A^2)_{\bf 650},
                  (A^\prime A)_{\bf 2430}(A^2)_{\bf 2430}$.
  On the other hand, because of the completeness,
\bequ
  (A_1A_2)_{\bf 2430}(A_3A_4)_{\bf 2430} =
               \sum_{I={\bf 1,78,650,2430,2925}}
               \lambda_I(A_1A_4)_I(A_3A_2)_I.
\eequ
  Therefore, 
\bequ
  (A^\prime A)_{\bf 2430}(A^2)_{\bf 2430} =
               \sum_{I={\bf 1,650,2430}}
               \lambda_I(A^\prime A)_I(A^2)_I,
\eequ
which implies that the above three couplings are not independent, 
and it is sufficient to examine the first two. They are essentially 
described as ${\rm tr}A^\prime A{\rm tr}A^2$ and 
${\rm tr}A^\prime A^3$ in matrix language.
If the desirable coupling existed, it would apparently be included
only in $(A^\prime A)_{\bf 650}(A^2)_{\bf 650}$ and
${\rm tr}A^\prime A^3$.
Thus we can conclude that it does not exist if
${\rm tr}A^\prime A^3$ does not include $\sum_i x^\prime_i {x_i}^3$.

  From (\ref{A-matrix}), we find 
\beqn
  {\rm tr}A^\prime A &=& {4\over3}y^\prime y
                        +{\rm tr}_{16}\Ll B^\prime B
                              + {1\over2\sqrt3}B^\prime y
                              + {1\over2\sqrt3}y^\prime B
                              + {1\over12}y^\prime y \Rl  \nn\\
                     &&   +{\rm tr}_{10}\Ll C^\prime C
                              - {1\over\sqrt3}C^\prime y
                              - {1\over\sqrt3}y^\prime C
                              + {1\over3}y^\prime y \Rl   \nn\\
                     &=& \Ls {4\over3}+{16\over12}+{10\over3}\Rs
                                                      y^\prime y
                        +\Ls 16{1\over2^2}+ 2\Rs
                                         \sum_i x^\prime_i x_i \nn\\
                     &=& 6\Ls y^\prime y + \sum_i x^\prime_i x_i\Rs.
\eeqn
  Similarly,
\beqn
  {\rm tr}A^\prime A^3 &=& {16\over9}y^\prime y^3
                          +{\rm tr}_{16}\Ll B^\prime B^3
                                + 3{1\over12}\Ls B^\prime By^2
                                + y^\prime yB^2\Rs
                                + {1\over144}y^\prime y^3 \Rl  \nn\\
                      &&    +{\rm tr}_{10}\Ll C^\prime C^3
                                + 3{1\over3}\Ls C^\prime Cy^2
                                + y^\prime yC^2\Rs
                                + {1\over9}y^\prime y^3 \Rl  \nn\\
                       &=& {16\over9}y^\prime y^3  
                       +16\Ll{1\over2^4}\Ls3\sum_i x^\prime_i x_i
                                               \sum_i x_i^2
                                            - 2\sum_i x^\prime_i
                                               x_i^3\Rs\right.\nn\\
                        &&       + \left.{3\over12}{1\over2^2}
                                           \Ls\sum_i x^\prime_ix_iy^2
                                            + y^\prime y \sum_i
                                              x^\prime_ix_i\Rs
                               + {1\over144}y^\prime y^3 \Rl  \nn\\
                      &&    +\Ll 2\sum_i x^\prime_ix_i^3
                                + {3\over3}\Ls 2\sum_i
                                                x^\prime_ix_iy^2
                                            + y^\prime y 2\sum_i
                                              x^\prime_ix_i\Rs
                                + {10\over9}y^\prime y^3 \Rl  \nn\\
                      &=& 3\Ls y^\prime y + \sum_i
                                  x^\prime_ix_i  \Rs
                           \Ls y^2 + \sum_i x_i^2  \Rs  \nn\\
                      &=& {1\over12}{\rm tr}A^\prime A{\rm tr}A^2.
\eeqn
Thus, it is explicitly shown that 
 the desirable coupling does not exist because of 
 the group theoretical cancellation between the contributions from 
 the ${\rm tr}_{\rm 16}$ part and the ${\rm tr}_{\rm 10}$ part.\\

There are several solutions, and the simplest one is to use the term
${\overline \Phi}A^\prime A^3 \Phi$.
At first glance, it seems to have no effect, because
$\Phi {\overline \Phi}$ is written as
\begin{equation}
  \Phi {\overline \Phi} = \left(
  \begin{array}{ccc}
    \VEV{{\overline \Phi}\Phi} & 0 & 0 \\
    0 & 0_{16} & 0 \\
    0 & 0 & 0_{10} \\
  \end{array}
\right).
\end{equation}
  However this form is a special combination of
$(\Phi {\overline \Phi})_{\bf 1}, (\Phi {\overline \Phi})_{\bf 78}
$ and $(\Phi {\overline \Phi})_{\bf 650}$.
  In fact, we have 
\begin{eqnarray}
  \left(
  \begin{array}{ccc}
    \VEV{{\overline \Phi}\Phi} & 0 & 0 \\
    0 & 0_{16} & 0 \\
    0 & 0 & 0_{10} \\
  \end{array}
  \right) &=& {\VEV{{\overline \Phi}\Phi} \over 54} \Ll
  2\Ls
  \begin{array}{ccc}
    1 & 0 & 0 \\
    0 & {\bf1}_{16} & 0 \\
    0 & 0 & {\bf1}_{10} \\
  \end{array}
  \Rs + 3\Ls
  \begin{array}{ccc}
    4 & 0 & 0 \\
    0 & {\bf1}_{16} & 0 \\
    0 & 0 & -2\times{\bf1}_{10} \\
  \end{array}
  \Rs \right. \nn \\
  &&+ \left.\Ls
  \begin{array}{ccc}
    40 & 0 & 0 \\
    0 & -5\times{\bf1}_{16} & 0 \\
    0 & 0 & 4\times{\bf1}_{10} \\
  \end{array}
  \Rs\right],
\end{eqnarray}
where the three matrices on the r.h.s. are proportional to the SO(10) 
singlets of ${\bf 1}, {\bf 78}$ and ${\bf 650}$, respectively. 
Since the interactions for each 
representation have independent couplings, generically the cancellation
is not realized with no fine-tuning.

There are several other solutions for this problem.
The essential ingredient is the interaction between $A'A^3$ and 
some other operator whose VEV breaks \E6, 
because the cancellation
is due to a feature of the \E6 group. 
We now introduce some of these solutions.
\begin{itemize}
  \item  Allowing the higher-dimensional term $A^\prime A^5$. 
  Since $\VEV{A^2}$ breaks \E6, the cancellation can be avoided,
  which can be shown by a straightforward calculation.
  Since the number of solutions of the $F$-flatness conditions 
  increases, it becomes less natural to obtain a DW VEV. 
  But the number of vacua is still finite.
  \item  Introducing additional adjoint Higgs fields $B^\prime$ and 
       $B$, and giving $B$ the VEV pointing to an SO(10) singlet direction.
         Then $B$ plays the same role as the above 
         $(\Phi {\cc \Phi})_{\bf 78}$.
         Examining the superpotential
   \begin{equation}
      W=B'B+\bar\Phi B'\Phi,
   \end{equation}
         the desired VEV $\VEV{{\bf 1}_B}\neq 0$ and 
         $\VEV{{\bf 45}_B}= 0$ is easily obtained.
\end{itemize}

\chapter{Operators that induce mass matrices}
\label{E6HiggsSpectrum}

In this appendix, we give the operators that induce the mass matrices
of superheavy particles in \E6 models.

First, we examine the operator matrix $O_{24}$ 
of ${\bf 24}$ in SU(5), which induces the mass matrices
$M_I\ (I=X,G,W)$, 
\begin{equation}
O_{{\bf24}}=\bordermatrix{
 I\backslash \bar I  &    {\bf45}_A(-1)   &     {\bf45}_{A'}(4)           \cr
{\bf45}_A(-1)                &     0      &      A'A \cr
{\bf45}_{A'}(4)   &  A'A   &  {A'}^2       \cr
},
\end{equation}
where the numbers in the parentheses denote typical charges.

Next, we examine the operator matrix $O_{{\bf10}}$ of ${\bf 10}$ in SU(5),
which induces the mass matrices $M_I\ (I=Q, U^c, E^c)$, 
\begin{equation}
\bordermatrix{
I\backslash \bar I  &{\bf\cc{16}}_{\bar \Phi}(2) & 
{\bf\cc{16}}_{\bar C}(-2)&{\bf\cc{16}}_{A}(-1) &{\bf45}_{A}(-1) &
{\bf\cc{16}}_{\bar C'}(8)   & 
                    {\bf\cc{16}}_{\bar A'}(4) &  {\bf45}_{A'}(4)  \cr
{\bf16}_\Phi(-3) & 0 & 0 & 0& 0& \bar C'A\Phi  & \bar\Phi A'A\Phi
         & 0 \cr
{\bf16}_C(-6)& 0 & 0 &0 & 0& \bar C'AC &  0  & 0 \cr
{\bf16}_A(-1) & 0 & 0 & 0& 0& \bar C'A\Phi   & \bar \Phi A'A\Phi
   & 0 \cr
{\bf45}_{A}(-1) & 0 & 0 & 0 & 0& \bar C'AC  & 0 & 
       A'A  \cr
{\bf16}_{C'}(7) & \bar\Phi A C' & \bar CAC' &
        \bar \Phi AC' & \bar CAC' & 
        \bar C'C' & \bar\Phi A'C' & 
        \bar CA'C'  \cr
{\bf16}_{A'}(4) & \bar \Phi A'A\Phi & 0 & 
        \bar\Phi A'A\Phi & 0&\bar C'A'\Phi &  
        {A'}^2& \bar C{A'}^2\Phi  \cr
{\bf45}_{A'}(4) & 0 & 0 &0 & A'A
        & \bar C'A'C &  
        \bar\Phi {A'}^2C 
        & {A'}^2 \cr
},
\end{equation}
where we have given only one example, even if there are several 
corresponding operators.

Finally, we examine the operator matrix $O_{\bf\bar5}$ of ${\bf 5}$ and ${\bf \bar 5}$
in SU(5), which induces the mass matrices $M_I\ (I=L,D^c)$, 
\begin{equation}
O_{\bf\bar5}=\left(
    \begin{matrix} 0 & 0 & A_{\bf\bar5} \cr  
                   B_{\bf\bar5} & C_{\bf\bar5} & D_{\bf\bar5} \cr 
                   E_{\bf\bar5} & F_{\bf\bar5} & G_{\bf\bar5} \cr
    \end{matrix}\right),
\end{equation}
\begin{equation}
A_{\bf\bar5}=\bordermatrix{
 I\backslash \bar I   &  {\bf10}_{C'}(7)
                    & {\bf10}_{\bar C'}(8) &
                    {\bf\cc{16}}_{\bar C'}(8) &
                    {\bf\cc{16}}_{A'}(4)  \cr
{\bf10}_\Phi(-3) & C'A\Phi^2 & \bar C'(A+Z)\Phi & 
0  & 0 \cr
{\bf10}_C (-6) & 0 & \bar C'(A+Z)C & 0 & 0 \cr
{\bf16}_C(-6) & 0 & 0 & \bar C'(A+Z)C
   & 0 \cr
{\bf16}_A(-1)  & 0 & \bar C'AC & \bar C'A\Phi
   & A'A \cr
}, 
\end{equation}
\begin{equation}
B_{\bf\bar5}=
\bordermatrix{
 I\backslash \bar I  &{\bf10}_\Phi(-3) & {\bf10}_C(-6)& 
                    {\bf\cc{16}}_{\bar C}(-2) &{\bf\cc{16}}_{A}(-1)   \cr
{\bf10}_{\bar C} (-2)&
0 & 0& 0 & 0
 \cr
{\bf16}_\Phi (-3)& 0 & 0 &  0 & 0  \cr
{\bf10}_{\bar \Phi}(2) & 
0 & 0& 0 & 
\bar\Phi^2A^2\bar C   \cr
},
\end{equation}
\begin{equation}
C_{\bf\bar5}=\bordermatrix{
 I\backslash \bar I   &  {\bf10}_{\bar \Phi} (2)
                    & {\bf10}_{\bar C}(-2) &
                    {\bf\cc{16}}_{\bar \Phi} (2)\cr
{\bf10}_{\bar C} (-2) & \bar\Phi^2\bar C & 0 & 0 \cr
{\bf16}_\Phi (-3) & 0 & 0 & 0 \cr
{\bf10}_{\bar \Phi} (2) &
\bar\Phi^3 & \bar\Phi^2\bar C &
\bar\Phi^2\bar C \cr
},
\end{equation}
\begin{equation}
D_{\bf\bar5}=\bordermatrix{
 I\backslash \bar I   &  {\bf10}_{C'} (7)
                    & {\bf10}_{\bar C'} (8)&
                    {\bf\cc{16}}_{\bar C'} (8)&
                    {\bf\cc{16}}_{A'}(4)  \cr
{\bf10}_{\bar C} (-2) & \bar C(A+Z)C' &
\bar C'A\bar C\bar\Phi& \bar C'A\bar C^2 & 
\bar C^2A'A\bar\phi \cr
{\bf16}_\Phi (-3) & 0 & \bar C'A\bar\Phi C\Phi& 
   \bar C'(A+Z)\Phi
   & \bar\Phi A'A\Phi \cr
{\bf10}_{\bar \Phi} (2) &
\bar\Phi(A+Z)C' & \bar C'A\bar\Phi^2 &
\bar C'A\bar\Phi\bar C & 
 \bar \Phi^2A'\bar C \cr
}, 
\end{equation}
\begin{equation}
E_{\bf\bar5}=\bordermatrix{
 I\backslash \bar I  & {\bf10}_\Phi(-3)&{\bf10}_C (-6)&{\bf\cc{16}}_{\bar C}(-2)&
{\bf\cc{16}}_{A}(-1)\cr
{\bf10}_{C'}(7) & C'A\Phi^2 & 0 & 0 & \bar CAC' \cr
{\bf10}_{\bar C'}(8) &\bar C'(A+Z)\Phi &\bar C'(A+Z)C &
\bar C'(A+Z)\bar C^2 & \bar C'A\bar\Phi\bar C \cr
{\bf16}_{C'}(7) & 0 & 0 &\bar C(A+Z)C' & \bar\Phi AC' \cr
{\bf16}_{A'}(4)& 0 & 0 & 0 & A'A \cr
}, 
\end{equation}
\begin{equation}
F_{\bf\bar5}=\bordermatrix{
 I\backslash \bar I  & {\bf10}_{\bar \Phi}(2) &{\bf10}_{\bar C}(-2) &
{\bf\cc{16}}_{\bar \Phi}(2) \cr
{\bf10}_{C'}(7) &\bar\Phi(A+Z)C' & \bar C(A+Z)C' & 
   \bar \Phi\bar C C'\Phi \cr
{\bf10}_{\bar C'}(8) & \bar C'A\bar\Phi^2 & 
       \bar C'A\bar C\bar \Phi & 
       \bar C'A\bar\Phi\bar C \cr
{\bf16}_{C'} (7)& \bar\Phi^2AC'C &
      \bar C\bar\Phi AC'C & \bar\Phi(A+Z)C' \cr
{\bf16}_{A'} (4)& 0 & 0 & \bar\Phi A'A\Phi \cr
}, 
\end{equation}
\begin{equation}
G_{\bf\bar5}=\bordermatrix{
 I\backslash \bar I  & {\bf10}_{C'} (7)
                    & {\bf10}_{\bar C'}(8)  &
                    {\bf\cc{16}}_{\bar C'}(8) &
                    {\bf\cc{16}}_{A'}(4)  \cr
{\bf10}_{C'}(7)  & {C'}^2\Phi & 
  \bar C'C' &
  \bar C'\bar C C'\Phi  & \bar CA'C' \cr
{\bf10}_{\bar C'}(8)  &\bar C'C' & 
(\bar C')^2\bar \Phi & (\bar C')^2\bar C & 
\bar C'A'\bar C\bar\Phi \cr
{\bf16}_{C'} (7)& {C'}^2C & 
  \bar C'\bar\Phi C'C & \bar C'C' &  
\bar\Phi A'C' \cr
{\bf16}_{A'}(4)& C'A'\Phi C & 
\bar C'A'C & \bar C'A'\Phi &  
{A'}^2 \cr
}.
\end{equation}


\begin{thebibliography}{99}
\bibitem{U(1)}    E. Witten,  Phys. Lett. B {\bf 149}, 351 (1984);
                M. Dine, N. Seiberg, and E. Witten,
                  Nucl. Phys. B {\bf 289}, 589 (1987);
                J.J. Atick, L.J. Dixon, and A. Sen,
                  Nucl. Phys. B {\bf 292}, 109 (1987);
                M. Dine, I. Ichinose, and N. Seiberg,
                  Nucl. Phys. B  {\bf 293}, 253 (1987).
\bibitem{IR}  L. Ib\'a\~nez and G.G. Ross,
                  Phys. Lett. B{\bf 332} (1994) 100;\\
                  P. Bin\'etruy and P. Ramond,
                  Phys. Lett. B{\bf 350} (1995) 49;\\
                  E. Dudas, S. Pokorski and C.A. Savoy,
                  Phys. Lett. B{\bf 356} (1995) 45;\\
                  P. Bin\'etruy, S. Lavignac and P. Ramond,
                  Nucl. Phys. B{\bf 477} (1996) 353.
\bibitem{GS}    M. Green and J. Schwarz,
                  Phys. Lett. {\bf B149} (1984),117.

\bibitem{maekawa} N. Maekawa, Prog. Theor. Phys. {\bf 106} (2001)401; 
                    arXiv:hep-ph/0110276.
\bibitem{E6} N. Maekawa and T. Yamashita, Prog. Theor. Phys. {\bf 107},
                  1201 (2002).
\bibitem{reduced}  N. Maekawa and T. Yamashita, 
                 Prog. Theor. Phys. {\bf 110}, 93 (2003).
\bibitem{FlippedSO(10)}  N. Maekawa and T. Yamashita, 
                       Phys. Lett. B {\bf 567}, 330 (2003).
\bibitem{sliding}    N. Maekawa and T. Yamashita, 
                     Phys. Rev. D {\bf 68}, 055001 (2003).
\bibitem{1-loop} N. Maekawa, Prog. Theor. Phys. {\bf 107}, 597 (2002).
\bibitem{2-loop}  N. Maekawa and T. Yamashita, 
                  Prog. Theor. Phys. {\bf 108}, 719 (2002). 
\bibitem{NGCU}  N. Maekawa and T. Yamashita, 
                  Phys. Rev. Lett. {\bf90} (2003) 121801. 
\bibitem{BM}       M. Bando and N. Maekawa, Prog. Theor. Phys. {\bf 106}
                   (2001) 1255.
\bibitem{horiz}  N. Maekawa, Phys. Lett. B {\bf 561}, 273 (2003); 
                   arXiv:hep-ph/0304076; hep-ph/0402224. 
\bibitem{HorizHiggs}  N. Maekawa and T. Yamashita, 
                      JHEP {\bf 0407}, 009 (2004). 


\bibitem{atmos} Super-Kamiokande Collaboration,
                    Phys. Lett. B {\bf 436}, 33 (1998);
                    Phys. Rev. Lett. {\bf 81}, 1562 (1998); 
                K2K Collaboration,
                    Phys.Rev.Lett. {\bf 90} 041801 (2003); 
                    arXiv:hep-ex/0411038.
\bibitem{solar}   Super-Kamiokande Collaboration, Phys. Rev. Lett. 
                  {\bf 86}, 5656 (2001); Phys. Lett. B {\bf 539},
                  179 (2002);
                  SNO Collaboration, Phys. Rev. Lett. {\bf 89}, 
                  011301 (2002); 
                  {\bf 89}, 011302 (2002); {\bf 92}, 181301 (2004);
                  KamLAND Collaboration, Phys. Rev. Lett. {\bf 90}, 
                  021802 (2003); arXiv:hep-ex/0406035.



\bibitem{SKproton} Super-Kamiokande Collaboration, Phys. Rev. Lett.
                  {\bf 81}, 3319 (1998); {\it ibid} {\bf 83}, 1529 (1999).

\bibitem{SUSY} Joseph D. Lykken, arXiv:hep-th/9612114. 
\bibitem{Martin} Stephen P. Martin, arXiv:hep-ph/9709356. 

\bibitem{mu} N. Maekawa, Phys. Lett. {\bf B521} (2001) 42.


\bibitem{ProtonCancel} B. Bajc, P. F. Perez and G. Senjanovic, 
                       Phys.Rev. D{\bf 66}  075005 (2002).


\bibitem{KakizakiProton} M. Kakizaki and M. Yamaguchi, 
                         JHEP {\bf 0206},032 (2002). 


\bibitem{GIFT} K. Inoue, A. Kakuto and T. Takano,
                    Prog. Theor. Phys. {\bf 75} (1986) 664;\\
                  A. Anselm and A. Johansen, 
                    Phys. Lett. {\bf B200}, (1988) 331;\\ 
                  A. Anselm, Sov. Phys. JETP {\bf67}, (1988) 663; \\
                  Z.G. Berezhiani and G. Dvali, 
                    Sov. Phys. Lebedev. Inst. Rep. {\bf5}, (1989)55;\\
                  Z.G. Berezhiani, C. Csaki, and L. Randall, 
                    Nucl. Phys. {\bf B44}, (1995) 61;\\
                  M. Bando and T. Kugo, 
                    Prog. Theor. Phys. {\bf 109},(2003)87. 
\bibitem{orbifold}  Y. Kawamura, Prog. Theor. Phys. {\bf 103},(2000)613;
                   {\it ibid.} {\bf 105} (2001) 691;
                   {\it ibid.} {\bf 105} (2001) 999;
                   L.J. Hall and Y. Nomura, Phys. Rev. D {\bf64}, 055003 (2001)
                   ;{\it ibid} {\bf65}, 125012 (2002);{\it ibid} {\bf 66},
                    075004 (2002).

\bibitem{DW}       S. Dimopoulos and F. Wilczek, NSF-ITP-82-07;
                   M. Srednicki, Nucl. Phys. {\bf B202} (1982) 327.

\bibitem{SS}   E. Witten, Phys. Lett. {\bf B105} (1981) 267;\\
               D.V. Nanopoulos and K. Tamvakis, 
                 Phys. Lett. {\bf B113}, (1982)151. 
\bibitem{Sen}  S. Dimopoulos and H. Georgi, 
                 Phys. Lett. {\bf 117B}, (1982) 287;\\
               K. Tabata, I.Umemura and K.Yamamoto, 
                 Prog. Theor. Phys. {\bf 71} (1984) 615;\\
               A. Sen, Phys. Lett. {\bf B148} (1984) 65.
\bibitem{su6}  S.M. Barr, Phys. Rev. D {\bf57} (1998) 190.
               G. Dvali Phys. Lett. {\bf B324} (1994) 59.


\bibitem{MP} H.Georgi, Phys. Lett. {\bf B108} (1982)283;\\
                  A. Masiero, D.V. Nanopoulos, K. Tamvakis and T.Yanagida,
                    Phys. Lett. {\bf 115} (1982) 380;\\
                  B. Grinstein, Nucl. Phys. {\bf B206} (1982) 387;
\bibitem{FlippedSU(5)} S. M. Barr, Phys. Lett. {\bf B112} (1982) 219;\\ 
                     I. Antoniadis, J. Ellis, J. Hagelin and 
                       D.V. Nanopoulos, Phys. Lett. {\bf B194} (1987) 231.
                       ; {\it ibid.} {\bf B205}, (1988) 459.


\bibitem{FN}      C.D. Froggatt and H.B. Nielsen,
                  Nucl. Phys. {\bf B147}, 277 (1979).


\bibitem{Murayama} H. Murayama and A. Pierce, Phys. Rev. D {\bf65}, 
                     055009 (2002).
\bibitem{Goto}     T. Goto and T. Nihei, Phys. Rev. D {\bf59}, 115009 (1999).
\bibitem{lattice} JLQCD Collaboration, S. Aoki et al., Phys. Rev.
                  D {\bf70}, 111501 (2004).

\bibitem{babu}    K.S. Babu and S.M. Barr, Phys. Rev. D {\bf65}, 095009 (2002).




\bibitem{BarrRaby} S.M. Barr and S. Raby, 
                   Phys. Rev. Lett. {\bf 79} (1997) 4748.


\bibitem{GlobalFit} M. Maltoni, T. Schwetz, M.A. Tortola and 
                    J.W.F. Valle, 
                    New J.Phys.{\bf6},122 (2004). 




\bibitem{ETwist}  M.~Bando and T.~Kugo, 
                    Prog. Theor. Phys. {\bf 101}, 1313 (1999). \\
                  M.~Bando, T.~Kugo and K.~Yoshioka, 
                    Prog. Theor. Phys. {\bf 104},211 (2000). 








\bibitem{su3}     N. Maekawa and Q. Shafi, Prog. Theor. Phys. {\bf 109}, 279 
                  (2003).

\bibitem{discrete} F. Wilczek and A. Zee, 
                     Phys. Lett. {\bf B70}, 418 (1977), 
                     (Erratum-ibid.{\bf72B}, 504 (1978));
                   S. Pakvasa and H. Sugawara,
                     Phys. Lett. {\bf B73}, 61 (1978) ;
                   A. De Rujula, H. Georgi and S.L. Glashow,
                     Annals Phys. {\bf109}, 258 (1977). 
\bibitem{abel}   J.K. Elwood, N. Irges, P. Ramond
                   Phys. Lett. {\bf B413}, 322 (1997) ; 
                 N. Irges, S. Lavignac, and P. Ramond, 
                   Phys. Rev. D {\bf58}, 035003(1998);
                 C.H. Albright and S. Nandi, Mod. 
                   Phys. Lett. {\bf A11}, 737 (1996); 
                   Phys. Rev. D {\bf 53}, 2699 (1996); 
                 Q. Shafi and Z. Tavartkiladze, 
                   Phys. Lett. {\bf B459}, 563 (1999); 
                   {\it ibid} {\bf B482}, 145 (2000); 
                   {\it ibid} {\bf B487}, 145 (2000); 
                   Nucl. Phys. {\bf B573}, 40 (2000);
                 Y. Nomura, T. Sugimoto, 
                   Phys. Rev. D {\bf61}, 093003 (2000);
                 K.-I. Izawa, K. Kurosawa, Y.Nomura, T.Yanagida ,
                   Phys. Rev. D {\bf60}, 115016 (1999).
\bibitem{nonabel} Z. Berezhiani, Phys. Lett. {\bf B150}, 177 (1985);
                  T. Blazek, S. Raby, and K. Tobe, 
                    Phys. Rev. D {\bf 62}, 055001 (2000);
                  R. Kitano and Y. Mimura, 
                    Phys. Rev. D {\bf 63}, 016008 (2001);
                  G.G. Ross and L. Velasco-Sevilla, 
                    Nucl. Phys. {\bf B653},3 (2003);
                  S. Raby, Phys. Lett. B {\bf 561}, 119 (2003);
                  M.-C. Chen and K.T. Mahanthappa, 
                    Phys. Rev. D {\bf 68}, 017301 (2003).
\bibitem{FCNC}   F. Borzumati and A. Masiero, Phys. Rev. Lett. {\bf 57}, 961
                 (1986); R. Barbieri, L.J. Hall, and A. Strumia, Nucl. Phys.
                 B {\bf 445}, 219 (1995); J. Hisano, T. Moroi, K. Tobe, 
                 M. Yamaguchi, and T. Yanagida, Phys. Lett. B {\bf 357}, 
                 579 (1995);
                 J. Hisano, T. Moroi, K. Tobe, and M. Yamaguchi,
                  Phys. Rev. D {\bf 53}, 2442 (1996); J. Hisano and D. Nomura,
                  Phys. Rev. D {\bf 59}, 116005 (1999);
                 J. Sato and K. Tobe, Phys. Rev. D {\bf 63}, 116010 (2001);
                 A. Masiero, S.K. Vempati, O. Vives, Nucl. Phys. B 
                 {\bf 649}, 189 (2003).
\bibitem{Sdiscrete} D.B. Kaplan and M. Schmaltz, 
                      Phys. Rev. D {\bf 49}, 3741 (1994);
                    L. J. Hall and H. Murayama, 
                      Phys. Rev. Lett. {\bf 75}, 3985 (1995) 
                    R. Dermisek, S. Raby, 
                      Phys. Rev. D {\bf 62} 015007 (2000);
                   K. Hamaguchi, M. Kakizaki and M. Yamaguchi,
                     Phys. Rev. D{\bf68}, 056007 (2003); 
                   Tatsuo Kobayashi, Jisuke Kubo and Haruhiko Terao,
                     Phys. Lett. {\bf B568}, 83 (2003). 
\bibitem{Sabel}   Y. Nir and N. Seiberg, Phys. Lett. {\bf B309}, 337 (1993);
                  Y. Nir and G. Raz, Phys. Rev. D {\bf 66}, 035007 (2002).
\bibitem{Snonabel} M. Dine, A. Kagan, and R. Leigh, Phys. Rev. D {\bf 48},
                   4269 (1993); A. Pomarol and D. Tommasini, Nucl. Phys.
                   {\bf B466}, 3 (1996); R. Barbieri, G. Dvali, and L.J. Hall,
                   Phys. Lett. {\bf B377}, 76 (1996);
                   R. Barbieri and L.J. Hall, Nuovo Cim. A {\bf 110}, 1 (1997);
                   K.S. Babu and S.M. Barr, Phys. Lett. {\bf B387}, 87 (1996);
                   R. Barbieri, L.J. Hall, S. Raby, and A. Romanino,
                     Nucl. Phys. {\bf B493}, 3 (1997);
                   Z. Berezhiani, Phys. Lett. {\bf B417}, 287 (1998);
                   G. Eyal, Phys. Lett. {\bf B441}, 191 (1998);
                   R. Barbieri, P. Creminelli, and A. Romanino, 
                     Nucl. Phys. {\bf B559}, 17 (1999);
                   S.F. King and G.G. Ross, 
                     Phys. Lett. {\bf B520}, 243 (2001).                   
\bibitem{GGMS}  F. Gabbiani, E. Gabrielli, A. Masiero and L. Silvestrini 
                  Nucl. Phys. {\bf B477}, 321 (1996).
\bibitem{SW}      S.K. Soni and H.A. Weldon, Phys. Lett. {\bf B126}, 215 (1983).
\bibitem{Okada}    R. Barbieri, L.J. Hall, and A. Romanino, Phys. Lett. 
                   {\bf B401}, 47 (1997);
                  T. Goto, Y. Okada, Y. Shimizu, T. Shindou, and M. Tanaka,
                  Phys. Rev. D {\bf 66}, 035009 (2002).
\bibitem{KMY}     Y. Kawamura, H. Murayama, and M. Yamaguchi,
                  Phys. Lett. {\bf B324}, 52 (1994);Phys.Rev. D {\bf 51}, 
                  1337 (1995).

\end{thebibliography}
\end{document}